%% file: paper.tex
\documentclass[sigconf]{acmart}
\usepackage{multirow}
\usepackage[usenames,dvipsnames]{xcolor}
\usepackage[breakable, theorems, skins]{tcolorbox}
\usepackage{booktabs,tabularx,array}
\newcolumntype{Y}{>{\raggedright\arraybackslash}X}
\newcolumntype{L}[1]{>{\raggedright\arraybackslash}p{#1}}
\usepackage[english]{babel}
\usepackage{lscape}
\usepackage{longtable}
\usepackage{ragged2e}
\usepackage[table]{xcolor}

\DeclareRobustCommand{\mybox}[2][gray!20]{
\begin{center}

\begin{tcolorbox}[
        breakable,
        left=10pt,
        right=10pt,
        top=8pt,
        bottom=8pt,
        colback=#1,
        colframe=#1,
        width=0.9\linewidth,
        enlarge left by=0mm,
        boxsep=0pt,
        arc=0pt,outer arc=0pt,
        ]
        #2
\end{tcolorbox}
\end{center}

\vspace{-0em}
}
\AtBeginDocument{%
  }
\usepackage{array}
\setcopyright{acmlicensed}
\copyrightyear{2027}
\acmYear{2027}
\acmDOI{XXXXXXX.XXXXXXX}
\acmConference[CHI '27]{ACM Conference on Human Factors in Computing Systems}{2027}{Pittsburgh, PA}
\acmISBN{978-1-4503-XXXX-X/2018/06}

\begin{document}

\title{Self-Care and Mental Health: Mapping Over A Decade of HCI Interventions}

\author{Anna M. Fang}
\affiliation{%
  \institution{Carnegie Mellon University, Human-Computer Interaction Institute}
  \city{Pittsburgh}
  \state{Pennsylvania}
  \country{USA}
}
\email{annadfang@gmail.com}

\author{Tony Wang}
\authornote{Second and third authors contributed equally.}
\affiliation{%
  \
  \institution{Cornell University, \\Department of Information Science}
  \city{Ithaca}
  \state{New York}
  \country{USA}
}
\email{yw2567@cornell.edu}

\author{Jenny Fu}
\authornotemark[1]
\affiliation{%
  \
  \institution{Columbia University, \\Data Science Institute}
  \city{New York}
  \state{New York}
  \country{USA}
}
\email{xf2318@columbia.edu}

\renewcommand{\shortauthors}{Fang, Wang \& Fu}

\begin{abstract}
Technology increasingly supports self-care for understanding and improving one's own mental health. HCI is at the center of the turn towards self-care technology, yet we lack an account of who these interventions serve, what practices they support, how technology mediates those practices, and assumptions underlying design for self-care. In order to characterize the current landscape and inform future research, we analyzed 91 SIGCHI papers that contribute HCI interventions for mental health self-care from the ACM Digital Library from 2015 through June 2026. Then, we conducted an interpretive synthesis to surface six \textit{orientations of self-care}, which describe how HCI self-care interventions constitute care through shared assumptions regarding self, care, and technology. Overall, our work provides an interconnected vocabulary for positioning HCI mental health self-care, highlights changing responsibilities of care towards users, and discusses implications for providing a more situated account of HCI self-care technology in addressing the 'general' user.
\end{abstract}



\received{September 2026}

\begin{teaserfigure}
 \centering
 \includegraphics[width=0.7\textwidth]{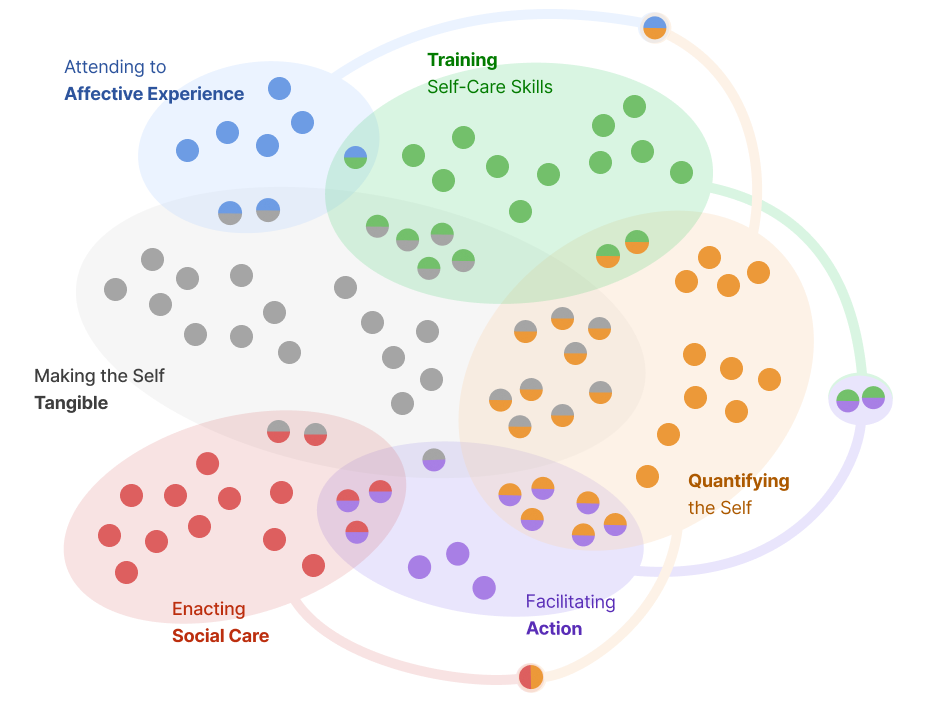}
  \caption{\textbf{Six Orientations of HCI Interventions for Mental Health \& Self-Care}, surfaced in our scoping review of 91 SIGCHI papers from 2015 to mid-2026. Papers in external rings account for overlaps that could not be represented spatially.}
  \Description{}
  \label{fig:teaser}
\end{teaserfigure}

\maketitle

\input{sections/01-intro}
\input{sections/02-background}

\input{sections/03-methods}
\input{sections/04-results}
\input{sections/05-framework}

\input{sections/06-discussion}
\input{sections/07-conclusion}

\bibliographystyle{ACM-Reference-Format}
\bibliography{ref}

% Flush main-text floats and use full-width, multipage appendix tables.
\clearpage
\onecolumn
\appendix
\section{ACM DL Query}
\label{app:acm-query}
\begingroup
\small
\ttfamily
\raggedright
([Title: "mental health" OR "mental wellbeing" OR "mental well-being" OR  "psychological health" OR "psychological wellbeing" OR "psychological well-being" OR "emotional health" OR "emotional wellbeing" OR "emotional well-being"]\par
OR [Abstract: "mental health" OR "mental wellbeing" OR "mental well-being" OR  "psychological health" OR "psychological wellbeing" OR "psychological well-being" OR "emotional health" OR "emotional wellbeing" OR "emotional well-being"]\par
OR [Keywords: "mental health" OR "mental wellbeing" OR "mental well-being" OR  "psychological health" OR "psychological wellbeing" OR "psychological well-being" OR "emotional health" OR "emotional wellbeing" OR "emotional well-being"])\par
AND\par
([Title: "intervention" OR "application" OR "app" OR "treatment" OR "care" OR "probe" OR "prototype" OR "system" OR "artifact" OR "chatbot" OR "agent" OR "tool" OR "interaction" OR "design" OR "device" OR "interface"] \par
OR [Abstract: "intervention" OR "application" OR "app" OR "treatment" OR "care" OR "probe" OR "prototype" OR "system" OR "artifact" OR "chatbot" OR "agent" OR "tool" OR "interaction" OR "design" OR "device" OR "interface"]\par
OR [Keywords: "intervention" OR "application" OR "app" OR "treatment" OR "care" OR "probe" OR "prototype" OR "system" OR "artifact" OR "chatbot" OR "agent" OR "tool" OR "interaction" OR "design" OR "device" OR "interface"]) \par
\par
\endgroup
\medskip
\textit{in ACM DL, SIGCHI-sponsored venues, full papers, since 01/01/2015}\\
Search conducted through June 17, 2026.

\clearpage
\section{Data Extraction Template}
\label{app:data-extraction}

\begingroup
\small
\setlength{\LTleft}{0pt}
\setlength{\LTright}{0pt}
\begin{longtable}{@{}*{3}{>{\raggedright\arraybackslash}p{\dimexpr(\linewidth-4\tabcolsep)/3\relax}}@{}}
\caption{Data Extraction Template}\label{tab:data-extraction}\label{tab:placeholder}\\
\toprule
\textbf{Descriptive Characteristics} & \textbf{Contribution} & \textbf{Intervention Approach} \\

\midrule

\textit{(a)} Venue \newline
\textit{(b)} Year \newline
\textit{(c)} First Author Country Affil.
&
\textit{(d)} Contribution Statement \newline
\textit{(e)} Core Contribution Category \newline
\textit{(f)} Target User Group \newline
\textit{(g)} Participant Inclusion/Exclusion \newline
\textit{(h)} Target Mental Health Issue
&
\textit{(i)} Short Description of Intervention \newline
\textit{(j)} Intervention Functions \newline
\textit{(k)} Self-Care Activity \newline
\textit{(l)} Platform
\\
\bottomrule
\end{longtable}
\endgroup

\section{Paper coding for primary findings}
\label{app:paper-coding}

\begingroup
\small
\setlength{\tabcolsep}{4pt}
\renewcommand{\arraystretch}{0.9}
\setlength{\LTleft}{0pt}
\setlength{\LTright}{0pt}
\setlength{\LTpre}{6pt}
\setlength{\LTpost}{8pt}
\begin{longtable}{@{}>{\raggedright\arraybackslash}p{0.35\linewidth}>{\raggedleft\arraybackslash}p{0.06\linewidth}>{\raggedright\arraybackslash}p{\dimexpr0.59\linewidth-4\tabcolsep\relax}@{}}
\caption{Core Contribution}\label{tab:core-contribution}\\
\rowcolor{black}
\multicolumn{3}{c}{\color{white}\fontsize{9.5pt}{11pt}\selectfont\bfseries Core Contribution} \\[2pt]
\textbf{Sub-theme} & \textbf{$N$} & \textbf{Citations} \\
\midrule
\endfirsthead
\toprule
\multicolumn{3}{c}{\textbf{Core Contribution (continued)}} \\
\midrule
\textbf{Sub-theme} & \textbf{$N$} & \textbf{Citations} \\
\midrule
\endhead
\midrule
\multicolumn{3}{r}{\footnotesize\textit{Continued on next page}} \\
\endfoot
\bottomrule
\endlastfoot
Addressing Underserved Populations \& Contexts & 32 & \cite{aldaweesh2026hasn,antle2019design,antle2019evaluating,10.1145/3706598.3713269,10.1145/3706598.3713485,choi2026daily,chow2023feeling,collins2022covid,elvitigala2021stressshoe,guo2025,10.1145/3706598.3713699,jin2024exploring,kim2023routineaid,kitson2023co,mccarren2026exploring,miri2022far,park2024collaborative,10.1145/3571884.3597142,potapov2020lifemosaic,10.1145/3718084,rubin2015towards,ryu2020simple,seo2024chacha,simm2016anxiety,speer2021mindfulnest,sun2025conversations,surani2026co,thieme2016challenges,10.1145/3544548.3581319,wei2025,10.1145/3613904.3642369,10.1145/3706598.3714277} \\
Scalability & 5 & \cite{antle2019evaluating,doherty2019engagement,elvitigala2021stressshoe,simm2016anxiety,yu2025creatively} \\
Engagement \& Adherence & 15 & \cite{barkercanler2024flexible,10.1145/3772318.3790933,10.1145/3544548.3581188,konrad2015finding,10.1145/3772318.3791591,matheus2024ommie,10.1145/3613904.3642662,park2021wrote,roo2017inner,schroeder2018pocket,10.1145/3613904.3642761,soler2024arcadia,spitale2025vita,10.1145/3772318.3791817,yu2017stresstree} \\
Personalization and Context-Awareness & 22 & \cite{barkercanler2024flexible,bhattacharjee2023investigating,chandra2026tech,10.1145/3706598.3713269,delatorre2025sonora,huang2015emotion,10.1145/3706598.3713730,10.1145/3613904.3642766,konrad2015finding,10.1145/3715336.3735795,10.1145/3699761,rohani2020mubs,10.1145/3706598.3713883,spitale2025vita,10.1145/3544548.3581319,vossen2024effect,10.1145/3643834.3661570,wagener2022mood,10.1145/3772318.3791817,zhang2026asafeplace,10.1145/3706598.3713453,10.1145/3745900.3746099} \\
Shifting to Everyday Contexts & 14 & \cite{barkercanler2024flexible,10.1145/3706598.3713485,10.1145/3772318.3790933,kim2025lino,koch2021taking,lee2019caring,10.1145/3613904.3642662,10.1145/3643834.3660702,sakel2024social,10.1145/3706598.3713883,tan2023mindful,10.1145/3643834.3661570,wagener2023selvreflect,yan2022emoglass} \\
Translating Clinical Care Beyond Clinical Settings & 9 & \cite{hamid2022you,10.1145/3613904.3642937,10.1145/3772318.3790406,matthews2015situ,rohani2020mubs,schroeder2018pocket,soler2024arcadia,10.1145/3613904.3643054,thieme2016challenges} \\
Safety and Trust & 7 & \cite{10.1145/3706598.3713485,10.1145/3613904.3642937,kim2022prediction,10.1145/3715336.3735795,surani2026co,10.1145/3772318.3791817,10.1145/3613904.3642369} \\
Enhancing Self-Care Activity through Novel Interaction & 14 & \cite{cai2023listen,choi2026daily,10.1145/3800645.3813019,lee2017designing,lee2019caring,10.1145/3772318.3791615,10.1145/3563657.3595998,10.1145/3715336.3735845,prpa2018attending,rajcic2020mirror,rasch2024mind,schneeberger2021stress,woo2024notalone,zhang2026vrcalmplus} \\
\end{longtable}

\par\endgroup

\begingroup
\small
\setlength{\tabcolsep}{4pt}
\renewcommand{\arraystretch}{0.9}
\setlength{\LTleft}{0pt}
\setlength{\LTright}{0pt}
\setlength{\LTpre}{6pt}
\setlength{\LTpost}{8pt}
\begin{longtable}{@{}>{\raggedright\arraybackslash}p{0.35\linewidth}>{\raggedleft\arraybackslash}p{0.06\linewidth}>{\raggedright\arraybackslash}p{\dimexpr0.59\linewidth-4\tabcolsep\relax}@{}}
\caption{Target User Group}\label{tab:target-user-group}\\
\rowcolor{black}
\multicolumn{3}{c}{\color{white}\fontsize{9.5pt}{11pt}\selectfont\bfseries Target User Group} \\[2pt]
\textbf{Population} & \textbf{$N$} & \textbf{Citations} \\
\midrule
\endfirsthead
\rowcolor{black}
\multicolumn{3}{c}{\color{white}\fontsize{9.5pt}{11pt}\selectfont\bfseries Target User Group (continued)} \\[2pt]
\textbf{Population} & \textbf{$N$} & \textbf{Citations} \\
\midrule
\endhead
\midrule
\multicolumn{3}{r}{\footnotesize\textit{Continued on next page}} \\
\endfoot
\bottomrule
\endlastfoot
People with depression & 3 & \cite{10.1145/3613904.3642937,10.1145/3715336.3735795,rohani2020mubs} \\
People with personality disorder & 1 & \cite{thieme2016challenges} \\
Patients with schizophrenia spectrum disorders & 1 & \cite{10.1145/3613904.3642369} \\
People with bipolar disorder & 1 & \cite{matthews2015situ} \\
People with panic disorder & 1 & \cite{rubin2015towards} \\
People with disordered eating or eating disorders & 2 & \cite{10.1145/3706598.3713485,10.1145/3613904.3643054} \\
Complex, difficult-to-treat disorders & 1 & \cite{schroeder2018pocket} \\
Mental health patients & 1 & \cite{soler2024arcadia} \\
People with ADHD & 1 & \cite{park2024collaborative} \\
Autistic people & 3 & \cite{kim2023routineaid,miri2022far,simm2016anxiety} \\
People with learning disability & 1 & \cite{thieme2016challenges} \\
Deaf and hard of hearing & 1 & \cite{choi2026daily} \\
Children and youth & 10 & \cite{antle2019design,antle2019evaluating,kitson2023co,miri2022far,park2024collaborative,10.1145/3571884.3597142,potapov2020lifemosaic,seo2024chacha,speer2021mindfulnest,sun2025conversations} \\
Young adults & 1 & \cite{bhattacharjee2023investigating}\\
Older adults & 7 & \cite{guo2025,jin2024exploring,mccarren2026exploring,ryu2020simple,surani2026co,wei2025,10.1145/3706598.3714277} \\
Women & 5 & \cite{aldaweesh2026hasn,doherty2019engagement,10.1145/3613904.3643054,sun2025conversations,thieme2016challenges} \\
Non-western or Non-WEIRD & 2 & \cite{aldaweesh2026hasn,sun2025conversations} \\
Low SES & 2 & \cite{antle2019design,antle2019evaluating} \\
University students & 10 & \cite{chandra2026tech,10.1145/3706598.3713269,hamid2022you,kim2022prediction,lee2017designing,10.1145/3699761,10.1145/3718084,vossen2024effect,woo2024notalone,yu2025creatively} \\
Workers & 3 & \cite{chow2023feeling,elvitigala2021stressshoe,10.1145/3544548.3581319} \\
Caregivers & 1 & \cite{10.1145/3706598.3713699} \\
General population & 43 & \cite{barkercanler2024flexible,cai2023listen,collins2022covid,delatorre2025sonora,10.1145/3772318.3790933,10.1145/3544548.3581188,huang2015emotion,10.1145/3800645.3813019,10.1145/3706598.3713730,10.1145/3613904.3642766,kim2025lino,koch2021taking,konrad2015finding,lee2019caring,10.1145/3772318.3790406,10.1145/3772318.3791591,matheus2024ommie,10.1145/3772318.3791615,10.1145/3563657.3595998,10.1145/3613904.3642662,10.1145/3715336.3735845,park2021wrote,10.1145/3643834.3660702,prpa2018attending,rajcic2020mirror,rasch2024mind,roo2017inner,sakel2024social,schneeberger2021stress,10.1145/3613904.3642761,10.1145/3706598.3713883,spitale2025vita,tan2023mindful,10.1145/3643834.3661570,wagener2022mood,wagener2023selvreflect,10.1145/3772318.3791817,yan2022emoglass,yu2017stresstree,zhang2026asafeplace,zhang2026vrcalmplus,10.1145/3706598.3713453,10.1145/3745900.3746099} \\
\end{longtable}

\par\endgroup

\begingroup
\small
\setlength{\tabcolsep}{4pt}
\renewcommand{\arraystretch}{0.9}
\setlength{\LTleft}{0pt}
\setlength{\LTright}{0pt}
\setlength{\LTpre}{6pt}
\setlength{\LTpost}{8pt}
\begin{longtable}{@{}>{\raggedright\arraybackslash}p{0.35\linewidth}>{\raggedleft\arraybackslash}p{0.06\linewidth}>{\raggedright\arraybackslash}p{\dimexpr0.59\linewidth-4\tabcolsep\relax}@{}}
\caption{Mental Health Topics}\label{tab:mental-health-topics}\\
\rowcolor{black}
\multicolumn{3}{c}{\color{white}\fontsize{9.5pt}{11pt}\selectfont\bfseries Mental Health Topics} \\[2pt]
\textbf{Target issue} & \textbf{$N$} & \textbf{Citations} \\
\midrule
\endfirsthead
\toprule
\multicolumn{3}{c}{\textbf{Mental Health Topics (continued)}} \\
\midrule
\textbf{Target issue} & \textbf{$N$} & \textbf{Citations} \\
\midrule
\endhead
\midrule
\multicolumn{3}{r}{\footnotesize\textit{Continued on next page}} \\
\endfoot
\bottomrule
\endlastfoot
General Well-being & 17 & \cite{aldaweesh2026hasn,bhattacharjee2023investigating,hamid2022you,10.1145/3706598.3713699,jin2024exploring,kim2023routineaid,lee2019caring,10.1145/3772318.3791591,park2024collaborative,10.1145/3571884.3597142,potapov2020lifemosaic,10.1145/3613904.3642761,spitale2025vita,surani2026co,yu2025creatively,10.1145/3706598.3713453,10.1145/3745900.3746099} \\
Emotion: Awareness and Reflection & 17 & \cite{barkercanler2024flexible,cai2023listen,10.1145/3706598.3713269,choi2026daily,huang2015emotion,10.1145/3800645.3813019,kim2025lino,10.1145/3772318.3790406,10.1145/3699761,10.1145/3613904.3642662,10.1145/3715336.3735845,park2021wrote,rajcic2020mirror,rasch2024mind,10.1145/3706598.3713883,wagener2023selvreflect,yan2022emoglass} \\
Emotion: Regulation & 16 & \cite{choi2026daily,10.1145/3544548.3581188,huang2015emotion,kitson2023co,koch2021taking,10.1145/3772318.3791615,miri2022far,10.1145/3715336.3735845,rasch2024mind,schroeder2018pocket,soler2024arcadia,speer2021mindfulnest,thieme2016challenges,10.1145/3643834.3661570,wagener2022mood,yan2022emoglass} \\
Stress/Relaxation & 15 & \cite{10.1145/3706598.3713269,chow2023feeling,delatorre2025sonora,elvitigala2021stressshoe,10.1145/3706598.3713730,10.1145/3613904.3642766,kim2022prediction,konrad2015finding,10.1145/3563657.3595998,schneeberger2021stress,sun2025conversations,10.1145/3544548.3581319,yu2017stresstree,zhang2026asafeplace,zhang2026vrcalmplus} \\
Mindfulness or Bodily Awareness & 14 & \cite{chandra2026tech,10.1145/3772318.3790933,10.1145/3544548.3581188,koch2021taking,mccarren2026exploring,10.1145/3563657.3595998,prpa2018attending,roo2017inner,schroeder2018pocket,10.1145/3613904.3643054,tan2023mindful,thieme2016challenges,10.1145/3772318.3791817,yu2017stresstree} \\
Emotion: Expression & 10 & \cite{cai2023listen,choi2026daily,guo2025,kim2025lino,lee2017designing,park2021wrote,seo2024chacha,sun2025conversations,10.1145/3643834.3661570,wagener2023selvreflect} \\
Anxiety & 9 & \cite{antle2019design,antle2019evaluating,matheus2024ommie,10.1145/3718084,rubin2015towards,ryu2020simple,schroeder2018pocket,simm2016anxiety,vossen2024effect} \\
Social Connectedness or Loneliness & 8 & \cite{collins2022covid,guo2025,10.1145/3563657.3595998,10.1145/3643834.3660702,sakel2024social,wei2025,woo2024notalone,10.1145/3706598.3714277} \\
Depression & 7 & \cite{doherty2019engagement,10.1145/3613904.3642937,10.1145/3715336.3735795,rohani2020mubs,ryu2020simple,schroeder2018pocket,vossen2024effect} \\
Bipolar & 1 & \cite{matthews2015situ} \\
Eating Disorders & 1 & \cite{10.1145/3706598.3713485} \\
Schizophrenia & 1 & \cite{10.1145/3613904.3642369} \\
\end{longtable}
\par\endgroup

\begingroup
\small
\setlength{\tabcolsep}{4pt}
\renewcommand{\arraystretch}{0.9}
\setlength{\LTleft}{0pt}
\setlength{\LTright}{0pt}
\setlength{\LTpre}{6pt}
\setlength{\LTpost}{8pt}
\begin{longtable}{@{}>{\raggedright\arraybackslash}p{0.35\linewidth}>{\raggedleft\arraybackslash}p{0.06\linewidth}>{\raggedright\arraybackslash}p{\dimexpr0.59\linewidth-4\tabcolsep\relax}@{}}
\caption{Activities}\label{tab:activities}\\
\rowcolor{black}
\multicolumn{3}{c}{\color{white}\fontsize{9.5pt}{11pt}\selectfont\bfseries Activities} \\[2pt]
\textbf{Activity} & \textbf{$N$} & \textbf{Citations} \\
\midrule
\endfirsthead
\rowcolor{black}
\multicolumn{3}{c}{\color{white}\fontsize{9.5pt}{11pt}\selectfont\bfseries Activities (continued)} \\[2pt]
\textbf{Activity} & \textbf{$N$} & \textbf{Citations} \\
\midrule
\endhead
\midrule
\multicolumn{3}{r}{\footnotesize\textit{Continued on next page}} \\
\endfoot
\bottomrule
\endlastfoot
Conversing and social support & 25 & \cite{aldaweesh2026hasn,bhattacharjee2023investigating,cai2023listen,10.1145/3706598.3713485,choi2026daily,collins2022covid,guo2025,10.1145/3613904.3642937,lee2019caring,10.1145/3715336.3735795,10.1145/3772318.3791591,mccarren2026exploring,park2021wrote,10.1145/3571884.3597142,ryu2020simple,schneeberger2021stress,seo2024chacha,10.1145/3613904.3642761,10.1145/3706598.3713883,spitale2025vita,surani2026co,vossen2024effect,wei2025,10.1145/3706598.3714277,10.1145/3706598.3713453} \\
(Self-)Tracking & 22 & \cite{barkercanler2024flexible,bhattacharjee2023investigating,10.1145/3706598.3713269,chow2023feeling,doherty2019engagement,elvitigala2021stressshoe,huang2015emotion,10.1145/3706598.3713730,10.1145/3613904.3642766,kim2025lino,kim2022prediction,lee2017designing,matthews2015situ,10.1145/3699761,10.1145/3613904.3642662,potapov2020lifemosaic,rohani2020mubs,rubin2015towards,sakel2024social,simm2016anxiety,yan2022emoglass,10.1145/3613904.3642369} \\
Meditation/Mindfulness & 13 & \cite{chandra2026tech,10.1145/3772318.3790933,koch2021taking,konrad2015finding,10.1145/3563657.3595998,prpa2018attending,10.1145/3718084,roo2017inner,10.1145/3613904.3643054,sun2025conversations,tan2023mindful,thieme2016challenges,10.1145/3772318.3791817} \\
Sensory and physiological stimulation & 13 & \cite{antle2019evaluating,10.1145/3544548.3581188,matheus2024ommie,miri2022far,10.1145/3643834.3660702,schneeberger2021stress,simm2016anxiety,soler2024arcadia,speer2021mindfulnest,thieme2016challenges,yu2017stresstree,zhang2026asafeplace,zhang2026vrcalmplus} \\
Journaling & 11 & \cite{kim2025lino,10.1145/3613904.3642937,10.1145/3772318.3791615,10.1145/3699761,10.1145/3715336.3735845,park2021wrote,seo2024chacha,10.1145/3706598.3713883,yan2022emoglass,yu2025creatively,10.1145/3745900.3746099} \\
Artistic/creative expression & 7 & \cite{10.1145/3800645.3813019,rajcic2020mirror,sun2025conversations,10.1145/3643834.3661570,wagener2022mood,wagener2023selvreflect,zhang2026asafeplace} \\
Auditory (music, calming sounds) & 6 & \cite{cai2023listen,choi2026daily,delatorre2025sonora,jin2024exploring,koch2021taking,10.1145/3613904.3643054} \\
Treatment program & 4 & \cite{konrad2015finding,schroeder2018pocket,10.1145/3544548.3581319,yan2022emoglass} \\
Goal setting and planning & 3 & \cite{kim2023routineaid,konrad2015finding,park2024collaborative} \\
Gaming & 3 & \cite{antle2019design,hamid2022you,soler2024arcadia} \\
Embodied and metaphorical actions & 3 & \cite{kitson2023co,10.1145/3772318.3790406,rasch2024mind} \\
Reminiscence & 2 & \cite{thieme2016challenges,10.1145/3706598.3714277} \\
Informational resources & 2 & \cite{10.1145/3706598.3713699,woo2024notalone} \\
\end{longtable}

\par\endgroup

\begingroup
\small
\setlength{\tabcolsep}{4pt}
\renewcommand{\arraystretch}{0.9}
\setlength{\LTleft}{0pt}
\setlength{\LTright}{0pt}
\setlength{\LTpre}{6pt}
\setlength{\LTpost}{8pt}
\begin{longtable}{@{}>{\raggedright\arraybackslash}p{0.35\linewidth}>{\raggedleft\arraybackslash}p{0.06\linewidth}>{\raggedright\arraybackslash}p{\dimexpr0.59\linewidth-4\tabcolsep\relax}@{}}
\caption{Six Orientations of Self-Care in HCI}\label{tab:self-care-orientations}\\
\rowcolor{black}
\multicolumn{3}{c}{\color{white}\fontsize{9.5pt}{11pt}\selectfont\bfseries Six Orientations of Self-Care in HCI} \\[2pt]
\textbf{Orientation} & \textbf{$N$} & \textbf{Citations} \\
\midrule
\endfirsthead
\rowcolor{black}
\multicolumn{3}{c}{\color{white}\fontsize{9.5pt}{11pt}\selectfont\bfseries Six Orientations of Self-Care in HCI (continued)} \\[2pt]
\textbf{Orientation} & \textbf{$N$} & \textbf{Citations} \\
\midrule
\endhead
\midrule
\multicolumn{3}{r}{\footnotesize\textit{Continued on next page}} \\
\endfoot
\bottomrule
\endlastfoot
Making the Self Tangible & 34 & \cite{antle2019design,antle2019evaluating,cai2023listen,choi2026daily,10.1145/3772318.3790933,10.1145/3544548.3581188,10.1145/3800645.3813019,kim2025lino,10.1145/3613904.3642937,kitson2023co,lee2017designing,10.1145/3772318.3790406,10.1145/3772318.3791615,10.1145/3699761,10.1145/3715336.3735845,park2021wrote,prpa2018attending,rajcic2020mirror,rasch2024mind,roo2017inner,schneeberger2021stress,soler2024arcadia,10.1145/3706598.3713883,speer2021mindfulnest,10.1145/3613904.3643054,sun2025conversations,thieme2016challenges,10.1145/3643834.3661570,wagener2022mood,wagener2023selvreflect,yu2017stresstree,yu2025creatively,zhang2026asafeplace,zhang2026vrcalmplus} \\
Quantifying the Self & 29 & \cite{barkercanler2024flexible,chandra2026tech,10.1145/3706598.3713269,chow2023feeling,doherty2019engagement,elvitigala2021stressshoe,10.1145/3772318.3790933,10.1145/3544548.3581188,huang2015emotion,10.1145/3706598.3713730,10.1145/3613904.3642766,kim2023routineaid,kim2022prediction,matthews2015situ,10.1145/3699761,10.1145/3613904.3642662,potapov2020lifemosaic,prpa2018attending,rohani2020mubs,roo2017inner,rubin2015towards,sakel2024social,simm2016anxiety,soler2024arcadia,speer2021mindfulnest,yan2022emoglass,10.1145/3613904.3642369,yu2017stresstree,zhang2026vrcalmplus} \\
Training Self-Care Skills & 21 & \cite{antle2019design,antle2019evaluating,chandra2026tech,chow2023feeling,hamid2022you,10.1145/3706598.3713699,kitson2023co,konrad2015finding,matheus2024ommie,mccarren2026exploring,miri2022far,park2021wrote,10.1145/3571884.3597142,10.1145/3718084,schneeberger2021stress,schroeder2018pocket,10.1145/3613904.3642761,spitale2025vita,tan2023mindful,vossen2024effect,10.1145/3772318.3791817} \\
Enacting Social Care & 18 & \cite{aldaweesh2026hasn,bhattacharjee2023investigating,10.1145/3706598.3713485,collins2022covid,guo2025,10.1145/3613904.3642937,lee2019caring,10.1145/3715336.3735795,10.1145/3772318.3790406,10.1145/3772318.3791591,10.1145/3643834.3660702,ryu2020simple,seo2024chacha,surani2026co,wei2025,woo2024notalone,10.1145/3706598.3714277,10.1145/3706598.3713453} \\
Facilitating Actions & 14 & \cite{bhattacharjee2023investigating,10.1145/3706598.3713485,elvitigala2021stressshoe,huang2015emotion,10.1145/3706598.3713730,kim2022prediction,mccarren2026exploring,10.1145/3772318.3791615,park2024collaborative,rohani2020mubs,seo2024chacha,10.1145/3544548.3581319,10.1145/3613904.3642369,10.1145/3745900.3746099} \\
Attending to Affective Experience & 10 & \cite{delatorre2025sonora,jin2024exploring,koch2021taking,10.1145/3563657.3595998,10.1145/3643834.3660702,simm2016anxiety,thieme2016challenges,vossen2024effect,zhang2026asafeplace,10.1145/3706598.3714277} \\
\end{longtable}

\par\endgroup

\clearpage
\section{PRISMA-ScR Checklist}
\label{app:prisma-scr}
\begingroup
\footnotesize
\setlength{\tabcolsep}{4pt}
\renewcommand{\arraystretch}{1.1}
\setlength{\LTleft}{0pt}
\setlength{\LTright}{0pt}
\begin{longtable}{@{}L{0.17\linewidth}L{0.05\linewidth}L{0.49\linewidth}L{\dimexpr0.29\linewidth-6\tabcolsep\relax}@{}}
\caption{PRISMA-ScR Checklist}\label{tab:prisma-scr}\\
\toprule
\textbf{Section} & \textbf{No.} & \textbf{PRISMA-ScR checklist item} & \textbf{Report Location} \\
\midrule
\endfirsthead
\multicolumn{4}{c}{\textbf{PRISMA-ScR Checklist (continued)}} \\
\toprule
\textbf{Section} & \textbf{No.} & \textbf{PRISMA-ScR checklist item} & \textbf{Report Location} \\
\midrule
\endhead
\midrule
\multicolumn{4}{r}{\textit{Continued on next page}} \\
\endfoot
\bottomrule
\endlastfoot

\textbf{TITLE} & 1 &
Identify the report as a scoping review. &
Introduction (Section 1), Abstract. We make an exception here due to a primary focus of our paper being a mapping framework emergent--but not part of--the review. We identify the scoping review in Abstract, Introduction, keywords, and throughout the paper. \\

\midrule
\textbf{ABSTRACT} & 2 &
Provide a structured summary that includes, as applicable: background, objectives, eligibility criteria, sources of evidence, charting methods, results, and conclusions related to the review questions and objectives. &
Abstract (Page 1) \\

\midrule
\textbf{INTRODUCTION} & 3 &
Describe the rationale for the review in the context of what is already known. Explain why the review questions/objectives lend themselves to a scoping review approach. &
Introduction (Section 1), Background (Section 2) \\

& 4 &
Provide an explicit statement of the questions and objectives being addressed, with reference to their key elements (e.g., population or participants, concepts, and context) or other relevant elements used to conceptualize the review questions and/or objectives. &
Introduction (Section 1) \\

\midrule
\textbf{METHODS} & 5 &
Indicate whether a review protocol exists; state if and where it can be accessed and, if available, provide registration information, including the registration number. &
Methods (Section 3) \\

& 6 &
Specify characteristics of the sources of evidence used as eligibility criteria (e.g., years considered, language, and publication status), and provide a rationale. &
Methods: Stage 2 (Section 3.2) \\

& 7 &
Describe all information sources in the search (e.g., databases with dates of coverage and contact with authors to identify additional sources), as well as the date the most recent search was executed. &
Methods: Stage 2 (Section 3.2) \\

& 8 &
Present the full electronic search strategy for at least one database, including any limits used, such that it could be repeated. &
Methods: Stage 2 (Section 3.2) and Appendix~\ref{app:acm-query} \\

& 9 &
State the process for selecting sources of evidence (i.e., screening and eligibility) included in the scoping review. &
Methods: Stage 3 (Section 3.3) \\

& 10 &
Describe the methods of charting data from the included sources of evidence (e.g., calibrated or tested forms, whether data charting was done independently or in duplicate) and any processes for obtaining and confirming data from investigators. &
Methods: Stage 4 (Section 3.4) \\

& 11 &
List and define all variables for which data were sought and any assumptions and simplifications made. &
Methods: Stage 4 (Section 3.4) \\

& 12 &
If done, provide a rationale for conducting a critical appraisal of included sources of evidence; describe the methods used and how this information was used in any data synthesis. &
N/A \\

& 13 &
Describe the methods of handling and summarizing the data that were charted. &
Methods: Stage 5 (Section 3.5) \\

\midrule
\textbf{RESULTS} & 14 &
Give numbers of sources of evidence screened, assessed for eligibility, and included in the review, with reasons for exclusions at each stage, ideally using a flow diagram. &
Methods (Figure \ref{fig:inclusion-diagram}) and Results (Section 4)\\

& 15 &
For each source of evidence, present characteristics for which data were charted and provide the citations. &
Results (Section 4) \\

& 16 &
If done, present data on critical appraisal of included sources of evidence (see Item 12). &
N/A \\

& 17 &
For each included source of evidence, present the relevant data that were charted that relate to the review questions and objectives. &
Results (Section 4) \\

& 18 &
Summarize and/or present the charting results as they relate to the review questions and objectives. &
Results (Section 4) \\

\midrule
\textbf{DISCUSSION} & 19 &
Summarize the main results, including an overview of concepts, themes, and types of evidence available; link to the review questions and objectives; and consider relevance to key groups. &
Framework (Section 5), Discussion (Section 6) \\

& 20 &
Discuss the limitations of the scoping review process. &
Discussion (Section 6) \\

& 21 &
Provide a general interpretation of the results with respect to the review questions and objectives, as well as potential implications and/or next steps. &
Discussion (Section 6) \\

\midrule
\textbf{FUNDING} & 22 &
Describe sources of funding for the included sources of evidence, as well as sources of funding for the scoping review. Describe the role of the funders of the scoping review. &
Acknowledgements (after Section 7 Conclusion) \\

\end{longtable}
\endgroup

\end{document}

%% file: sections/01-intro.tex
\section{Introduction}

A longstanding and fundamental part of mental health is \textit{self-care} -- the activities individuals undertake in understanding, maintaining, and improving their own health \cite{world2009self,levin1983self}. Self-care is particularly consequential for mental health given the substantial burden of mental health conditions worldwide; 1.2 billion people live with mental health conditions, while persistent barriers including cost, limited availability, and stigma continue to constrain access to formal services \cite{santomauro2026updated,who2025world,patel2018lancet,andrade2014barriers}. At the same time, mental health is increasingly understood not simply as the presence or absence of disorder, but as a multidimensional and ubiquitous aspect of everyday life \cite{world2022world,keyes2002mental, fang2025social}. Experiences such as stress, anxiety, loneliness, and emotion regulation challenges are transdiagnostic and common, preceding or occurring independently of clinical diagnoses \cite{keyes2002mental}. Thus, self-care for mental health has grown as a practice across the spectrum of mental health needs, from everyday coping and well-being to ongoing management of chronic or clinical conditions.

Human-Computer Interaction (HCI) is at the center of this shift towards self-care as digital technologies are increasingly intertwined with how people care for their own mental health outside traditional care settings \cite{ahmed2021mobile,nunes2015self}. Over several decades, HCI has developed mobile mental health applications, wearable sensing systems, conversational agents and AI-mediated therapy, and more, which are used independently for self-directed support~\cite{torous2025evolving,balcombe2022human}. Advances in large language models and embodied interaction have only further expanded the design space of HCI for self-care given greater ability for adaptive and personalized interactions~\cite{stade2024large,10.1145/3780045.3780052}. Collectively, this work positions HCI as an active mediator of how people practice self-care. 

Despite this rich breadth of work, we lack systematic understanding of how HCI has conceptualized and designed interventions for self-care and mental health. It remains unclear who and what these interventions commonly address, what forms of self-care activities they support, and what motivates these interventions. These questions are particularly consequential because (HCI) design values become embedded in self-care interventions \cite{sengers2005reflective}, which often reach people directly, outside clinical settings or supervision, and without evaluation standards expected in other fields regarding therapeutic efficacy. Therefore, comprehensive mapping of these interventions is necessary for shared understanding of the current landscape and which conceptions of self-care are prioritized and enacted.

Specifically, we address the following research questions:

\begin{enumerate} 
\item[RQ 1] \textbf{Who} and \textbf{what mental health needs} are addressed by HCI interventions for self-care \& mental health?
\item[RQ 2] What \textbf{self-care practices} are supported by HCI interventions for self-care \& mental health?
\item[RQ 3] What \textbf{core contributions} do papers frame their interventions around in regards to self-care for mental health?
\end{enumerate}

We answer these questions through a scoping review of 91 SIGCHI papers on HCI interventions for mental health self-care. Then, grounded in this synthesis, we surface ways that HCI conceptualizes self-care through \textbf{six orientations that map interventions' shared assumptions} of what aspects of the self are to be cared for, how to care for those aspects, and the role technology assumes in supporting that care. We therefore highlight how HCI design has approached the relationship between the \textit{self}, \textit{care}, and \textit{technology}.
% \begin{enumerate}

% \item[\textbf{RQ 5}] What are the different \textbf{orientations of self-care in HCI}, and what roles does technology assume in enacting these?
% \end{enumerate}

This paper makes two contributions: (1) an overview of HCI interventions for mental health self-care, characterizing target populations, target mental health topics, and self-care activities, and (2) six orientations of self-care that identify how technologies enact care: by \textit{making experience tangible, \textit{quantifying the self}, \textit{training self-care skills}, \textit{facilitating action}, \textit{enacting social care}, or \textit{attending directly to affective experience}}. Together, these contributions offer a vocabulary for positioning and navigating self-care research in HCI, and help the community collectively understand \textit{what} forms of mental health self-care HCI currently supports, \textit{for whom}, \textit{how}, and \textit{with what assumptions}. In doing so, we also highlight the forms of mental health self-care HCI enables, privileges, and leaves underexplored.

%% file: sections/02-background.tex
\section{Background}
\label{background}
In order to distinguish our contribution as well as ground key definitions used in our review, we draw on prior work in HCI for mental health as well as self-care literature from nursing and medical scholarship.

\subsection{Self-Care: Conceptualizations and Tensions}
\label{sec:rw-mentalhealth}

Self-care has existed from before many of today's healthcare systems were established \cite{mccormack2003examination}. The desire to engage in self-care is inherent to the human and health experience, and has persisted even as care has become increasingly oriented towards professional and acute settings \cite{martinez2021self,kickbusch1989self}. Inspired from women's and wellness movements, the focus of mental health has continued to shift towards increasing autonomy, support and community, and self-agency \cite{kickbusch1989self}. 

Within nursing and chronic-care research, self-care has commonly been understood as individuals' capacity to recognize and respond to their own health needs~\cite{riegel2021self, orem1995nursing}. However, the degree of independence that self-care requires varies across literature. While some definitions emphasize that self-care can be conducted alone or in coordination with others \cite{nunes2015self, who2026selfcare}, other literature implies that self-care is a lone activity; the recipient of care is the sole initiator and actor as an independent entity, without involvement from other people \cite{levin1983self, godfrey2011care}, and care is reliant only on the recipient's capabilities \cite{sousa12002conceptual}. Consistent throughout these varied framings of self-care, though, is that the individual is more than a recipient of care and instead takes on an active role in initiating and continuing their own ongoing care \cite{nunes2015self, levin2026self}. In our review, we accordingly treat self-care as a property of the user's agency in executing self-care activity, which may be considered anything from the management of diagnosed conditions to the maintenance or improvement of general well-being.
%; the same intervention practice (e.g., mood tracking) may even be positioned along this spectrum, from symptom management to a general wellness practice \cite{godfrey2011care,mamukashvili2023long,parsons2017home}.

We note that self-care is simultaneously a technical and normative consideration. Critical scholarship in sociology and public health has argued that the contemporary wellness industry can translate structural determinants of health -- inequality, overwork, discrimination, inadequate social infrastructure -- into individual problems amenable to personal optimization \cite{cederstrom2015wellness, crawford1980healthism}. Furthermore, although self-governed interventions are framed as personalized, they still encode designers' decisions and remain susceptible to biases -- for example, Western and clinical assumptions that may not translate across contexts~\cite{pendse2021can}. Schull~\cite{schull2016data} and others have raised concerns about the quantification and gamification of self-care, and the ``responsibilization'' of mental healthcare to the individual; technology can amplify, rather than mitigate, these dynamics by framing well-being as a set of metrics or goals to be achieved through self-reliant behaviors~\cite{lupton2016quantified,ruckenstein2017datafication}. As a result, (HCI) design of self-care technology encodes assumptions about -- and marks a shift in -- what individuals carry responsibility for.

\subsection{HCI Research in Self-Care and Mental Health}
\label{sec:related-reviews}

Driven both by the scale of the global mental health crisis and by rapid technical advancements~\cite{inkster2018empathy, torous2025evolving}, HCI has contributed to mental health research for several decades through a plethora of digital mental health interventions (DMHIs). These range from persuasive and behavior-change technologies~\cite{fogg2003persuasive,consolvo2006design}, to tracking and sensing technologies~\cite{wang2018tracking, sanches2019hci}, and, more recently, conversational agents for mental health support~\cite{li2023systematic}. The breadth of HCI in mental health can be seen through existing reviews and design frameworks that have addressed various technological approaches \cite{thieme2020machine,ahmed2021mobile}, populations \cite{10.1145/3805040, 10.1145/3780045.3780052}, therapeutic approaches \cite{knowles2014qualitative,balcombe2022human, coyle2007computers}, and design and evaluation strategies \cite{slovak2023designing, sanches2019hci}. Prior reviews have also discussed care from the perspective of ethico-political dimensions of HCI perspectives \cite{petterson2025expanding} and care's positioning in HCI (e.g., as management of conditions, as politics and advocacy) \cite{wang2026caring}; however, this prior work has not focused on self-care and mental health nor more direct relationships in HCI between self, care, and technological support.

HCI scholarship at the intersection of mental health, self-care, and technology spans various epistemological traditions, methods, and conceptions of self-care that necessitate additional frameworks for synthesis. For example, Kou et al. \cite{kou2019turn} noted that both personal informatics and online mental health communities research address concepts of the self in different ways, necessitating a broader framework for understanding the \textit{"turn to self"} in HCI. In terms of technology design, Slovak et al. \cite{slovak2023designing} surveyed HCI research on emotion regulation interventions---many of which are used in self-care contexts---to note three levels of considerations for technology designers: theory-informed decisions utilizing psychological frameworks, strategic intervention delivery, and practical decisions around design. The HCI community has also hosted a range of initiatives to understand how HCI in health can extend past method and technical boundaries \cite{agapie2024hcipublichealth, avellino2025envisioning, epstein2023symposium}; our review seeks to build upon such efforts by addressing conceptualizations across the field regarding self-care interventions.

Nunes et al.’s 2015 review of self-care technologies for chronic conditions provides the most direct precedent for examining self-care as an organizing concept within HCI~\cite{nunes2015self}. Their review examined how people use technology in self-care practices, analyzing the literature through roles, relationships, technologies, research methods, and design tensions. Beyond individual systems, self-care technology is rooted in negotiation of what users should do, what technology should do, and what role formal healthcare should retain \cite{nunes2015self}. Yet mental health and self-care research in HCI has had relatively limited attention in earlier literature; of the 29 studies reviewed across 30 years, Nunes et al.'s review only included three papers focused on mental health. We therefore provide a HCI review focused specifically on mental health and self-care, examining work published in the last decade since Nunes et al.'s review to examine the present landscape. This review is particularly timely given rapid expansion of conversational agents, AI, and immersive environments applied to mental health. Given the growing intersection of technology mediating self-care and consequently the inherent redefining of responsibilities that now lie with users and systems, self-care is now as much an HCI problem as a healthcare problem. 
%just saving space, could add back.
%Recent interventions inherently redefine what users should monitor, change, or mitigate and are worth a current review of the literature.

%% file: sections/03-methods.tex
\section{Methodology}
\label{methods}
\begin{figure*}
    \includegraphics[width=0.7\linewidth]{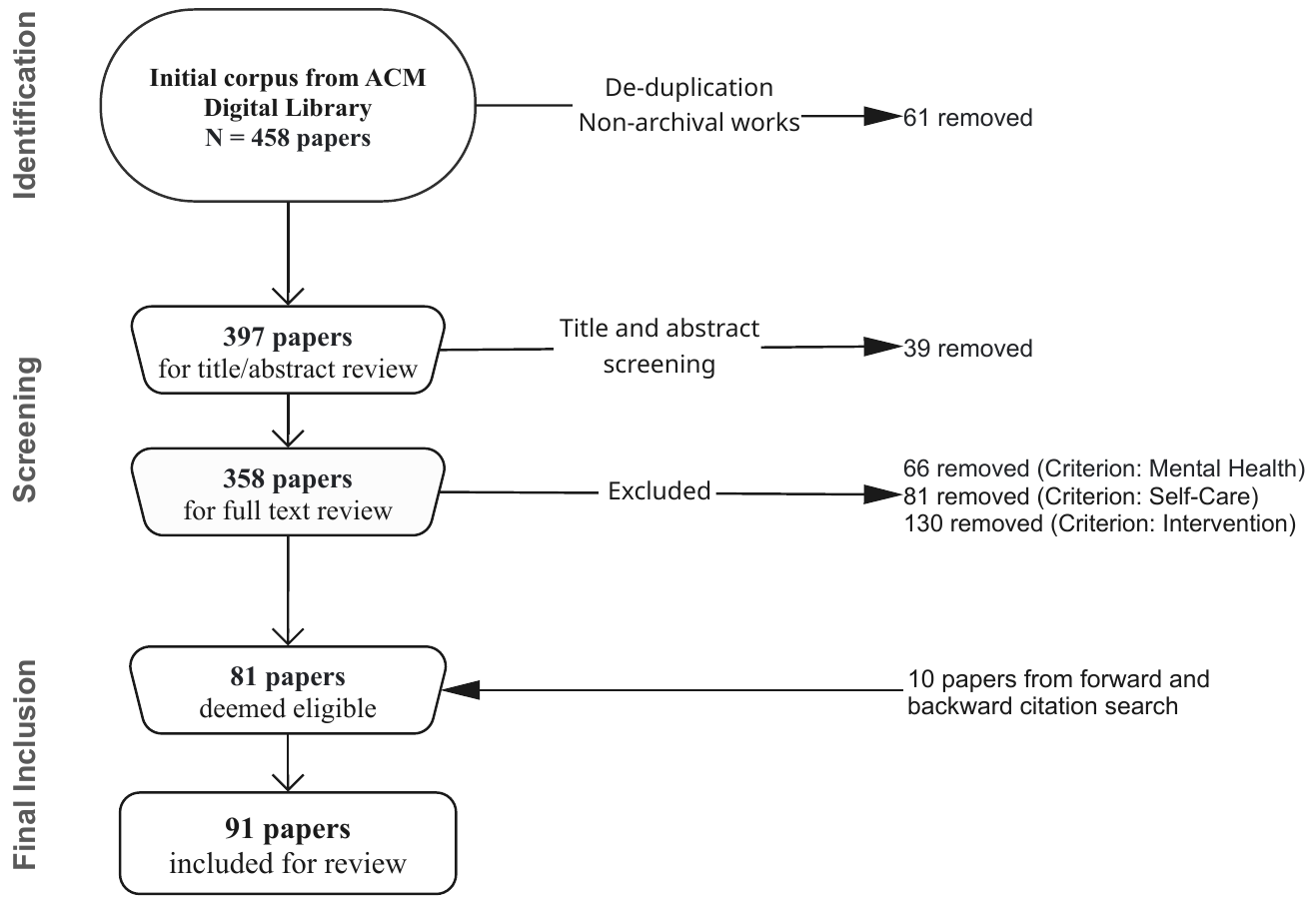}
    \caption{Diagram for each stage of identification, screening, and final inclusion. Eligibility criteria are listed sequentially in order of application.}
\label{fig:inclusion-diagram}
\end{figure*}

We conducted a scoping review to synthesize and map the landscape of HCI interventions to support mental health self-care. Scoping reviews are appropriate for emerging and interdisciplinary research areas, and are well-suited for research questions on research priorities, approaches, and gaps \cite{arksey2005scoping}. This review followed the PRISMA extension for scoping reviews (PRISMA-ScR) and standard procedure for scoping review \cite{tricco2018prisma, mak2022steps}. 
%commenting out for word count, can add back later
%We summarize this procedure in five stages: Identifying the Research Question, Identifying Relevant Studies, Selecting Relevant Literature, Charting the Data, and Collating, Summarizing, and Reporting Results. 
A completed PRISMA-ScR checklist is available in the appendix, and our work is pre-registered through the Center for Open Science\footnote{\url{https://osf.io/vxr8s/overview?view_only=d35754c675dc49c68fb31f56f1c768f0}}.

%Rather than evaluating intervention efficacy, our goal was to examine how HCI researchers conceptualize, design, and evaluate technologies for mental health self-care across diverse contexts and interaction paradigms; therefore, a literature review in its traditional definition~\cite{} was not appropriate.

\subsection{Stage 1: Identifying the Research Question}
We identified \textit{RQ1}, \textit{RQ2}, and \textit{RQ3} as initial questions from the onset of the project focused on why and for whom HCI centered its self-care interventions. These questions emerged out of gaps identified through initial project scoping and our team's experiences working in HCI for mental health. In addition, we contribute six orientations of HCI and mental health self-care, which emerged during the charting phase as we noticed meaningful but uncaptured elements of technological mediation and assumptions about mental health care and self across papers.

\subsection{Stage 2: Identifying Relevant Studies}

Papers that contribute interventions designed for self-care do not consistently describe themselves using the term ``self-care''. While there are other adjacent terms (e.g., ``self-directed'', ``home care''), our initial scan of papers quickly revealed that no clear set of terms was consistently used and reliably covered relevant papers. We therefore chose to not initially restrict our search to papers explicitly using the term self-care or adjacent labels, and instead gathered HCI research addressing mental health or well-being interventions broadly, followed by manual screening for relevance.

First, consistent with prior reviews on mental health \cite{thieme2020machine, balcombe2022human, nunes2015self}, we searched for general mental health and well-being keywords. We searched for mental health-related terms such as ``\textit{mental health}'', ``\textit{mental wellbeing}'', ``\textit{emotional wellbeing}'', and ``\textit{psychological health}''.  We also considered more specific keywords (e.g., `stress', `PTSD', `depression') but decided to not pre-determine or prioritize particular constructs of mental health. Second, we searched for intervention-related terms given our scope for HCI's interventions for self-care, drawing on prior work \cite{potts2025digital}. These terms included ``\textit{intervention}'', ``\textit{app}'', ``\textit{tool}'', ``\textit{system}'' and ``\textit{prototype}''. The full query is available in the appendix. Mental health and intervention-related terms were searched in titles, keywords, and abstracts. Full-text searching produced over 62,000 records, the huge majority of which mentioned mental health only incidentally.

We conducted our search in May and June 2026, and covered papers published in 2015 onwards (see Section \ref{background}). Given our focus on HCI interventions and following standard for scoping reviews focused on HCI contributions \cite{10.1145/3544548.3581332}, we identified articles from the ACM Digital Library to include all SIGCHI conferences and TOCHI, following prior methods from HCI reviews \cite{sanches2019hci, wang2026caring} to capture a manageable and representative sample of HCI contributions. Our initial search included papers from CHI, UbiComp, DIS, UIST, HRI, ICMI-MLMI, CSCW, IUI, IDC, UMAP, MobileHCI, VRST, C\&C, CUI, TEI, HAI, IMX, CHI PLAY, SCW, and TOCHI. We restricted our search to papers written in English and full research articles in order to gather sufficient detail for data extraction. We obtained 458 unique records through the keyword search and removed 61 papers that were incorrectly classified as full papers by ACM Digital Library or duplicates, leaving 397 papers for screening.

\subsection{Stage 3: Selecting Relevant Literature}
\label{methods:stage3}
\subsubsection{Initial Title and Abstract Screening}
In the first phase of screening, the first author screened papers by title and abstract to exclude papers that were clearly out of scope. Papers were excluded at this stage when it was clear that mental health was mentioned incidentally; these included papers focused on other types of health (e.g., diabetes, physical therapy) or technical contribution without focus on users (e.g., eye-tracking methods). In total, 39 papers were removed during title and abstract screening. 

\subsubsection{Eligibility}
Three authors reviewed the title, abstract, and full text of the remaining corpus to conduct eligibility screening. Eligibility criteria were developed iteratively by the research team through weekly meetings (one- to two-hours over five weeks) with the goal of identifying papers that designed and contributed an intervention for mental health self-care (see Table \ref{tab:eligibility-criteria}). During the initial rounds, all coders jointly reviewed paper abstracts to build a shared set of criteria that we then applied to a random set of 40 papers drawn from the screened papers (n=358). However, our initial calibrations produced insufficient independent agreement (proportionate agreement around 0.5), thus leading to further discussion on recurring ambiguities, refinement of inclusion criteria, and synchronous review of 6-10 papers in order to resolve disagreements. After this process and a final round of a new random set of 40 papers for independent coding, the team reached substantial agreement for Criterion: Mental Health (Fleiss' $\kappa=0.65$\footnote{Mental Health was met by a large portion of papers, so the presented agreement statistics is lower than raw agreement.}) and nearly perfect agreement for Criterion: Self-Care ($\kappa=0.83$) and Criterion: Intervention ($\kappa=0.85$). 

\begin{table*}[h]
\caption{Eligibility criteria and examples illustrating their boundaries.
Papers were included only when they met all three criteria.}
\label{tab:eligibility-criteria}

\setlength{\tabcolsep}{4pt}
\renewcommand{\arraystretch}{1}
\small
\begin{tabularx}{\textwidth}{
    @{}
    L{0.12\textwidth}
    L{0.29\textwidth}
    L{0.32\textwidth}
    L{0.22\textwidth}
    @{}
}
\toprule
\textbf{Criterion}
    & \textbf{Included}
    & \textbf{Excluded}
    & \textbf{Boundary case} \\
\midrule

\textbf{Mental Health}
    &
    The paper is focused on mental health and well-being, including both
    clinical and non-clinical mental health concerns.
    &
    We distinguish neurodevelopmental conditions and neurocognitive conditions from mental health conditions \cite{sachdev2009neurocognitive,reed2019icd11,who2024icd11}. We excluded papers if the focus was on concepts that may plausibly affect mental health (e.g., productivity) but do not centrally position mental health.
   
    &
    An intervention concerning autism is included when addressing a mental health concern (such as of autistic people) was central to its objective. \\

\addlinespace

\textbf{Self-Care}
    &
    The paper has design intent towards self-care, an individual's practices in understanding, maintaining, and improving their own health. Self-care is initiated and carried out by the end-user who it intends to benefit. 
    &
    We excluded papers about care primarily dependent on another person. Definitions vary on whether self-care can include collaboration with others (see Section \ref{background}); we did not exclude work solely due to incorporating other people or resources, but excluded if the care depended on another \textit{person's} active participation or co-regulation.
    &
    A chatbot that gives emotional support to a user may be included; a conversational interface that connects users to therapists is not.\\
\addlinespace

\textbf{Intervention}
    &
    The paper contributed a designed intervention -- such as a usable
    prototype, tool, or other system artifact -- and
    described its intended or observed interaction with users. 
    &
    We excluded papers that study use of existing infrastructure (e.g., helplines), as opposed to contributing the design of an intervention itself, given our focus and research questions.
    &
    A paper that contributes an emotional support chatbot system is included, while an observational study of how people use chatbots was not.\\

\bottomrule
\end{tabularx}
\end{table*}

%We also considered evidence of design intent, including whether authors framed the intervention as supporting self-guided care, designed it for individuals to independently manage their mental health, or evaluated its use or effects in everyday life outside of active caregiver support.

%This definition was iteratively refined through team discussion and application to ambiguous cases during eligibility screening; its implications and boundaries are discussed further in the Discussion.

% (probably already implied by the above) We excluded papers focused on technical development or system performance (e.g., developing or benchmarking new algorithms for conversational agents) rather than users’ experience of the intervention.

Using the above set of criteria, a total of 22.3\% of papers were full text reviewed by all three authors and the remaining were completed by the first author; the first author then brought 11 uncertain cases back to the research team to be resolved through discussion. Our final corpus consisted of 81 papers that met all three criteria. Forward and backward citation search was then conducted, which yielded 10 papers that passed screening and eligibility. Repeated citation searching on these newly included papers resulted in no additional eligible papers identified, producing a final corpus of \textbf{91 papers}.

\subsection{Stage 4: Charting the Data}
\label{methods:stage4}
Next, we created a charting template to extract relevant items from the final corpus (n=91) in order to answer our research questions. First, we identified descriptive characteristics to extract from each paper including publication year, venue, and country (of the first author’s listed affiliation). Next, three authors established a shared analytical framework that enabled us to collect and analyze standard information from each study according to our research questions: who and what mental health needs are accounted for in a paper's contributed intervention(s) (RQ1), the self-care activities conducted by users (RQ2), and the paper's core contribution (RQ3).

The final extraction template is presented in the appendix and consisted of the following dimensions. First, we recorded each paper's \textit{targeted user population}, \textit{participant population}, and \textit{target mental health issue} (e.g., depression, stress) that was being addressed by the paper's intervention. We coded the target user population distinctly from the participant population, as papers do not necessarily use the same criteria. We marked ``general'' when the population was either explicitly framed as such or inferred as such from a full text read when they imposed no population-specific boundary on the intervention’s design or intended users. Next, we summarized each paper's intervention functionalities and its \textit{intervention approach}. We first provided a \textit{summary of the intervention} in a short paragraph, then recorded an intervention's \textit{user activity} (e.g., journaling, meditation), \textit{technological affordances for supporting that activity} (e.g., recording user behavior, notifying an action), and \textit{the technological platform(s)} used. Lastly, we charted the paper's \textit{contribution} through a short summary of its contribution statement and any presented justification for focusing on self-directed care, particularly in contrast to other forms of care, e.g., clinical or peer support. Note that several of the above dimensions could be multi-coded, as an intervention may target multiple groups of users or contain a suite of functions.
%When necessary, we created categories within dimensions. For example, we agreed on a category for technological `function' (with codes such as recording behavior, teaching a skill) and a category for what the users do (with codes such as writing down emotions, practicing a new skill) as distinct within self-care activities. 

With the final template, three authors independently charted a random set of 10 papers of the final corpus (n=91). The first author then completed a draft of charting the remaining papers, cross-checked by second and third authors iteratively over multiple rounds. The team repeatedly discussed ambiguities, hypothetical, and boundary cases, as well as refined code definitions in synchronous weekly meetings until reaching consensus. 
 
%We aggregate these categories as general as, in the absence of a stated boundary, the intervention is positioned for an undifferentiated user. The first author led the interpretive coding and continually discussed emerging codes, interpretations, and ambiguities through weekly one- to two-hour meetings with the rest of the research team, who also cross-checked the codebook.

\subsection{Stage 5: Collating, Summarizing, and Reporting Results}
\label{methods:stage5}

\subsubsection{Descriptive Findings: Themes in HCI for Mental Health Self-Care}
We applied descriptive analysis to capture trends in publication year, venue, and country. To answer RQ1-RQ3, we conducted thematic analysis \cite{clarke2017thematic} to identify overarching themes and patterns in the charted data.
%For example, given a set of codes from Stage 4 for \textit{intervention approach} that include the function of technology (i.e. offer advice in response to user input) and activities users do (e.g. converse with a chatbot, self-disclose their mood), we could interpret this as a form of convesational support among \textit{intervention approaches}. 
We first familiarized ourselves with the final charted data and generated a set of meaningful second-order codes when necessary, then grouped codes into candidate themes that captured broader patterns across the data. We then iteratively refined these themes, defining each according to its central organizing concept and boundaries. Finally, we synthesized the themes into an overarching narrative to develop our findings. Additionally, we examined relationships across dimensions, such as mental health topics and technological affordances.

%Noticed something was emergent not addressed by initial findings/rqs/analysis, looking across columns of technological function, activity, core contribution. so then orientations emerged. authors synchronous discussion blah blah.

%technology's role on the self, acting on the self, assumptions

%jenny's paper quotation: We aimed to analyze 102 papers consistently, despite the fact that many of these papers discussed the concept of “self” and “AI” abstractly and vaguely.

\subsubsection{Emergent Findings: Six Orientations of Mental Health Self-Care in HCI}
During the charting phase, our team also discovered emergent, shared views across papers regarding the role of self and technology in HCI for self-care, which was not captured by RQ1-RQ3 nor explicitly stated in prior work. Here, we leveraged higher-level interpretive synthesis on how interventions configured the relationship among \textit{self}, \textit{care}, and \textit{technology} by examining across dimensions in the charted data. We conducted this synthesis in three stages. 

First, all authors observed existing codes and themes related to \textit{user activities},  \textit{technology functions},
%---two categories within the dimension of intervention approach---
and \textit{intervention description} of papers. Second, we generated care operationalization notes for each intervention. These described the care `operation' of the intervention in delivering a user activity, and both the aspects of self deemed important to care for as well as the technical affordances (e.g., surfaces data about behaviors, prompting an action) that made the care operation possible. Third, we clustered these notes into \textit{orientations of self-care} based on our shared interpretations of the relationship between user activities, technological functions, and the intended contribution to supporting care. This process was done by a team of three authors through synchronous walk-the-wall affinity diagramming, followed by discussions to resolve disagreement, comparison against the corpus, examining prototypical and boundary cases, verifying the orientations against the entire (n=91) corpus, and iterative refinement until team agreement on shared interpretation of the orientations \cite{lucero2015using}. We do not report IRR here due to the reflexive, interpretive, and recursive nature of forming this framework \cite{mcdonald2019reliability, bowman2023using}.

% \subsection{Positionality}
% \amfangcomment{if we have the room, given interpretive framework, we could add this}

%% file: sections/04-results.tex
\section{Results}

In this section, we characterize \textbf{when} and \textbf{where} the corpus was published (Section \ref{findings:desc-char}), \textbf{for whom interventions were designed for} and \textbf{for what mental health issues} (RQ1, Section \ref{findings:populations} and \ref{findings:mental-health-topics}), \textbf{what self-care activities were supported} (RQ2, Section \ref{findings:activities}), and framing of \textbf{core contributions} addressed by papers' interventions (RQ3, Section \ref{findings:contributions}). Coding for the 91 papers and all of the above findings is available in the appendix.

This analysis grounds our understanding of the landscape of HCI mental-health self-care, and we later present emergent, synthesized orientations of how HCI interventions enact different orientations toward the self and care (Section 5).

\subsection{Descriptive Characteristics}
\label{findings:desc-char}

\begin{figure*}[ht]
    \centering
    \includegraphics[width=0.7\linewidth]{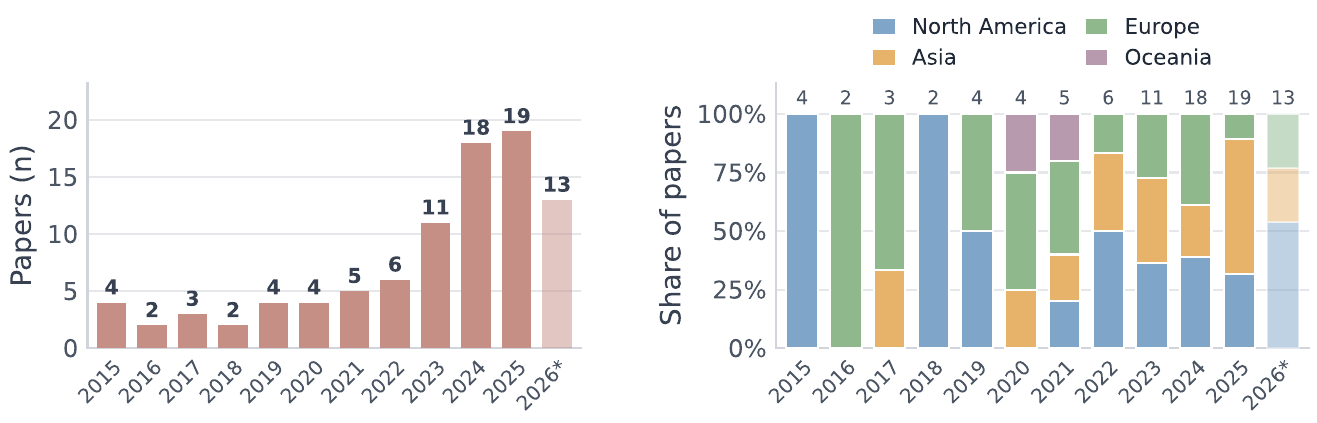}
    \includegraphics[width=0.6\linewidth]{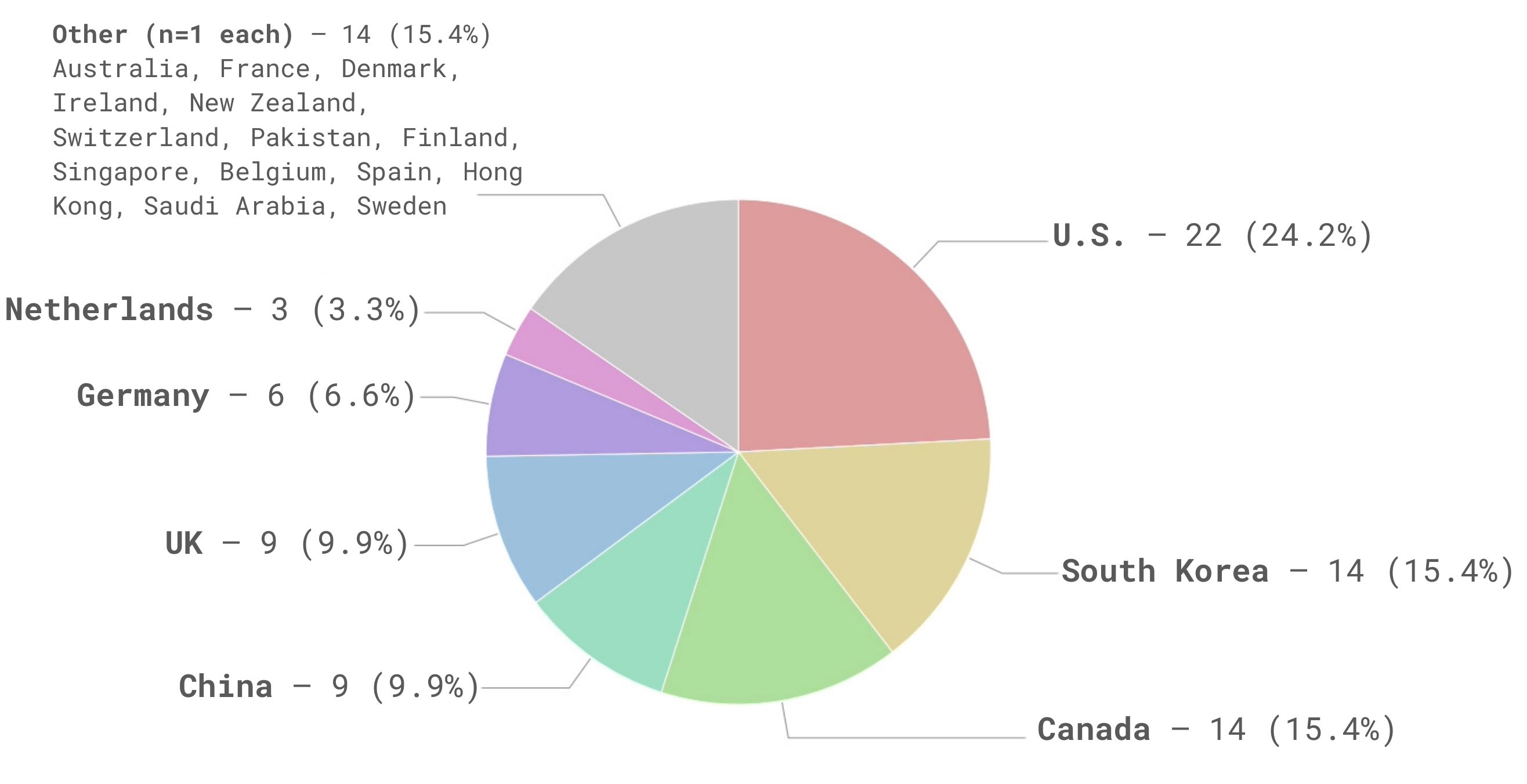}
    \caption{Count of papers in our corpus per year (top left) and regional proportions per year (top right); 2026 (partial year) values are visualized at 50\% opacity. Also shown is the count of papers from each country (bottom).}
    \label{fig:growth-total}
\end{figure*}

Our review of 91 papers spans from 2015 until June 2026. While people have long engaged in home monitoring, symptom tracking, and self-directed coping, the scope of HCI interventions regarding self-care for mental health has expanded rapidly (Figure \ref{fig:growth-total}). Little work before this time period focused on self-care for mental health conditions (\cite{nunes2015self}, see Section \ref{background}), indicating that this review is particularly timely. This growth is not solely attributed to overall publication growth; for example, the count of papers in our corpus at just CHI jumped from 1 paper in 2020 to 13 papers in 2025 while publication numbers at CHI grew about 64\% in the same time frame. 

The literature is primarily published at CHI (60.4\%) and DIS (17.6\%); all remaining venues contributed no more than three papers each. In terms of the first author’s affiliated country, the regional composition of the corpus has shifted over time; Asia contributed one paper in our corpus from 2015 to 2019 but a proportional share of 20\% to 47.4\% annually since 2020 (Figure \ref{fig:growth-total}). Overall, the corpus has first authors affiliated from 21 countries across North America (n=36, 39.6\%), Asia (n=27, 29.7\%), Europe (n=26, 28.6\%), and Oceania (n=2, 2.2\%). The United States is the most represented country, followed by South Korea and Canada (Figure \ref{fig:growth-total}).

\subsection{Who Interventions Are Designed For}
\label{findings:populations}
We present five themes of populations that interventions targeted: \textit{mental health disorders} (n=11, 12.1\%), \textit{disability} (n=6, 6.6\%), \textit{demographics} (n=22, 24.2\%), \textit{occupational and social roles} (n=14, 15.4\%), and \textit{general} (n=43, 47.3\%). A paper could fill multiple codes (e.g., children with disability as \textit{demographics} and \textit{disability} \cite{antle2019evaluating}). 

\subsubsection{Mental Health Disorders}

Interventions focused on people with depression (n=3), eating disorders (n=2), bipolar disorder (n=1), schizophrenia (n=1), panic disorder (n=1), and personality disorder (n=1). Two additional papers targeted people with ``complex, difficult-to-treat disorders and suicidal behavior'' \cite{schroeder2018pocket} and ``mental health patients'' broadly without specifying particular diagnosis \cite{soler2024arcadia}. Only one intervention was explicitly built for people with anxiety disorders (panic disorder \cite{rubin2015towards}).

\subsubsection{Disability}
A handful of papers contributed interventions for autistic people\footnote{While there is no universally preferred language, we use identity-first language in this work given its preference by many autistic people as indicating identity inseparable from personhood.} (n=3), people with learning disability (n=1), ADHD (n=1), or Deaf and Hard-of-Hearing (DHH) individuals (n=1). These papers examined disability-specific needs shaped by intersecting mental health experiences \cite{simm2016anxiety, kim2023routineaid, miri2022far}. Work by Thieme et al. \cite{thieme2016challenges} cuts across multiple, intersecting identities as it studies women in secure hospital units who experience borderline personality disorder and learning disability. Overall, however, the diversity of disability experiences and well-known intersectional nature with mental health \cite{angelini2025speculating, brinkman2023shifting} has been designed for by relatively few interventions in HCI. 

Note that we include these populations here for organizational purposes but fully acknowledge no universal preference among people on identifying as disabled considering both medical and social models, such as for d/Deaf people.

\subsubsection{Demographics}
Demographics designed for included age, gender, non-WEIRD, and socio-economic status. Age was the most common factor. Papers often focused on youth, including pre-adolescent children (n=6) to teenagers (n=4), motivated by the importance of teaching self-care skills to children \cite{speer2021mindfulnest, park2024collaborative, seo2024chacha} and prevalence of poor mental health \cite{kitson2023co, antle2019design, sun2025conversations}. Older adults (n=7) were also framed as in-need due to being socially isolated \cite{ryu2020simple, wei2025, guo2025exploring} and having unique design needs to overcome barriers in adopting self-care interventions \cite{surani2026co}. Women's mental health was the focus of five papers, including adolescent girls \cite{sun2025conversations}, young women \cite{10.1145/3613904.3643054}, women in non-WEIRD contexts \cite{aldaweesh2026hasn}, women in hospital care units \cite{thieme2016challenges}, and pregnant women \cite{doherty2019engagement}. 

\subsubsection{Occupational and Social Roles}
Papers focused on university students (n=10), workers (n=3), and caregivers (n=1). For university students, interventions acknowledged significant mental health struggles and academic pressure among college students \cite{10.1145/3699761,10.1145/3706598.3713269}, and context-specific stressful experiences, such as public speaking \cite{10.1145/3718084}. Interventions for workers were mostly intended to fit within work routines (e.g., browser plugins \cite{10.1145/3544548.3581319}, passive stress monitoring around the office \cite{elvitigala2021stressshoe}). Surprisingly, only one paper in our corpus contributed an intervention for care\textit{giver} self-support \cite{10.1145/3706598.3713699}.
%As a result, although many of these papers frame caring for mental health in these contexts away from productivity-focused well-being, HCI contributions to self-care still appear to be heavily embedded within a workplace lens.

\subsubsection{General or Unspecified  population}
Papers most commonly contributed interventions for a ``general population'' or provided no population-specific framing based on mental health experiences, demographics, or other descriptors. We aggregate these papers because neither establishes a population-specific boundary around the intervention, and the latter's absence of any specification implies the intervention as rhetorically applicable to a general or undifferentiated user.

We further examined the \textit{participant} population for papers in this theme, surfacing here how mental health experiences are included or excluded for interventions aimed at either general or unspecified people. In this theme, 15 (34.9\%) papers explicitly excluded people with clinical symptoms or diagnoses, while 44.2\% did not report any health status criteria or were too vague to be determined (`healthy participants'). We note that the papers that excluded people with diagnosis also sometimes extended to any \textit{prior} or suspected diagnosis, those who are in therapy or taking medication, or any mental health diagnosis broadly (even outside the work's mental health topic of focus) \cite{10.1145/3643834.3660702,schneeberger2021stress, 10.1145/3772318.3790933, 10.1145/3772318.3790406}. Papers justified these exclusions through safety or ethical considerations, or provided no rationale. Only one paper explicitly acknowledged that excluding clinical populations was in tension with its stated goal of designing for a general population \cite{10.1145/3643834.3661570}. Seven papers (16.3\%) included people regardless of mental health experiences, either articulating the decision to not screen or screening for mental health status but not excluding any participants based on this screening (e.g., \cite{10.1145/3706598.3713883}, \cite{10.1145/3772318.3791591}). Interestingly, two papers (4.7\%) were restricted to those with high levels of stress symptoms although they were for a general population \cite{10.1145/3613904.3642766, 10.1145/3706598.3713730}, citing that including those with low levels of stress would be unsuitable for a stress management study or introduce biased data. 

Overall, ``general population'' was the most common population but encompassed different approaches to clinical inclusion and exclusion. We discuss implications of this in Discussion regarding how 'general' self-care interventions in HCI may be frequently focusing on a distinct slice of 'healthy' individuals.

\subsection{What Mental Health Topics Interventions Are Designed For}
\label{findings:mental-health-topics}

\begin{table*}[t]
\centering
\small
\caption{Corpus (n=91) characteristics for Sections 4.2-4.4. Papers could be multi-coded to accurately represent their contents, since papers on mental health can often address multiple populations, mental health topics, and/or activities.}
\label{tab:descriptive-stats}

\begin{minipage}[t]{0.31\textwidth}
\centering
\textbf{Population}\\[2pt]
\begin{tabular}{@{}p{2.8cm}rr@{}}
\toprule
Theme & N & \% \\
\midrule
General                 & 43 & 47.3\% \\
Demographic Groups            & 22 & 24.2\% \\
Occup.\ \& Social Groups & 14 & 15.4\% \\
Mental Health Disorder  & 11 & 12.1\% \\
Disability              & 6  & 6.6\%  \\

\bottomrule
\end{tabular}
\end{minipage}
\hfill
\begin{minipage}[t]{0.31\textwidth}
\centering
\textbf{Mental Health Topic}\\[2pt]
\begin{tabular}{@{}p{2.9cm}rr@{}}
\toprule
Theme & N & \% \\
\midrule
General Well-Being             & 17 & 18.7\% \\
Emotional Awareness \& Reflection & 17 & 18.7\% \\
Emotional Regulation           & 16 & 17.6\% \\
Stress \& Relaxation           & 15 & 16.5\% \\
Mindfulness                    & 14 & 15.4\% \\
Emot. Expression           & 10 & 11.0\% \\
Anxiety                        & 9  & 9.9\%  \\
Social Connectedness & 8 & 8.8\% \\
Depression                     & 7  & 7.7\%  \\
Bipolar                        & 1  & 1.1\%  \\
Eating Disorders               & 1  & 1.1\%  \\
Schizophrenia         & 1  & 1.1\%  \\

\bottomrule
\end{tabular}
\end{minipage}
\hfill
\begin{minipage}[t]{0.37\textwidth}
\centering
\textbf{Self-Care Activity}\\[2pt]
\begin{tabular}{@{}p{3.8cm}rr@{}}
\toprule
Theme & N & \% \\
\midrule
Conversing \& Social Support & 25 & 27.5\% \\
Tracking                     & 22 & 24.2\% \\
Meditation                   & 13 & 14.3\% \\
Sensory Regulation           & 13 & 14.3\% \\
Journaling                   & 11 & 12.1\% \\
Artistic/Creative Expression & 7  & 7.7\%  \\
Music/Calming Sounds         & 6  & 6.6\%  \\
Treatment Programs           & 4  & 4.4\%  \\
Gaming                       & 3  & 3.3\%  \\
Embodied/Metaphorical Actions & 3 & 3.3\%  \\
Goal Setting                 & 3  & 3.3\%  \\
Reminiscence                 & 2  & 2.2\%  \\
Informational Resources      & 2  & 2.2\%  \\
\bottomrule
\end{tabular}
\end{minipage}
\end{table*}
We discovered 12 mental health topics that interventions target: \textit{anxiety} (n=9, 9.9\%), \textit{depression} (n=7, 7.7\%), \textit{stress and relaxation} (n=15, 16.5\%), \textit{mindfulness} (n=14, 15.4\%), \textit{social connectedness and loneliness} (n=8, 8.8\%), \textit{emotional expression} (n=10, 11\%), \textit{emotional awareness and reflection} (n=17, 18.7\%), \textit{emotional regulation} (n=16, 17.6\%), \textit{bipolar} (n=1, 1.1\%), \textit{eating disorders} (n=1, 1.1\%), \textit{schizophrenia} spectrum (n=1, 1.1\%), and \textit{general well-being} (n=17, 18.7\%). 22 papers (24.2\%) addressed more than one topic.

The most prevalent topics were broad, non-diagnostic forms of mental well-being. \textbf{General Well-Being} encompassed general emotional support without particular focus or following healthy habits \cite{10.1145/3706598.3713453,10.1145/3613904.3642761,10.1145/3745900.3746099}. \textbf{Emotional Awareness and Reflection} interventions were aimed at drawing users' attention to their emotions \cite{yan2022emoglass} or prompting reflection \cite{10.1145/3706598.3713883,rajcic2020mirror, 10.1145/3800645.3813019}. In contrast, \textbf{Emotional Regulation} interventions aimed to help users manage their emotional responses, often through self-soothing strategies. For example, \textit{FAR} uses tactile vibration to help self-regulation by following slow breathing, and \cite{kitson2023co} uses VR to train emotion regulation skills. \textbf{Emotional Expression} considered the expression of thoughts or emotions as good for mental health, whether through expressive writing, conversation, or digital or tangible art. While emotion-related topics co-occurred (e.g., 5 papers with Emotional Awareness and Reflection \& Emotional Regulation), they did not overlap substantially, indicating that these themes captured distinct forms of well-being. A smaller set of interventions addressed \textbf{Social Connectedness and Loneliness}. 

Papers addressing \textbf{Stress and Relaxation} focused on general stress rather than stress disorders (e.g., PTSD). Reducing \textbf{Depression} was also somewhat common but often focused on the general psychological experience as opposed to clinical symptoms of depression; similarly, \textbf{Anxiety} was targeted in nine papers with eight referencing anxiety vaguely/generally or as an everyday psychological experience -- with the exception of panic disorder in \cite{rubin2015towards}. As seen in Section \ref{findings:populations}, designing for clinically anxious people was also not common; anxiety is thus often treated as a general experience but not a clinical one. A smaller set of papers explicitly targeted clinical diagnosis, with 1 paper each for bipolar disorder \cite{matthews2015situ}, schizophrenia \cite{10.1145/3613904.3642369}, and eating disorders \cite{10.1145/3706598.3713485}.

Taken together, our analysis of intended populations and mental health topics shows that HCI has predominantly positioned mental health self-care around `everyday' well-being and psychological experiences.

\subsection{What Self-Care Activities Interventions Support}
\label{findings:activities}
% \begin{figure*}
% \includegraphics[width=\textwidth]{sankey.png}
% \caption{Sankey diagram showing how self-care activities support different mental health issues. We normalize to avoid double counting and visual distortion given multi-to-multi codes. As a result, each paper on the leftside column has flow distributed evenly between activities (e.g., a paper that tackled one mental health issue using two activities contributes only 1/2 flow, rather than a full flow, to its two edges).}
% \label{fig:sankey}
% \end{figure*}
% \label{findings:activities}

In addition to \textit{who} and \textit{what} interventions targeted, we examined what users \textit{do} -- in other words, the self-care activity that the user engages in for mental health benefit, rather than the technology or modality. For example, journaling was coded as such regardless of whether it was supported through a mobile application, conversational system, or VR environment. In this way, activity captures the \textit{self-care practice that the intervention facilitates} while remaining agnostic to the technology. Activities identified were: \textit{conversing and social support} (n=25, 27.5\%), \textit{tracking} (n=22, 24.2\%), \textit{meditation} (n=13, 14.3\%), \textit{sensory regulation} (n=13, 14.3\%), \textit{journaling} (n=11, 12.1\%), \textit{artistic and creative expression} (n=7, 7.7\%), \textit{listening to music or calming sounds} (n=6, 6.6\%), \textit{treatment programs} (n=4, 4.4\%), \textit{gaming} (n=3, 3.3\%), \textit{embodied or metaphorical actions} (n=3, 3.3\%), \textit{goal setting} (n=3, 3.3\%), \textit{reminiscence} (n=2, 2.2\%), and \textit{reading information resources} (n=2, 2.2\%).

\textbf{Conversing and social support} were the most common, largely driven by generative conversational agents. These interventions enabled users to converse with a simulated partner, disclose experiences to an agent, and engage in supportive dialogue. \textbf{Tracking and monitoring} facilitated users recording behaviors, emotions, physiological states, or other aspects of everyday experience. This drew from the established HCI tradition of personal and lived informatics in which systems help people reflect on personal data to develop self-knowledge and inform action \cite{li2010stage, rooksby2014personal}. Together, conversing and self-tracking represent two frequent paradigms of technologically mediated self-care: one being relational and interactional, and the other oriented towards a traditional, individualistic understanding of the self's state. Additionally, \textbf{meditation and mindfulness}, \textbf{sensory or physiological self-regulation}, and \textbf{journaling} appeared in many interventions across recent years, perhaps due to turn of third-wave HCI focusing on somatic, embodied approaches and meaning-making among one's own experiences \cite{loke2018somatic, bodker2006second}. Overall, the corpus includes considerable technological diversity through XR, text-based and embodied virtual agents, mobile apps, and more, but underlying activities supported by these systems are often recognizable self-care practices.

\subsection{What Contributions HCI Interventions Make to Self-Care}
\label{findings:contributions}

\begin{table*}[h]
\centering
\caption{Core contributions to mental health self-care identified across the corpus and framed by a paper, organized by theme.}
\label{tab:core-contributions}
\small

\begin{tabular}{
    >{\raggedright\arraybackslash}p{0.11\textwidth}
    >{\raggedright\arraybackslash}p{0.17\textwidth}
    >{\raggedright\arraybackslash}p{0.27\textwidth}
    >{\raggedright\arraybackslash}p{0.3\textwidth}
    >{\centering\arraybackslash}p{0.05\textwidth}
}
\hline
\textbf{Theme} & \textbf{Sub-Theme} & \textbf{Description} & \textbf{Prototypical Example} & \textbf{$N$} \\
\hline

\noalign{\vskip 4pt}
\multirow{2}{0.12\textwidth}{\textbf{Accessibility of Self-Care}}
& \textit{Addressing Underserved Populations \& Contexts} &
Adapts mental health self-care to populations whose needs are not met from conventional care and/or current technologies.
& \textit{CSESC} provides a culturally aware, Arabic-language chatbot for Saudi young women \cite{aldaweesh2026hasn}. & 32 \\
\noalign{\vskip 4pt}
& \textit{Scalability} &
Frames contribution as addressing barriers related to cost and availability.
& \textit{StressShoe} contributes a low-cost way to self-track stress using off-the-shelf, shoe-mounted sensors \cite{elvitigala2021stressshoe}. & 5 \\
\noalign{\vskip 4pt}
\hline
\noalign{\vskip 4pt}
\multirow{2}{0.11\textwidth}{\textbf{Supporting Engagement and Sustained Use}}
& \textit{Engagement \& Adherence} &
Addresses dropout in self-care interventions through engaging, interactive, or motivating designs.
& \textit{ARCADIA} uses mixed-reality and biofeedback to allow users to regulate emotions in a 'self-care garden', providing a creative and motivating way to address patients' declining adherence. \cite{soler2024arcadia}. & 15 \\

\noalign{\vskip 4pt}
& \textit{Personalization \& Context-Awareness} &
Contribution is aimed at incorporating users' preferences, characteristics, or contexts into adaptive DMHIs.
& A personalizable therapy chatbot where users can customize therapy style and personality, and input their intentions \cite{vossen2024effect}. & 22 \\
\noalign{\vskip 4pt}
\hline
\noalign{\vskip 4pt}
\multirow{3}{0.12\textwidth}{\textbf{Enabling Self-Care in Everyday Life}}
& \textit{Shifting to Everyday Contexts} &
Positions contribution as fundamentally shifting mental health towards everyday and in-the-moment needs.
& \textit{Mindful Moments} integrates mindfulness practice into smartglasses in order to bring `casual' mindfulness into realistic, day-to-day walking. \cite{tan2023mindful}. & 14 \\

\noalign{\vskip 4pt}
& \textit{Translating Clinical Care Beyond Clinical Settings} &
Translates evidence-based therapeutic models and grounded professional expertise into self-care technologies.
& \textit{MoodRhythm} translates clinical therapy for bipolar disorder into a mobile app, and shows the value of low-level understanding through real-world design alongside clinicians \& patients \cite{matthews2015situ}. & 9 \\

\noalign{\vskip 4pt}
& \textit{Safety \& Trust} &
Addresses reliability, privacy, ethical concerns, and risks associated with self-directed (and increasingly AI-mediated) mental health support.
& \textit{MESA-Bot} attempt to mitigate older adult concerns in adopting mental health chatbots through addressing accessibility, privacy, and security \cite{surani2026co}. & 7 \\
\noalign{\vskip 4pt}
\hline
\noalign{\vskip 4pt}
\multirow{3}{0.12\textwidth}{\textbf{Enhancing Self-Care Activities}}
& \textit{Enhancing Self-Care Activity through Novel Interaction}
&
Introduces new interaction mechanisms for self-care activities; expands the ways in which self-care activities can be enacted.
& \textit{Mind Mansion} helps user cope with negative thoughts through interacting with a messy apartment in VR, where mess (puddles, dirt) is labeled with negative thoughts and cleaned \cite{rasch2024mind}. & 14 \\

\hline
\end{tabular}

\vspace{2pt}
\parbox{\textwidth}{\footnotesize
\textit{Note}: Contribution categories were not mutually exclusive. Papers could be coded to multiple themes, so totals exceed corpus n=91.
}
\end{table*}

Authors framed their contribution and the gap addressed by their interventions in eight ways, which we present in four themes: \textit{Accessibility of Self-Care} (n=34, 37.4\%), \textit{Supporting Engagement and Sustained Use} (n=33, 36.3\%), \textit{Enabling Self-Care in Everyday Life} (n=28, 30.8\%), and \textit{Enhancing Self-Care Activities} (n=14, 15.4\%) (Table \ref{tab:core-contributions}). In presenting these contributions, we also discuss how each of the above themes surfaces new responsibilities for ``self'' (the user) and systems in the landscape of self-care. While we aimed to identify each paper's primary framing of its contribution, we assigned multiple codes when a single category was insufficient (n=24, 26.4\%).

\subsubsection{Accessibility of Self-Care}
\label{findings:contributions:accessibility}
Many papers in our corpus delivered interventions that aimed to fill a gap in \textit{addressing underserved populations and contexts} (n=32, 35.2\%). These populations were often those in-need due to intersectional struggles with mental health -- for example, \cite{kim2023routineaid} and \cite{simm2016anxiety} addressed autistic people, who face high rates of anxiety and further daily functioning struggles from poor mental health. Fewer papers addressed enhancing \textit{scalability}. In contrast to addressing particular populations, these papers attempted to reduce broader barriers such as cost and availability through, for example, designing interventions with low-cost, off-the-shelf commercial hardware \cite{antle2019evaluating} or able to be deployed in low-SES communities \cite{antle2019design}. 

Overall, papers in this theme often motivate their use of DMHIs by its ability to make healthcare available to more people given limitations of cost and personnel in conventional care. Implicitly, papers then shift responsibility to designers and their systems to make that care (1) sufficiently effective for diverse needs and (2) cost-effective and available. Our findings reveal that addressing underserved community needs is a common contribution motivation, but goals of scalability are comparatively addressed less. While cost is frequently cited across our corpus as a motivation for DMHIs, relatively few papers treated it as an explicit design goal -- perhaps due to accessibility being assumed as simply inherent to DMHIs.

\subsubsection{Supporting Engagement and Sustained Use}
\label{findings:contributions:engagement}
In conventional care, providers and institutions help tailor treatment and sustain participation through appointments and therapeutic alliance; in self-care, this responsibility can shift to users, who sustain their own engagement with systems that continually make participation compelling. Papers thus positioned contributions around \textit{adherence} (n=15, 16.5\%), to design interactive or motivating interventions to address dropout -- a persistent challenge across digital mental health and self-care \cite{torous2018clinical, banos2022current}. Relatedly, \textit{personalization \& context-awareness} (n=22, 24\%) made self-care more relevant to users, accounting for day-to-day, in-the-moment changes given that self-care is not bounded to formal settings. This occurred through (a) \textit{user personalization}, where users directly indicate their preferences (e.g., \cite{10.1145/3706598.3713453,10.1145/3715336.3735795}); (b) \textit{automatic personalization}, where systems adapted based on users' characteristics or behaviors (e.g., \cite{konrad2015finding,10.1145/3772318.3791817}); and (c) \textit{context-awareness}, where systems adapted to situational contexts (e.g., \cite{huang2015emotion,10.1145/3699761,10.1145/3706598.3713730}).

\subsubsection{Enabling (Safe) Self-Care in Everyday Life}
Although all papers in the corpus concerned self-care outside conventional treatment to some extent, 14 papers (15.4\%) explicitly shaped the intervention around \textit{shifting to everyday contexts}, transforming formal practices into fundamentally ``casual'' ones \cite{tan2023mindful} and supporting mental health conditions but outside formal care \cite{10.1145/3706598.3713485}. While conventional care is largely organized around scheduled encounters and designated treatment settings, self-care relies on users to decide when, where, and under what circumstances care is needed. Because of the flexibility of self-care DMHIs, systems also have to translate existing practices to be available outside these formal settings. Some papers focused on the \textit{translation of clinical care} (n=9, 10\%) into digital resources that users can now access independently \cite{schroeder2018pocket}, or demonstrated usefulness of clinician involvement but for designing self-directed care treatment \cite{thieme2016challenges}. However, independent use with less continuous oversight also raises questions of \textit{appropriate, trustworthy, and safe support}. 7 papers (7.7\%) contributed interventions that were centrally designed around addressing privacy \cite{surani2026co}, explainability \cite{kim2022prediction}, trust \cite{surani2026co,10.1145/3715336.3735795}, and ethics in AI-driven conversational and predictive technologies \cite{10.1145/3706598.3713485,10.1145/3613904.3642937,10.1145/3613904.3642369,10.1145/3772318.3791817}. 

\subsubsection{Enhancing Self-Care Activity through Novel Interaction} 
Some interventions (n=14, 15.4\%) expanded what people could do as self-care by applying \textit{novel interaction} and modalities to existing practices. Examples included combining generative AI with journaling \cite{10.1145/3745900.3746099}, using predictive analysis to provide counterfactual stress-coping \cite{10.1145/3772318.3791615}, and using generative co-creation to support reflection \cite{10.1145/3800645.3813019}. Whereas other themes expanded \textit{who} can access self-care or \textit{where} it can occur, these papers expanded \textit{what people can do}, positioning technology as a means of transforming practices through which self-care is achieved. While providers conventionally select and facilitate established therapeutic or reflective activities and help users interpret their experiences, self-care technologies can create new representations and activities; unlike conventional care, though, these are not subject to the same clinical requirements in being `prescribed'.

%% file: sections/05-framework.tex
\section{Six Orientations of Mental Health Self-Care in HCI}
\label{framework}

\begin{table*}[h]
\sffamily
\caption{Six orientations represent recurring forms of care enacted by technology.}
\label{tab:self-care-framework}
\centering
\small
\setlength{\tabcolsep}{3pt}

\begin{tabularx}{\textwidth}{
    @{}
    >{\raggedright\arraybackslash}p{0.1\textwidth}
    >{\raggedright\arraybackslash}p{0.14\textwidth}
    >{\raggedright\arraybackslash}p{0.14\textwidth}
    >{\raggedright\arraybackslash}p{0.18\textwidth}
    >{\raggedright\arraybackslash}p{0.19\textwidth}
    >{\raggedright\arraybackslash}p{0.19\textwidth}
    @{}
}

\toprule
\textbf{Orientation} &
\textbf{Aspect of Self To Care For} &
\textbf{Care Operation} &
\textbf{Common Technology Functions} &
\textbf{Defining Boundary} &
\textbf{Prototypical Example} \\
\toprule

\textbf{Making the Self Tangible}&
Internal thoughts and experiences &
Externalizing and re-interpreting \textit{internal experience}. &
Reflecting physiological states; allowing free-form expression; embodying the user's state; generating new representation of user's state&
Representations function primarily as materials for expression, interpretation, or meaning-making. & 
\textit{Inner Garden} is a sandbox that changes landscape according to the user's breathing and heart rate \cite{roo2017inner}. \\

\addlinespace
\midrule
\addlinespace
\textbf{Quantifying the Self} &
Data, patterns &
Measuring and surfacing \textit{data, states, and patterns}.&
Allowing users to self-record data; reflecting physiological state; aggregating data; sensing or predicting state &
Calculates and surfaces evidence about the self's patterns, in order to provide knowledge. & \textit{MoodRhythm} allows bipolar patients to track basic routines like waking and meal times to track how they meet routine \& mental health goals \cite{matthews2015situ}. \\

\addlinespace
\midrule
\addlinespace
\textbf{Enacting Social Care}&
Social and relational &
Mediating or simulating \textit{social interaction}. &
Mimicking a person; simulating social interaction&
A human or human-like social interaction is central to the care mechanism. &
\textit{Vincent} is a chatbot that the user must care for, as a way to create self-compassion \cite{lee2019caring}.
\\
\addlinespace
\midrule
\addlinespace
\textbf{Training Self-Care Skills}&
Capabilities &
Instructing or providing practice for a \textit{skill}. &
Educating about a skill; guiding users through a practice &
The intended outcome is developing a capability that the user can subsequently apply. & \textit{Pocket Skills} is a conversational interface that teaches and supports the practice of core DBT skills \cite{schroeder2018pocket}.\\

\addlinespace
\midrule
\addlinespace

\textbf{Facilitating Action}&
Direct actions &
Prompting \textit{actions}. &
Notifying to do an action; guiding users through a practice &
Technology supports actions or behaviors that move the user towards an existing or suggested goal. & \textit{Home Sweet Office} nudges workers to conduct brief stress-reduction activities during the workday \cite{10.1145/3544548.3581319}.\\
\addlinespace
\midrule
\addlinespace

\textbf{Attending to Affective Experience} &
Affect and felt experience &
Regulating and soothing \textit{felt experience}. &
Evoking an emotion; directly regulating or soothing; simulating or mimicking a person&
The object of care is the user's immediate emotional experience, typically for direct soothing. & \textit{Awedyssey} is a VR experience built to trigger awe in order to support user's emotional and mental well-being \cite{10.1145/3563657.3595998}.\\

\bottomrule
\end{tabularx}

\vspace{3pt}
\parbox{\textwidth}{\footnotesize
\textit{Note.} Orientations are not mutually exclusive. Technology functions are the most frequently coded items to guide understanding of the orientations, rather than an exhaustive list.
}
\end{table*}

Next, we surface six recurring orientations that describe how technological mediation is expected to enable care across the papers in our corpus. These orientations reflect recurring normative configuration of three related elements: (1) the \textit{aspect of the self} treated as the object of care, (2) the \textit{operation through which care is expected} to occur, (3) the \textit{mediating role supported by technology}. They emerged through cross-dimensional analysis of our charting data, drawing on dimensions including intervention goals, self-care activities, and technological functions (Section \ref{methods:stage5}).

The six orientations are: \textbf{Making the Self Tangible} (n=34, 37.4\%), \textbf{Quantifying the Self} (n=29, 31.9\%), \textbf{Training Self-Care Skills} (n=21, 23.1\%), \textbf{Enacting Social Care} (n=18, 19.8\%), \textbf{Facilitating Action} (n=14, 15.4\%), and \textbf{Attending to Affective Experience} (n=10, 11\%), presented in Table \ref{tab:self-care-framework} and Figure \ref{fig:teaser}. The orientations represent interpretations of how the reviewed papers frame and operationalize care through intervention design.

The orientations address a different analytical question from the descriptive categories in Section 4. While an activity describes what the user does and mental health topics serve as goals, an \textit{orientation} reflects shared assumptions about how technological mediation is expected to constitute care. For example, conversation may provide an opportunity to rehearse a skill or to experience a supportive relationship; these interactions share an activity but address different aspects of self through different care operations. Conversely, a conversational interface and an immersive environment may both support the training of self-care skills. 
As such, interventions often grouped together by activity, platform, or population can enact different orientations. This framework deviates from prior models \cite{nunes2015self, wang2026caring, sanches2019hci, slovak2023designing} in offering a new lens to explicitly view the shared assumptions and relationship in mental health self-care technologies between \textit{the self, care, and technology}. 

\begin{figure*}[h]
    \centering
    \includegraphics[width=\linewidth]{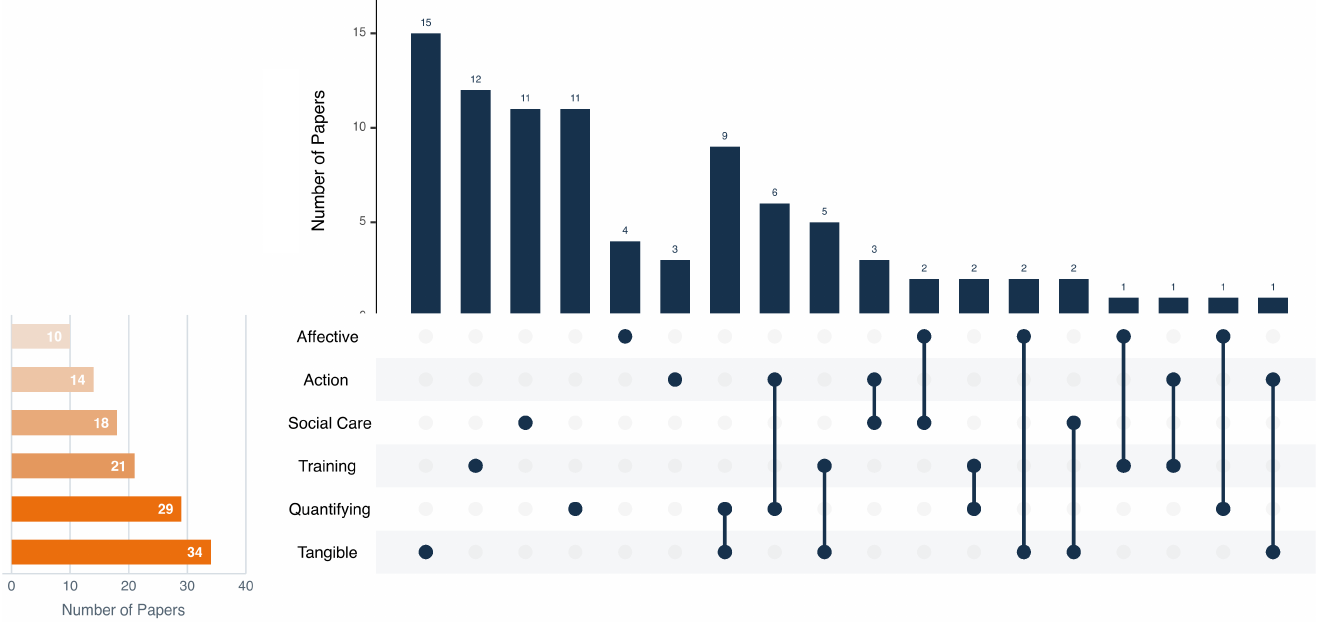}
    \caption{Upset diagram showing distribution of papers among the six orientations of HCI for self-care. }
    \label{fig:upset}
\end{figure*}

In presenting the orientations below, we describe how interventions address the aspect of self that a person seeks self-care for, the care operation, defining boundaries of each orientation, and prototypical examples (Table \ref{tab:self-care-framework}). Note that these orientations were not intended to be mutually exclusive, although we focused on labeling an orientation only when its care logic was central to an intervention. 61.5\% of these interventions had one orientation, while the remaining addressed two. Both a Venn (Figure \ref{fig:teaser}) and Upset diagram (Figure \ref{fig:upset}) show the distribution of papers across orientations. We also highlight example interventions that address multiple orientations in Figure \ref{fig:overlapping-orientations}.
\subsection{Orientation 1: \textit{Making the Self Tangible}}
\label{framework: tangible}

\mybox{\small{\textit{``[MoodShaper] allows users to autonomously create a virtual environment with the purpose of expressing emotions...by providing means for users to manipulate visual representations of negative emotions in VR''} \cite{10.1145/3643834.3661570}}}
%saving for word count: ``\textit{Our approach aims to provide users with a way to engage with their thoughts by cleaning a virtual apartment, which serves as a visual representation of the cognitive process of sorting one’s thoughts''} \cite{rasch2024mind} \newline 

A prominent assumption throughout our corpus is that a person may have \textit{internal experiences}, such as thoughts and emotions, that are difficult to articulate or engage with. As a result, a person cares for themselves by turning self-experience into a perceptible, interactive form; this interaction allows for better expression, meaning-making, and/or regulation. Technology makes this self-directed engagement possible by providing expressive materials or transforming an internal state into visual, written, auditory, embodied, or physical form.  Note that we use \textit{tangible} broadly -- the resulting form did not need to be a physical object, but could instead be visual or written form, embodied, or auditory form.

We identified two ways that technology supported people's re-interpretation of their self: (1) system-mediated transformation that uses embodied and interactive approaches to transform the person's state into a `tangible' form or (2) user-mediated transformation, where technology provides the space and `materials' for the user to transform their own state into a new `tangible' form (e.g., language, art). An example of the former is \textit{Mirror Ritual} that estimated facial expression and generated a poem in response in order to allow for better self-reflection by the user \cite{rajcic2020mirror}, and an example of the latter is \textit{Mood Worlds} that allowed users to paint their emotions into 3D environments to better visualize their emotional well-being \cite{wagener2022mood}. These processes may also be serialized as seen in \textit{NoRe}, which asked users to first articulate their experience into a journal with an emotion regulation intention, after which the system re-interpreted that entry into music \cite{10.1145/3715336.3735845}. 

This orientation entails a particular conception of the self, where externalization is the necessary process by which internal state can be meaningfully reflected upon. Since a person may not have convenient means to transform their state into engaging, interactive forms, they may turn to DMHIs. This often draws from tangible and embodied interaction, as well as somaesthetics of third-wave HCI \cite{bodker2006second, loke2018somatic}. Tangible interaction often sought to move information beyond screen-based representations by giving it perceptible and manipulable form; meaning is produced through bodily or `physical' engagement \cite{dourish2001action}. Interventions in this theme extend these traditions through technological mediation that provides the materials by which the self can interact with internal experience through (virtual) objects \cite{rasch2024mind}, dynamic landscapes and physical movement \cite{prpa2018attending, roo2017inner}, and drawings, poems, or music \cite{wagener2022mood, wagener2023selvreflect, zhang2026asafeplace}. 

\subsubsection{Function of Technology} 
The core technological functions in this orientation concerned making internal experience perceptible. 12 papers did so through technical affordances to \textit{reflect the user’s physiological state}, translating signals such as breathing or arousal into feedback that could be seen, heard, felt, or manipulated. Other papers allowed users to \textit{free-form express} thoughts, feelings, or experiences through an open expressive medium such as art (n=9), directly transformed and \textit{embodied the user's experience} into an interactive (often digital) object to manipulate (n=7), or \textit{AI-generated representations} of the user's emotions, such as into a poem or music (n=4). 

\subsection{Orientation 2: \textit{Quantifying the Self}}
\label{framework: quantifying}
\vspace{-1em}
\mybox{\small{\textit{``[DeepStress] visualizes self-tracked data such as stress changes over time, and provides summarized stress information for each context...DeepStress facilitated data-driven self-reflection, enabling users to pinpoint stressful contexts.''} \cite{10.1145/3613904.3642766}}}

Interventions categorized in \textsf{Quantifying the Self} similarly rely on self-understanding for internal parts of the self, but instead for users to gain knowledge on \textit{data and patterns} and inform subsequent choices. Whereas \textsf{Making the Self Tangible} primarily renders experience into an expressive form for meaning-making, interventions for \textsf{Quantifying the Self} instead view the self-state as data evidence from which factual measurement and knowledge about the self's patterns can be made. This orientation carries an epistemic claim: by centering the recording and pattern-making of self-data, the intervention positions the self as knowable through aggregation and observation over time. Interventions use experiences such as mood, sleep, medication, and other routines as states to be documented and made sense of through pattern. Note that quantification as used in this work can refer more broadly to structuring of experience as evidence for patterns, rather than only numerical measurement. As a result, interventions that allow users to record a log of qualitative data can also carry the same assumptions in this orientation, where patterns are to be extracted and aggregated across time or context.

\textsf{Quantifying the Self} closely draws from personal informatics and the ``Quantified Self'', which frame personal data as resources for self-knowledge \cite{lupton2016quantified}.
%, as well as Foucault's concept of turning the person into an object of observation \cite{foucaultTechnologiesSelfSeminar1988}.  
Interventions in this orientation position technology as an instrument that records these states and makes them available for the user's observation, either as care in itself or to prompt future action. The underlying assumption of these interventions is that user-led or system-led recording and surfacing of data patterns could reveal something consequential about the person; interventions thus decide what measurements and patterns count as meaningful knowledge about the self.

\subsubsection{Function of Technology} 
In \textsf{Quantifying the Self}, 14 papers were centered around \textit{facilitating users to self-track data} in order to surface patterns, while 3 papers had interventions focused on contributing \textit{sensing} or passive recording methods. Other functionalities present included \textit{aggregating information} into summaries or visualizations for users to view (n=4) and \textit{prediction} (such as of relapse or an oncoming mental health episode) based on the user's data (n=4).

\begin{figure*}[htbp]
    \centering
    \includegraphics[width=0.85\linewidth]{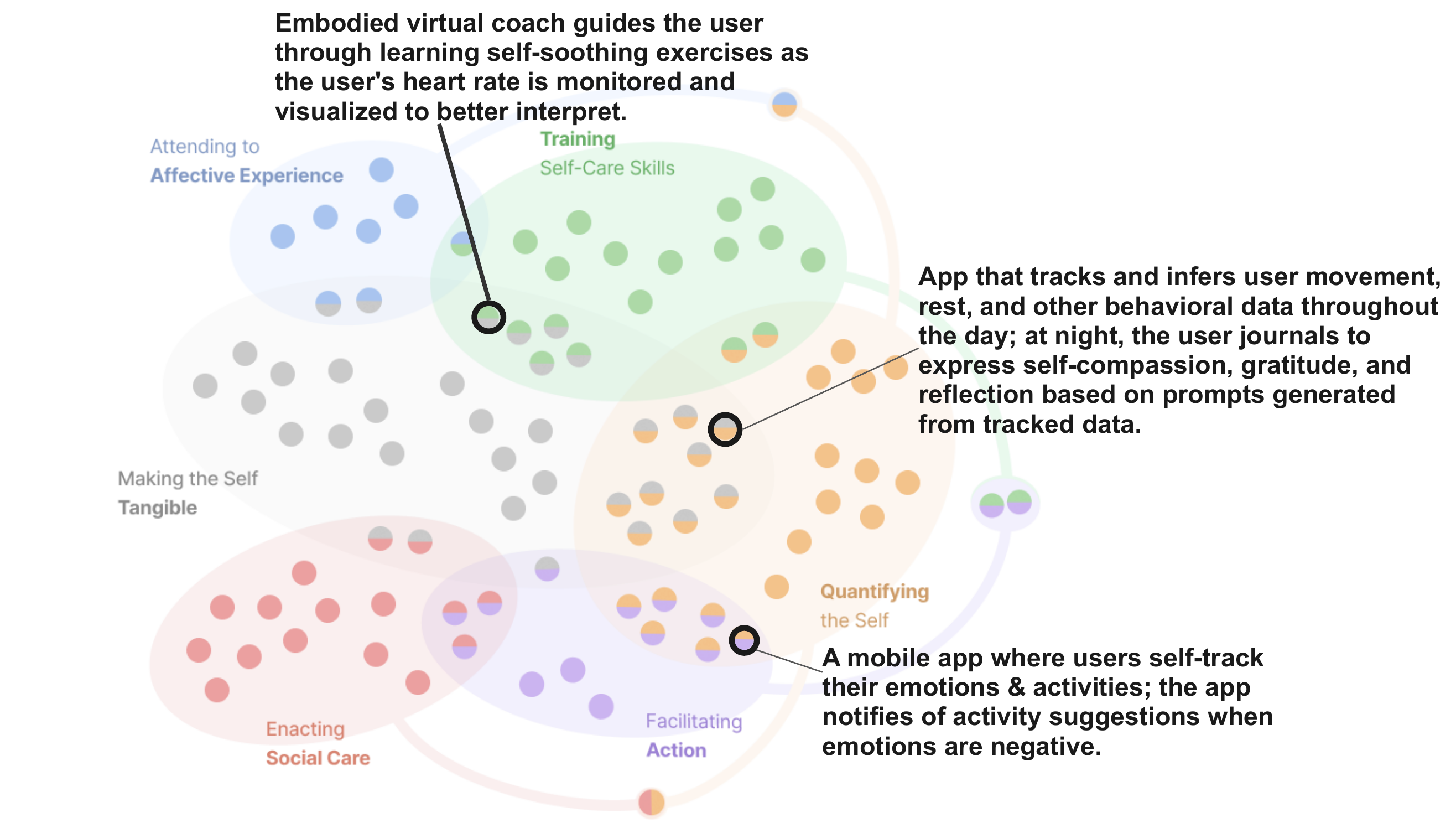}
    \caption{Examples of papers that touched on frequently overlapping orientations.}
    \label{fig:overlapping-orientations}
\end{figure*}

Conversely, an intervention that relies on data is not necessarily carrying the same assumptions as \textsf{Quantifying the Self} if data of the self is not treated as underlying patterns to be known. Mirror Ritual \cite{rajcic2020mirror} and DeepStress \cite{10.1145/3613904.3642766} illustrate the distinction; while Mirror Ritual detects facial expression to generate a poem and offer a representation with which the user can interpret their experience and emotion, DeepStress organizes self-tracked stress information across time and context to help users identify stressful situations. We interpret the former as emphasizing the expressive and interpretive possibilities of a representation for the self-state, and the latter as emphasizing patterns made available through recorded data. The distinction therefore concerns the role that a representation plays in care, rather than whether the system uses data; of course, interventions may also perform both care actions (Figure \ref{fig:overlapping-orientations}). Papers in this orientation did often touch on other self-care orientations as well. For example, six papers also notified or nudged the user to do some action (\textsf{Facilitating Action}, see Section \ref{framework: actions}) based on patterns identified through this quantification.

\subsection{Orientation 3: \textit{Training Self-Care Skills}}
\label{framework: training}
\vspace{-1em}
\mybox{\small{\textit{``mobile application that leverages the power of storytelling ... interactive scenarios and contextualised storylines can not only educate users about the psychological concepts that underpin the process of CBT but also enable the internalisation of these learned concepts'' }\cite{hamid2022you}}}
\vspace{-1em}
%"Youth Compass is a structured and interactive web-based intervention program developed to support adolescent psychological flexibility and well-being by taking the participants through exercises and activities over five weeks" \cite{10.1145/3571884.3597142}
In interventions focused on \textsf{Training Self-Care Skills}, the self is conceived as having a set of \textit{capabilities} that can be developed through learning and practice. Rather than treating well-being as a state to be represented or measured, these interventions position the self as an agent whose ability to care for itself can be cultivated. Therefore, the user pursues these interventions to develop their future capacity to (independently) recognize, regulate, or respond to difficult experiences. 

The self-care abilities being trained ranged widely, from mindfulness, to regulatory skills, to cognitive restructuring or reappraisal. The mechanism and interaction dimensions used to teach these capabilities also vary. For example, gaming and storytelling were used in works such as \cite{hamid2022you} but with the aim of \textit{teaching} students how to identify irrational and automatic thoughts, while other interventions use modalities such as VR to help people learn the skill of (and enact) cognitive reappraisal skills \cite{kitson2023co}. Training was thus broader than presenting didactic material but instead also included technology that provided rehearsal or feedback \cite{10.1145/3718084, schroeder2018pocket}. Technology thus takes on a pedagogical role in mediating care by building knowledge \cite{zimmerman2002becoming,bandura1982self}. Naturally, an intervention encodes a view of which capabilities are relevant to competent self-care and how those capabilities should be performed. This authority has historically sat with clinicians and evidence-based frameworks; however, technological interventions may encode these frameworks within their design. Overall, the self is not simply something to care for, but rather something that can be taught to care for itself.

\subsubsection{Function of Technology} 
The dominant function that technology filled was directly \textit{educating about a skill}, including informational support (n=13). These systems explained therapeutic concepts, modeled a technique, and provided instructional content. Other papers guided users through exercises or behaviors step-by-step. Such guidance supported \textsf{Training Self-Care Skills} when aimed at developing a capability for subsequent use, and \textsf{Facilitating Action} when aimed at just prompting or supporting an action (Section \ref{framework: actions}); papers could receive both orientations if both care logics were central. This orientation also often overlapped with Making the Self Tangible when interventions translated signals like breathing into a visible form that could also be manipulated by trained exercises (e.g., deep breathing). 

%Other functions appeared less frequently and typically supported the training process. For example, systems incorporated user-recorded data or expression, but with the end purpose of teaching and training a skill. For these interventions that may have been coded as these other orientation, we additionally coded them as \textit{Training Self-Care Skills} when these functions were explicitly organized toward acquiring or strengthening a self-care capability.

\subsection{Orientation 4: \textit{Facilitating Action}}
\label{framework: actions}
\vspace{-1em}

\mybox{\small{``\textit{enables children to set their own task goals, receive reminders about these goals, implement them, and confirm their progress}'' \cite{park2024collaborative}}}

While the self has goals and ideals, acting out behaviors to reach these goals may require support and reminders \cite{fogg2003persuasive,consolvo2006design}. In other words, the self is an intentional agent whose goals may not be achieved through desired or \textit{direct actions} without mediated care. In interventions that rely on \textsf{Facilitating Action}, technology provides an external agent of care by directing the user toward executing an action; therefore, people care for themselves by actually externalizing part of the self's executive function to technology. 

Interventions in this category varied in who established the desired action. For example, the user may define desired actions and the intervention serves to remind the user to execute those actions \cite{park2024collaborative, elvitigala2021stressshoe}, or responsibility lies in the intervention designers or system recommendations to predefine beneficial behaviors \cite{rohani2020mubs}. In either direction, though, the system determines moments that warrant action and the means by which those actions are triggered as desirable response. Care therefore targets the gap between \textit{wanting or intending} and \textit{enacting}.

\subsubsection{Function of Technology}
Interventions in \textsf{Facilitating Action} address a gap between intention and enactment by \textit{notifying or nudging users towards some action} (n=9) and guiding users through a practice step-by-step (n=3). This nudging can occur through reminders, goal structuring, and behavioral recommendation.

Interventions that addressed \textsf{Facilitating Action} often also addressed other orientations; it is relatively rare for a contribution to be solely focused on notification or nudging without also involving other aspects -- such as quantification -- to guide that notification or training as the end goal. For example, Home Sweet Office periodically prompted online workers to complete one of 160 stress-management activities \cite{10.1145/3544548.3581319}, \textit{StressShoe} inferred stress from foot movement to then trigger user-defined interventions \cite{elvitigala2021stressshoe}, and \textit{CounterStress} helped users view stress data and imagine alternative behaviors in order to nudge users towards alternative coping habits \cite{10.1145/3706598.3713730}. In all cases, technology actively forwarded the user toward a behavioral action. \textsf{Quantifying The Self} and \textsf{Facilitating Action} often overlapped, forming a loop in which recorded or inferred states are surfaced but also guide a prompted intervention to address inferred mental health states.

\subsection{Orientation 5: \textit{Enacting Social Care}}
\label{framework: social}
\mybox{\small{``\textit{an in-depth study on how LLM-driven agents, with their contextual awareness, understanding, and generation capabilities, can influence self-disclosure in older adults.}'' \cite{guo2025exploring}}}

While our review is focused on self-care interventions, the self is not necessarily always conceived as a lone individual; instead, as theories establish \cite{bickmore2005establishing,murnane2018personal}, the self has fundamentally \textit{social and relational} aspects, and requires care to its aspects of connection, belonging, and social roles. Therefore, interventions oriented towards \textsf{Enacting Social Care} focus on the assumption that people's social instincts are core to their mental health. This orientation complicates a classic individual understanding of self-care; while the user remains the initiator and central actor in the self-care practice, the care itself draws upon relational resources and addresses a self shaped through interactions and relationships, or social actions; relational experiences are relevant to self-care even when another person is not directly involved. Note, though, that this orientation should be interpreted in relation to our eligibility criteria, which excluded care dependent on active human co-regulation.

\subsubsection{Function of Technology} 

Interventions achieved caring for the `social' part of self through relational interaction. However, since we exclude papers reliant on co-regulation, almost all interventions in this orientation focused on \textit{simulation and mimicry of people and social behaviors}. The prominence of simulated interaction within this orientation should be understood in relation to our eligibility criteria, which exclude care dependent on another person’s active co-regulation. Within this scope, these interventions position social interaction as a care mechanism through technology but don't require another person's availability  \cite{andrade2014barriers}. This finding reveals how designers incorporate relational forms of support into independent, self-directed interventions.

\subsection{Orientation 6: \textit{Attending to Affective Experience}}
\label{framework: affective}
\vspace{-1em}
\mybox{\small{``\textit{people will engage in physically comforting activities (e.g., cuddling) with a home companion robot...and will provide care.}'' \cite{10.1145/3643834.3660702}}}
People care for their \textit{felt experience} with interventions in \textsf{Attending to Affective Experience}, seeking to soothe or alter how they feel in the moment. Unlike orientations in which users seek to understand themselves or train a capability, the interaction itself becomes the resource through which the person attends to an immediate affective need. Technology supplies sensory or interactive conditions for comfort, while the user initiates, receives, and regulates their engagement with that experience.
While all self-care, in some form, is about making the user feel better, interventions in this orientation tend to deliver immediate emotional experience, and interacting or selecting these interventions is itself the \textit{direct} and \textit{primary} act of care that a user chooses for helping themselves. 
For example, \textit{SnuggleBot} used a social robot that combined warmth, softness, and movement to provide physical comfort through cuddling \cite{10.1145/3643834.3660702}.

HCI has a longstanding engagement with affect and experience-centered design \cite{picard1997affective, mccarthy2007technology}. Affective interaction research has challenged an informational model of the self by emphasizing emotion as subjective, embodied, and produced through interaction \cite{boehner2005affect}. Overall, this orientation relies on systems to give care in the most direct, immediate manner and directly shapes the user's affect. 
%The orientation can also be read through the lens of emotional regulations, with focus on how emotional experience are shaped through changes to situations and interpretations \cite{gross1998emerging}.

\subsubsection{Functions of Technology}
The two dominant functions directly reflected the theme’s affective mechanism. Five papers \textit{directly regulated} the user -- for example, by delivering soothing sensory stimulation -- and five \textit{evoked an emotion or memory} through mediation like music. 

The boundary of this orientation therefore depends on the role that affect and the directness of care plays. Unlike \textsf{Quantifying the Self}, where affect becomes information about the self, or \textsf{Making the Self Tangible}, where internal experience is externalized for reflection, this orientation treats the experience itself as the object of care. An emotion tracker treats affect as information to be made knowable, while a metaphorical representation may make an emotion interpretable. An affective intervention, by contrast, has technology assuming the role of delivering comfort as the affective medium. 

\subsection{Summary and Reflection on the Six Orientations}
While mental health and self-care technology can be defined by modality, population, therapeutic approach, and activity \cite{ahmed2021mobile, thieme2020machine, 10.1145/3780045.3780052}, interventions across these factors may also share assumptions of ``who'' the self is to be cared for and what technology's role is in enacting care. The above orientations instead connect an intervention’s observable functions to an underlying care logic regarding what part of the self is consequential and what change is understood as care. This common structure makes it possible to compare interventions that appear technically similar but enact different forms of care, as well as interventions that appear technically different but address care and authority in similar ways. 

These orientations are distinct but not mutually exclusive; overlaps show how interventions configure several kinds of technological care work. The lens that this framework offers can be illustrated through examples. DMHIs for journaling, for example, may appear similar -- all presumably enable users to write and express their thoughts or emotions. However, our presented orientations instead distinguish interventions according to how technological mediation enacts care. A journaling system that provides an open written space is a user-mediated transformation within \textsf{Making the Self Tangible}, but a system that also aggregates entries to show patterns is \textsf{Quantifying the Self}. Another intervention might guide the user through their writing, such as through cognitive restructuring \textsf{Training Self-Care Skills}. A second group of examples is biofeedback interventions using heart rate signals for self-care \cite{sanches2019hci}. Systems that transform these signals into interactive representations for the user to make meaning of are \textsf{Making the Self Tangible}, but those that use these measurements to trigger recommendations are \textsf{Facilitating Action}. A third example are virtual and conversational agents \cite{li2023systematic}. While interface can appear similar, a conversational agent that helps the user practice and build a skill such as through virtual coaching (\textsf{Training Self-Care Skills}) carries distinct assumptions of caring for the self from one that socializes with the user (\textsf{Enacting Social Care}) or offers direct comfort (\textsf{Attending to the Affective Experience}). Consequently, our presented six orientations offer a new lens to view self-care technology, through the relationship between assumptions of \textit{self}, the \textit{care} operation by technology, and the \textit{technological functions} that enable that care.

%% file: sections/06-discussion.tex
\section{Discussion}
Through our analysis of 91 SIGCHI papers published over ten years regarding interventions for mental health self-care, we discovered distinct design and analytic layers of mapping self-care interventions: \textit{for whom} and \textit{for what needs} they are designed (Section \ref{findings:populations}, Section \ref{findings:mental-health-topics}), \textit{what the user does} (Section \ref{findings:activities}), \textit{what contribution} the intervention addresses (Section \ref{findings:contributions}), and \textit{with what assumptions} (Section \ref{framework}). This layered account offers a vocabulary for positioning and evaluating future interventions in mental health self-care. Taken together, our work shows key design decisions that can serve to redistribute care work among users, systems, and designers (Section \ref{findings:contributions}) and different relationships between self, care, and technology in self-care interventions (Section \ref{framework}). Below, we discuss implications for aligning design and evaluation with care assumptions, supporting agency and responsibility in self-care, and implications for situating HCI intervention bounds when addressing ``general'' users.

\subsection{Moving Towards Orientations of Care in Positioning HCI Self-Care Interventions}
The six orientations offer a lens for researchers to position their work through the underlying accounts of how technology is expected to enact care. Making this account explicit clarifies an intervention’s theory of care, i.e., the sequence of assumptions connecting an interaction with a system to a user's anticipated mental health benefit. As reviewed, orientations cut across activity and technical approach, and instead ask what aspects of self that a user seeks care for and what care constitutes.

Surfacing care logic in this way can help align evaluation with intervention intent. Usability and acceptability, commonly evaluated in mental health technology research \cite{inal2020usability, catania2024evaluation}, do not necessarily align with the aspect of the self that a user aims to care for and how technology is expected to meet those goals (e.g., learning and applying subsequent skills for \textsf{Training Self-Care Skills}). These distinct orientations and evaluation needs naturally require distinct forms of evidence. For example, evaluating \textsf{Quantifying the Self} may require examining whether the knowledge users derive from system representations is meaningful and appropriately grounded; for \textsf{Training Self-Care Skills}, evaluation might examine whether users can apply learned skills beyond the intervention; and for \textsf{Attending to Affective Experience}, immediate changes in affect may be most relevant. 

The six orientations can also support generative reflection for researchers, as choice of an intervention's technology functions should align with underlying definitions of care and aspects of self to care for (Table \ref{tab:self-care-framework}). However, we acknowledge that our orientations are a first step from the existing corpus, and new orientations may also emerge in the future. The orientations can also help researchers examine how similar technologies enact different forms of care. A journaling intervention, for example, might support expression through \textsf{Making the Self Tangible}, derive patterns through \textsf{Quantifying the Self}, or guide cognitive restructuring through \textsf{Training Self-Care Skills}. Choosing among these functions changes both the anticipated benefit and technology's role in producing it. Researchers can therefore use the orientations to ask whether adding a function supports the intended care or introduces a different care logic that requires justification. When an intervention combines orientations, articulating their relationship can clarify whether they complement one another or introduce tensions, such as between meaning-making and datafication of the self. Overall, these orientations offer sensitizing concepts for articulating and examining deliberate choices about the care an intervention provides.

\subsection{Agency and Responsibility in Self-Care Technology}
\label{discussion:agency}
Our findings in Section \ref{findings:contributions} highlighted how self-care interventions can reconfigure responsibilities. As papers in our corpus focused on expanding access to underserved populations, promoting sustained engagement for adherence, and translating existing practices into more accessible forms, these all served as a type of response to gaps in conventional mental health resources \cite{santomauro2026updated,who2025world,patel2018lancet,andrade2014barriers}. These interventions can redistribute work otherwise shared among individuals, professionals, and support networks, placing greater demands on users and embedding some care decisions in system design \cite{nissenbaum2001computer, nissenbaum1997accountability, sengers2005reflective}. Consequently, self-care technology relies on users to be \textit{active}, \textit{responsible}, and \textit{technically competent} \cite{oudshoorn2011telecare}.

As articulated by Mol's work on patient choice, giving users independent access to health interventions can create a `logic of choice' that positions users as responsible for appropriate action and any downstream effects from that choice, while `logic of care' is instead a design for the ongoing, continuously changing practice of care \cite{mol2008logic}. Our work that surfaces how self-care interventions reconfigure responsibility inherently leads to implications on whether interventions offer logic of choice versus logic of care implications for users, and whether users are supported by shared responsibility and responsiveness \cite{mol2008logic}. We connect this idea to an example from our corpus, where 36.3\% of papers contributed self-care interventions aimed at improving engagement and sustained adherence to mental health interventions (Section \ref{findings:contributions:engagement}). In conventional care, sustaining participation and adapting care often involves ongoing relational work, e.g., noticing changes, interpreting disengagement, renegotiating goals, and adjusting practices \cite{norcross2018new}. Self-care interventions redistribute parts of this work into technology features (e.g., reminders, motivational features) but leaves much of it up to the user -- not only to initiate and engage in an intervention, but at times even to manually configure the treatment intervention itself (Section \ref{findings:contributions:engagement}). Similarly, discontinued use -- a common barrier cited as limiting effectiveness of DMHIs -- can be interpreted primarily as a user's failure to adhere to an available intervention (`logic of choice'), or evidence that the care support requires fundamental adjustment (`logic of care') \cite{elish2019moral, norcross2018new, horvath1993role}. 

Because of reconfiguring responsibilities, self-care technology can thus make users bear both risk of harm from ill-fit interventions and blame for failing to manage their care \cite{brown2019against, ruse2024left}. However, while users are active participants in their self-care, this still occurs within conditions structured by an intervention and its designers. For example, systems that transform inner experience into external representations hold implicit governance over how to surface signals and its effects, and systems that teach self-care skills can assume pedagogical authority (Section \ref{framework}). Conventional care resources also enforce conditions and make consequential judgments; however, self-care technology can be portrayed as more agentic for the user while operating outside contexts in which care decisions can be more fundamentally updated. Thus, technological responsibility can be less accountable, visible, and malleable \cite{nissenbaum1997accountability}. This is an echo of Nissenbaum's work from 1997 (which we now apply to the self-care context) that technological mediation means that people are simultaneously ``\textit{assured of answerability}'' when the reality is ``an \textit{obscuring} of lines of accountability'' \cite{nissenbaum1997accountability}. 

Given this, we want to surface in this work that \textit{self-care interventions can be user-directed but not necessarily user-centered}. Beyond explicitly designing for aspects of self, care, and technology's role (as in Section \ref{framework}), designers may also consider how an intervention performs ongoing `logic of care' and the degree of true user agency. This can be closely tied to an orientation -- e.g., for \textsf{Quantifying the Self}, resistance may be tied to both recording and surfacing of data such as being able to question or alter interpretations -- so our work may be a helpful initial step in this direction. Thus, the goal is to acknowledge that expanding user agency by turning care interventions into self-directed, self-initiated practices is also accompanied by designing for the new relations and obligations through which care now becomes possible. Researchers should engage in iterative reflection during intervention development on what ``responsibilities'' are required (i.e., users needing to recognize needs, interpret outputs, evaluate recommendations, sustain engagement) and how the design supports them. Designs might support variable capacity through adjustable levels of guidance, opportunities to revise or contest system interpretations, involvement of trusted others, and pathways to other resources. The appropriate support will depend on the intervention’s orientation and intended context. Relatedly, reporting dropout, refusal, and discontinued use is helpful as evidence about the relationship between a person and an intervention. Since one of the biggest barriers to self-care interventions is dropout \cite{banos2022current}, transparent acknowledgement can aid in understanding unmet accessibility needs or excessive self-management demands in intervention design and evaluation.

\subsection{Reconsidering the “General” User}
\label{discussion:general-user}
47.3\% of papers in our corpus either explicitly described their intended users as a ``general population'' or did not establish a population-specific boundary based on any health, demographic, social, or other characteristic. We group these papers descriptively but also distinguish the consequences of broad targeting and unspecified scope below. 

As reviewed in Section \ref{findings:populations}, papers coded as ``general population'' often excluded participants in their study based on clinical symptoms, diagnoses, therapy involvement, or medication, or did not report enough information to definitively determine whether mental health conditions was part of inclusion/exclusion criteria. In a domain such as mental health where well-being moves along a continuum between healthy, distressed, and recovery \cite{keyes2002mental,patel2018lancet}, and at a time when over a billion people worldwide currently live with a mental health disorder \cite{santomauro2026updated}, exclusion constructs a paradoxical general user. Some papers in our corpus also excluded people who have \textit{prior} diagnosis or treatment, but half of all people are expected to experience a mental health disorder in their lifetime \cite{santomauro2026updated}. As a result, this evaluation then represents only a narrow portion of the stated population. For papers that do not specify enough information on target user population or participant health criteria, it is then difficult to infer the intervention’s evidentiary boundaries; however, mental health interventions require some account of whom they are designed to support and under what conditions they can be used safely.  

A common rationale for restricting participation in self-care interventions is to protect participants when researchers cannot provide clinical oversight, especially since there are known harms possible in mental health or care technology \cite{lawrence2024opportunities,wang2026care}. While we can appreciate this, we also identify two problems with this rationale. First, exclusion \textit{shifts rather than resolves} uncertainty. Exclusion is the field's primary safety strategy \cite{taher2023safety}, but there are ethical risks in excluding people with moderate or severe symptoms from digital mental health research \cite{mccall2021reconsidering}; namely, this constrains ecological validity and leaves safety least understood for people who may nevertheless encounter the technology, most in-need of alternative resources, and have particular design needs and considerations that HCI methods could well address \cite{taher2023safety, biagianti2017developing, thieme2016challenges}. This tension is especially consequential for self-care interventions, which is often justified precisely \textit{by} their availability outside formal care and without continuous oversight (see Section \ref{findings:contributions:accessibility}). In some cases, papers do not define any intended population, which also limits shared understanding for the applicability of the research. Second, the \textit{normalization} of this exclusion in HCI mental health interventions forms a consequential construction of the ``general'' or ``healthy'' user. Exclusion along these lines can position help-seeking and diagnosis as evidence of a far more nuanced and changing mental state, risk further stigma by using help-seeking behaviors as evidence of ``illness'' that interventions avoid evaluation on, and collapses people into binary distinction of mentally healthy vs. unhealthy. This pattern reveals an implicit model of a person who can partake in technologically mediated self-care, counterintuitively to the framing of self-care in many of these works as valuable for being wholly accessible.

Beyond being responsible and active (Section \ref{discussion:agency}), the user is also imagined as \textit{sufficiently stable}, \textit{self-aware}, and \textit{resourced} to perform self-care work. This echoes critiques of DMHIs that frame distress as universal while representing a narrow and privileged version of `everyone' \cite{parker2018mental}. Our findings in Section 4 also reflect elements of this as -- beyond user -- HCI may also form implicit judgment on which mental health experiences are a problem of `everyone' and which continue to be medicalized (e.g., anxiety, Section \ref{findings:mental-health-topics}).

For future mental health self-care research, we suggest that researchers provide an account of an intervention's intended range and evidentiary boundaries. This includes mental health inclusion/exclusion for target population, recruitment population, and analyzed sample, and reporting diagnosis, treatment status, or symptom severity as part of participant screening to give context to their perspectives, without excluding them entirely. Screening participants for mental health experience can help us understand mental health status as a variable to consider intervention effects, but studies should include people along the spectrum of symptoms if the intervention is meant to be applied as such. Self-care capacity and needs also can change over time; interventions then need to explicitly dictate conditions under which an intervention is appropriate, rather than presume static mental health states.

\subsection{Limitations}
This review has several limitations. First, we searched selected ACM HCI venues and TOCHI rather than broader digital health literature. Our findings therefore characterize how self-care has been designed and discussed only within the HCI and SIGCHI corpus. Venues in medical informatics and digital psychiatry are likely to address patient populations and define scope of their interventions, for example, and may not be subject to our discussion on general population framing. Thus, our findings in this work are best read as a pattern of \textit{HCI}'s approach to self-care rather than self-care technology literature as a whole. Similarly, this review is restricted in scope to English-language work. Second, our search strategy operationalized the review’s scope through papers that positioned themselves as addressing mental (or psychological and/or emotional) health in their titles, abstracts, or keywords. Consequently, the corpus may not include interventions addressing more specific experiences when authors did not situate that work explicitly within mental health or well-being. However, forward and backward citation searching helps to mitigate this risk. Third, relevant papers did not consistently identify their work as “self-care.” Although our eligibility criteria were grounded in prior definitions and refined through team discussion, determining whether an intervention was sufficiently user-initiated and self-directed required interpretation. The six orientations are likewise an interpretive synthesis rather than categories explicitly named by paper authors. They are intended as a useful analytic vocabulary, not an exhaustive ontology. Lastly, we acknowledge that the orientation framework in Section \ref{framework} emerged during analysis rather than in the original preregistration.

%% file: sections/07-conclusion.tex
\section{Conclusion}
In this paper, we conducted a scoping review of 91 SIGCHI papers on HCI interventions for mental health self-care published over the last decade. We identified five themes of user populations that interventions are designed for, 12 mental health topics that interventions target, and 13 types of self-care activities presented in our corpus' interventions. Additionally, we presented an emergent framework outlining six orientations in how HCI positions self-care interventions, which each outline assumptions on the aspect of self that people require care for, what care constitutes, and the role that technology plays to support care. Lastly, we discussed how our work provides a layered vocabulary to position future work, and implications for the redistribution of agency and responsibility for users in self-care interventions.

%% file: ref.bib
@book{world2022world,
  title={World mental health report: Transforming mental health for all},
  author={World Health Organization},
  year={2022},
  publisher={World Health Organization}
}

@article{nunes2015self,
  title={Self-care technologies in HCI: Trends, tensions, and opportunities},
  author={Nunes, Francisco and Verdezoto, Nervo and Fitzpatrick, Geraldine and Kyng, Morten and Gr{\"o}nvall, Erik and Storni, Cristiano},
  journal={ACM Transactions on Computer-Human Interaction (TOCHI)},
  volume={22},
  number={6},
  pages={1--45},
  year={2015},
  publisher={ACM New York, NY, USA}
}

@misc{who2026selfcare,
  author       = {{World Health Organization}},
  title        = {Self-care for Health and Well-being},
  year         = {2026},
  month        = jun,
  url          = {https://www.who.int/news-room/fact-sheets/detail/self-care-health-interventions},
  note         = {Accessed August 16, 2026}
}

@article{tricco2018prisma,
  title={PRISMA extension for scoping reviews (PRISMA-ScR): checklist and explanation},
  author={Tricco, Andrea C and Lillie, Erin and Zarin, Wasifa and O'Brien, Kelly K and Colquhoun, Heather and Levac, Danielle and Moher, David and Peters, Micah DJ and Horsley, Tanya and Weeks, Laura and others},
  journal={Annals of internal medicine},
  volume={169},
  number={7},
  pages={467--473},
  year={2018},
  publisher={American College of Physicians}
}

@article{patel2018lancet,
  title={The Lancet Commission on global mental health and sustainable development},
  author={Patel, Vikram and Saxena, Shekhar and Lund, Crick and Thornicroft, Graham and Baingana, Florence and Bolton, Paul and Chisholm, Dan and Collins, Pamela Y and Cooper, Janice L and Eaton, Julian and others},
  journal={The lancet},
  volume={392},
  number={10157},
  pages={1553--1598},
  year={2018},
  publisher={Elsevier}
}

@inproceedings{pendse2021can,
  title={“Can I not be suicidal on a Sunday?”: understanding technology-mediated pathways to mental health support},
  author={Pendse, Sachin R and Sharma, Amit and Vashistha, Aditya and De Choudhury, Munmun and Kumar, Neha},
  booktitle={Proceedings of the 2021 CHI Conference on Human Factors in Computing Systems},
  pages={1--16},
  year={2021}
}

@article{stade2024large,
  title={Large language models could change the future of behavioral healthcare: a proposal for responsible development and evaluation},
  author={Stade, Elizabeth C. and Stirman, Shannon Wiltsey and Ungar, Lyle H. and Boland, Cody L. and Schwartz, H. Andrew and Yaden, David B. and Sedoc, Jo{\~a}o and DeRubeis, Robert J. and Willer, Robb and Eichstaedt, Johannes C.},
  journal={npj Mental Health Research},
  volume={3},
  number={1},
  pages={12},
  year={2024},
  doi={10.1038/s44184-024-00056-z}
}

@article{li2023systematic,
  title={Systematic review and meta-analysis of AI-based conversational agents for promoting mental health and well-being},
  author={Li, Han and Zhang, Renwen and Lee, Yi-Chieh and Kraut, Robert E and Mohr, David C},
  journal={NPJ Digital Medicine},
  volume={6},
  number={1},
  pages={236},
  year={2023},
  publisher={Nature Publishing Group UK London}
}

@book{oudshoorn2011telecare,
  title={Telecare technologies and the transformation of healthcare},
  author={Oudshoorn, Nelly},
  year={2011},
  publisher={Springer}
}

@inproceedings{bowman2023using,
  title={Using thematic analysis in healthcare HCI at CHI: A scoping review},
  author={Bowman, Robert and Nadal, Camille and Morrissey, Kellie and Thieme, Anja and Doherty, Gavin},
  booktitle={Proceedings of the 2023 CHI conference on human factors in computing systems},
  pages={1--18},
  year={2023}
}

@article{mcdonald2019reliability,
  title={Reliability and inter-rater reliability in qualitative research: Norms and guidelines for CSCW and HCI practice},
  author={McDonald, Nora and Schoenebeck, Sarita and Forte, Andrea},
  journal={Proceedings of the ACM on human-computer interaction},
  volume={3},
  number={CSCW},
  pages={1--23},
  year={2019},
  publisher={ACM New York, NY, USA}
}

@article{wang2018tracking,
  title={Tracking Depression Dynamics in College Students Using Mobile Phone and Wearable Sensing},
  author={Wang, Rui and Wang, Weichen and daSilva, Alex and Huckins, Jeremy F. and Kelley, William M. and Heatherton, Todd F. and Campbell, Andrew T.},
  journal={Proceedings of the ACM on Interactive, Mobile, Wearable and Ubiquitous Technologies},
  volume={2},
  number={1},
  pages={43},
  year={2018},
  doi={10.1145/3191777}
}

@inproceedings{consolvo2006design,
  title={Design Requirements for Technologies that Encourage Physical Activity},
  author={Consolvo, Sunny and Everitt, Katherine and Smith, Ian E. and Landay, James A.},
  booktitle={Proceedings of the SIGCHI Conference on Human Factors in Computing Systems},
  pages={457--466},
  year={2006},
  publisher={ACM},
  doi={10.1145/1124772.1124840}
}

@book{levin2026self,
  title={Self-care: Lay initiatives in health},
  author={Levin, Lowell S and Katz, Alfred H and Holst, Erik},
  year={2026},
  publisher={Taylor \& Francis}
}

@book{dourish2001action,
  title={Where the action is: the foundations of embodied interaction},
  author={Dourish, Paul},
  year={2001},
  publisher={MIT press}
}

@article{bickmore2005establishing,
  title={Establishing and maintaining long-term human-computer relationships},
  author={Bickmore, Timothy W and Picard, Rosalind W},
  journal={ACM Transactions on Computer-Human Interaction (TOCHI)},
  volume={12},
  number={2},
  pages={293--327},
  year={2005},
  publisher={ACM New York, NY, USA}
}

@book{picard1997affective,
  title={Affective computing},
  author={Picard, Rosalind W},
  year={1997},
  publisher={The MIT press}
}

@article{taher2023safety,
  title={The safety of digital mental health interventions: systematic review and recommendations},
  author={Taher, Rayan and Hsu, Che-Wei and Hampshire, Chloe and Fialho, Carolina and Heaysman, Clare and Stahl, Daniel and Shergill, Sukhi and Yiend, Jenny},
  journal={JMIR mental health},
  volume={10},
  number={1},
  pages={e47433},
  year={2023},
  publisher={JMIR Publications Inc., Toronto, Canada}
}

@inproceedings{boehner2005affect,
  title={Affect: from information to interaction},
  author={Boehner, Kirsten and DePaula, Rog{\'e}rio and Dourish, Paul and Sengers, Phoebe},
  booktitle={Proceedings of the 4th decennial conference on Critical computing: between sense and sensibility},
  pages={59--68},
  year={2005}
}

@article{parker2018mental,
  author  = {Parker, Lisa and Bero, Lisa and Gillies, Donna and Raven, Melissa
             and Mintzes, Barbara and Jureidini, Jon and Grundy, Quinn},
  title   = {Mental Health Messages in Prominent Mental Health Apps},
  journal = {The Annals of Family Medicine},
  year    = {2018},
  volume  = {16},
  number  = {4},
  pages   = {338--342},
  doi     = {10.1370/afm.2260}
}

@article{mccall2021reconsidering,
  title={Reconsidering the ethics of exclusion criteria in research on digital mental health interventions},
  author={McCall, Hugh C and Hadjistavropoulos, Heather D and Loutzenhiser, Lynn},
  journal={Ethics \& Behavior},
  volume={31},
  number={3},
  pages={171--180},
  year={2021},
  publisher={Taylor \& Francis}
}

@article{murnane2018personal,
  title={Personal informatics in interpersonal contexts: towards the design of technology that supports the social ecologies of long-term mental health management},
  author={Murnane, Elizabeth L and Walker, Tara G and Tench, Beck and Voida, Stephen and Snyder, Jaime},
  journal={Proceedings of the ACM on Human-Computer Interaction},
  volume={2},
  number={CSCW},
  pages={1--27},
  year={2018},
  publisher={ACM New York, NY, USA}
}

@article{clarke2017thematic,
  title={Thematic analysis},
  author={Clarke, Victoria and Braun, Virginia},
  journal={The journal of positive psychology},
  volume={12},
  number={3},
  pages={297--298},
  year={2017},
  publisher={Taylor \& Francis}
}

@inproceedings{zhang2026asafeplace,
  title={ASafePlace: User-Led Personalization of VR Relaxation via an Art Therapy Activity},
  author={Zhang, Chuyang and Yu, Bin and Wang, Yuchao and Yuan, Mansi and Wang, Wanqi and Je, Seungwoo and An, Pengcheng},
  booktitle={Proceedings of the 2026 CHI Conference on Human Factors in Computing Systems},
  pages={1--24},
  year={2026}
}

@inproceedings{wagener2023selvreflect,
  title={SelVReflect: A guided VR experience fostering reflection on personal challenges},
  author={Wagener, Nadine and Reicherts, Leon and Zargham, Nima and Bart{\l}omiejczyk, Natalia and Scott, Ava Elizabeth and Wang, Katherine and Bentvelzen, Marit and Stefanidi, Evropi and Mildner, Thomas and Rogers, Yvonne and others},
  booktitle={Proceedings of the 2023 CHI Conference on Human Factors in Computing Systems},
  pages={1--17},
  year={2023}
}

@article{crawford1980healthism,
  title={Healthism and the medicalization of everyday life},
  author={Crawford, Robert},
  journal={International journal of health services},
  volume={10},
  number={3},
  pages={365--388},
  year={1980},
  publisher={SAGE Publications Sage CA: Los Angeles, CA}
}

@book{cederstrom2015wellness,
  title={The wellness syndrome},
  author={Cederstr{\"o}m, Carl and Spicer, Andr{\'e}},
  year={2015},
  publisher={John Wiley \& Sons}
}

@book{lupton2016quantified,
  title={The quantified self},
  author={Lupton, Deborah},
  year={2016},
  publisher={John Wiley \& Sons}
}

@article{ruckenstein2017datafication,
  title={The datafication of health},
  author={Ruckenstein, Minna and Sch{\"u}ll, Natasha Dow},
  journal={Annual review of anthropology},
  volume={46},
  number={1},
  pages={261--278},
  year={2017},
  publisher={Annual Reviews}
}

@article{schull2016data,
  title={Data for life: Wearable technology and the design of self-care},
  author={Sch{\"u}ll, Natasha Dow},
  journal={BioSocieties},
  volume={11},
  number={3},
  pages={317--333},
  year={2016},
  publisher={Springer}
}

@article{godfrey2011care,
  title={Care of self--care by other--care of other: The meaning of self-care from research, practice, policy and industry perspectives},
  author={Godfrey, Christina M and Harrison, Margaret B and Lysaght, Rosemary and Lamb, Marianne and Graham, Ian D and Oakley, Patricia},
  journal={International Journal of Evidence-Based Healthcare},
  volume={9},
  number={1},
  pages={3--24},
  year={2011},
  publisher={Wiley Online Library}
}

@article{sousa12002conceptual,
  title={Conceptual analysis of self-care agency},
  author={Sousa$^1$, Valmi D},
  year={2002}
}

@article{kickbusch1989self,
  title={Self-care in health promotion},
  author={Kickbusch, Ilona},
  journal={Social Science \& Medicine},
  volume={29},
  number={2},
  pages={125--130},
  year={1989},
  publisher={Elsevier}
}

@article{mccormack2003examination,
  title={An examination of the self-care concept uncovers a new direction for healthcare reform.},
  author={McCormack, Dianne},
  journal={Nursing Leadership (Toronto, Ont.)},
  volume={16},
  number={4},
  pages={48--62},
  year={2003}
}

@article{arksey2005scoping,
  title={Scoping studies: towards a methodological framework},
  author={Arksey, Hilary and O'malley, Lisa},
  journal={International journal of social research methodology},
  volume={8},
  number={1},
  pages={19--32},
  year={2005},
  publisher={Taylor \& Francis}
}

@article{riegel2021self,
  title={Self-care research: where are we now? Where are we going?},
  author={Riegel, Barbara and Dunbar, Sandra B and Fitzsimons, Donna and Freedland, Kenneth E and Lee, Christopher S and Middleton, Sandy and Stromberg, Anna and Vellone, Ercole and Webber, David E and Jaarsma, Tiny},
  journal={International journal of nursing studies},
  volume={116},
  pages={103402},
  year={2021},
  publisher={Elsevier}
}

@article{nissenbaum2001computer,
  title={How computer systems embody values},
  author={Nissenbaum, Helen},
  journal={Computer},
  volume={34},
  number={3},
  pages={120--119},
  year={2001},
  publisher={IEEE}
}

@article{elish2019moral,
  title={Moral crumple zones: Cautionary tales in human-robot interaction (pre-print)},
  author={Elish, Madeleine Clare},
  journal={Engaging Science, Technology, and Society (pre-print)},
  year={2019}
}

@article{nissenbaum1997accountability,
  title={Accountability in a computerized society},
  author={Nissenbaum, Helen},
  journal={Human values and the design of computer technology},
  pages={41--64},
  year={1997},
  publisher={Cambridge University Press Cambridge}
}

@article{brown2019against,
  title={Against moral responsibilisation of health: Prudential responsibility and health promotion},
  author={Brown, Rebecca CH and Maslen, Hannah and Savulescu, Julian},
  journal={Public Health Ethics},
  volume={12},
  number={2},
  pages={114--129},
  year={2019},
  publisher={Oxford University Press}
}

@article{biagianti2017developing,
  title={Developing digital interventions for people living with serious mental illness: perspectives from three mHealth studies},
  author={Biagianti, Bruno and Hidalgo-Mazzei, Diego and Meyer, Nicholas},
  journal={Evidence Based Mental Health},
  volume={20},
  number={4},
  pages={98--101},
  year={2017},
  publisher={BMJ Publishing Group Ltd, Royal College of Psychiatrists and British~…}
}

@article{catania2024evaluation,
  title={Evaluation of digital mental health technologies in the United States: systematic literature review and framework synthesis},
  author={Catania, Julianna and Beaver, Steph and Kamath, Rakshitha S and Worthington, Emma and Lu, Minyi and Gandhi, Hema and Waters, Heidi C and Malone, Daniel C},
  journal={JMIR mental health},
  volume={11},
  number={1},
  pages={e57401},
  year={2024},
  publisher={JMIR Publications Inc., Toronto, Canada}
}

@article{inal2020usability,
  title={Usability evaluations of mobile mental health technologies: systematic review},
  author={Inal, Yavuz and Wake, Jo Dugstad and Guribye, Frode and Nordgreen, Tine},
  journal={Journal of medical Internet research},
  volume={22},
  number={1},
  pages={e15337},
  year={2020},
  publisher={JMIR Publications Toronto, Canada}
}

@inproceedings{petterson2025expanding,
  title={Expanding Care Conceptualizations: An Integrative Literature Review of Care in HCI},
  author={Petterson, Adrian and Mattka, Jocelyn and Chandra, Priyank},
  booktitle={Proceedings of the ACM SIGCAS/SIGCHI Conference on Computing and Sustainable Societies},
  pages={481--503},
  year={2025}
}

@article{ruse2024left,
  title={Left to their own devices: the significance of mental health apps on the construction of therapy and care: JN Ruse et al.},
  author={Ruse, Jesse N and Schraube, Ernst and Rhodes, Paul},
  journal={Subjectivity},
  volume={31},
  number={4},
  pages={410--428},
  year={2024},
  publisher={Springer}
}

@book{mol2008logic,
  title={The logic of care: Health and the problem of patient choice},
  author={Mol, Annemarie},
  year={2008},
  publisher={Routledge}
}

@article{horvath1993role,
  title={The role of the therapeutic alliance in psychotherapy.},
  author={Horvath, Adam O and Luborsky, Lester},
  journal={Journal of consulting and clinical psychology},
  volume={61},
  number={4},
  pages={561},
  year={1993},
  publisher={American Psychological Association}
}

@article{norcross2018new,
  title={A new therapy for each patient: Evidence-based relationships and responsiveness},
  author={Norcross, John C and Wampold, Bruce E},
  journal={Journal of clinical psychology},
  volume={74},
  number={11},
  pages={1889--1906},
  year={2018},
  publisher={Wiley Online Library}
}

@book{fogg2003persuasive,
  title={Persuasive Technology: Using Computers to Change What We Think and Do},
  author={Fogg, B. J.},
  publisher={Morgan Kaufmann},
  year={2003},
  address={San Francisco, CA},
  isbn={9781558606432},
  doi={10.1016/B978-1-55860-643-2.X5000-8}
}

@article{inkster2018empathy,
  title={An Empathy-Driven, Conversational Artificial Intelligence Agent (Wysa) for Digital Mental Well-Being: Real-World Data Evaluation Mixed-Methods Study},
  author={Inkster, Becky and Sarda, Shubhankar and Subramanian, Vinod},
  journal={JMIR mHealth and uHealth},
  volume={6},
  number={11},
  pages={e12106},
  year={2018},
  doi={10.2196/12106}
}

@article{keyes2002mental,
  title={The mental health continuum: From languishing to flourishing in life},
  author={Keyes, Corey LM},
  journal={Journal of health and social behavior},
  pages={207--222},
  year={2002},
  publisher={JSTOR}
}

@article{torous2025evolving,
  title={The evolving field of digital mental health: current evidence and implementation issues for smartphone apps, generative artificial intelligence, and virtual reality},
  author={Torous, John and Linardon, Jake and Goldberg, Simon B and Sun, Shufang and Bell, Imogen and Nicholas, Jennifer and Hassan, Lamiece and Hua, Yining and Milton, Alyssa and Firth, Joseph},
  journal={World Psychiatry},
  volume={24},
  number={2},
  pages={156--174},
  year={2025},
  publisher={Wiley Online Library}
}

@inproceedings{kitson2023co,
  title={Co-designing a virtual reality intervention for supporting cognitive reappraisal skills development with youth},
  author={Kitson, Alexandra and Antle, Alissa N and Slovak, Petr},
  booktitle={Proceedings of the 22nd annual ACM interaction design and children conference},
  pages={14--26},
  year={2023}
}

@article{andrade2014barriers,
  title={Barriers to mental health treatment: results from the WHO World Mental Health surveys},
  author={Andrade, Laura Helena and Alonso, Jordi and Mneimneh, Zeina and Wells, J Elizabeth and Al-Hamzawi, Ali and Borges, Guilherme and Bromet, Evelyn and Bruffaerts, Ronny and De Girolamo, Giovanni and De Graaf, Ron and others},
  journal={Psychological medicine},
  volume={44},
  number={6},
  pages={1303--1317},
  year={2014},
  publisher={Cambridge University Press}
}

@article{levin1983self,
  title={Self-care in health},
  author={Levin, Lowell S and Idler, Ellen L},
  journal={Annual review of public health},
  volume={4},
  number={1},
  pages={181--201},
  year={1983},
  publisher={Annual Reviews 4139 El Camino Way, PO Box 10139, Palo Alto, CA 94303-0139, USA}
}

@techreport{world2009self,
  title={Self-care in the context of primary health care},
  author={World Health Organization and others},
  year={2009},
  institution={WHO Regional Office for South-East Asia}
}

@inproceedings{10.1145/3571884.3597142,
author = {Peltola, Johanna and Kaipainen, Kirsikka and Keinonen, Katariina and Kiuru, Noona and Turunen, Markku},
title = {Developing A Conversational Interface for an ACT-based Online Program: Understanding Adolescents’ Expectations of Conversational Style},
year = {2023},
isbn = {9798400700149},
publisher = {Association for Computing Machinery},
address = {New York, NY, USA},
url = {https://doi.org/10.1145/3571884.3597142},
doi = {10.1145/3571884.3597142},
booktitle = {Proceedings of the 5th International Conference on Conversational User Interfaces},
articleno = {1},
numpages = {16},
location = {Eindhoven, Netherlands},
series = {CUI '23}
}

@article{orem1995nursing,
  title={Nursing: Concepts of practice},
  author={Orem, Dorothea Elizabeth and Taylor, Susan G and Renpenning, Kathie McLaughlin},
  year={1995}
}

@article{martinez2021self,
  title={Self-care: A concept analysis},
  author={Mart{\'\i}nez, Nicole and Connelly, Cynthia D and P{\'e}rez, Alexa and Calero, Patricia},
  journal={International journal of nursing sciences},
  volume={8},
  number={4},
  pages={418--425},
  year={2021},
  publisher={Elsevier}
}

@article{coyle2007computers,
  title={Computers in talk-based mental health interventions},
  author={Coyle, David and Doherty, Gavin and Matthews, Mark and Sharry, John},
  journal={Interacting with computers},
  volume={19},
  number={4},
  pages={545--562},
  year={2007},
  publisher={OUP}
}

@article{mak2022steps,
  title={Steps for conducting a scoping review},
  author={Mak, Susanne and Thomas, Aliki},
  journal={Journal of graduate medical education},
  volume={14},
  number={5},
  pages={565--567},
  year={2022},
  publisher={The Accreditation Council for Graduate Medical Education}
}

@article{reed2019icd11,
  author  = {Reed, Geoffrey M. and First, Michael B. and Kogan, Charles S. and Hyman, Steven E. and Gureje, Oye and Gaebel, Wolfgang and Maj, Mario and Stein, Dan J. and Maercker, Andreas and Tyrer, Peter and Claudino, Ang{\'e}lica M. and Garralda, Elena and Salvador-Carulla, Luis and Ray, Rajiv and Lovell, Anne M. and Saxena, Shekhar},
  title   = {Innovations and changes in the {ICD}-11 classification of mental, behavioural and neurodevelopmental disorders},
  journal = {World Psychiatry},
  volume  = {18},
  number  = {1},
  pages   = {3--19},
  year    = {2019},
  doi     = {10.1002/wps.20611}
}

@article{sachdev2009neurocognitive,
  title={Neurocognitive disorders: Cluster 1 of the proposed meta-structure for DSM-V and ICD-11: Paper 2 of 7 of the thematic section:‘A proposal for a meta-structure for DSM-V and ICD-11’},
  author={Sachdev, P and Andrews, G and Hobbs, MJ and Sunderland, M and Anderson, TM},
  journal={Psychological Medicine},
  volume={39},
  number={12},
  pages={2001--2012},
  year={2009},
  publisher={Cambridge University Press}
}

@book{who2024icd11,
  author    = {{World Health Organization}},
  title     = {Clinical Descriptions and Diagnostic Requirements for ICD-11 Mental, Behavioural and Neurodevelopmental Disorders},
  year      = {2024},
  publisher = {World Health Organization},
  address   = {Geneva},
  isbn      = {978-92-4-007726-3},
  url       = {https://www.who.int/publications/i/item/9789240077263}
}

@article{ahmed2021mobile,
  title={Mobile applications for mental health self-care: A scoping review},
  author={Ahmed, Arfan and Ali, Nashva and Giannicchi, Anna and Abd-alrazaq, Alaa A and Ahmed, Mohamed Ali Siddig and Aziz, Sarah and Househ, Mowafa},
  journal={Computer Methods and Programs in Biomedicine Update},
  volume={1},
  pages={100041},
  year={2021},
  publisher={Elsevier}
}

@inproceedings{tan2023mindful,
  title={Mindful moments: Exploring on-the-go mindfulness practice on smart-glasses},
  author={Tan, Felicia Fang-Yi and Ram, Ashwin and Haigh, Chloe and Zhao, Shengdong},
  booktitle={Proceedings of the 2023 ACM Designing Interactive Systems Conference},
  pages={476--492},
  year={2023}
}

@inproceedings{elvitigala2021stressshoe,
  title={StressShoe: a DIY toolkit for just-in-time personalised stress interventions for office workers performing sedentary tasks},
  author={Elvitigala, Don Samitha and Scholl, Philipp M and Suriyaarachchi, Hussel and Dissanayake, Vipula and Nanayakkara, Suranga},
  booktitle={Proceedings of the 23rd international conference on mobile human-computer interaction},
  pages={1--14},
  year={2021}
}

@inproceedings{soler2024arcadia,
  title={Arcadia: A gamified mixed reality system for emotional regulation and self-compassion},
  author={Soler-Dominguez, Jose Luis and Navas-Medrano, Samuel and Pons, Patricia},
  booktitle={Proceedings of the 2024 CHI Conference on Human Factors in Computing Systems},
  pages={1--17},
  year={2024}
}

@inproceedings{aldaweesh2026hasn,
  title={''It Hasn’t Lived in Our Society”: Investigating Cultural Sensitivity in LLM Chatbots for Emotional Support},
  author={Aldaweesh, Sarah and Alelsheikh, Ghzal and Alhamed, Falwah and Van Kleek, Max and Shadbolt, Nigel},
  booktitle={Proceedings of the 2026 CHI Conference on Human Factors in Computing Systems},
  pages={1--20},
  year={2026}
}

@inproceedings{matthews2015situ,
  title={In situ design for mental illness: Considering the pathology of bipolar disorder in mhealth design},
  author={Matthews, Mark and Voida, Stephen and Abdullah, Saeed and Doherty, Gavin and Choudhury, Tanzeem and Im, Sangha and Gay, Geri},
  booktitle={Proceedings of the 17th International Conference on Human-Computer Interaction with Mobile Devices and Services},
  pages={86--97},
  year={2015}
}

@inproceedings{rasch2024mind,
  title={Mind mansion: exploring metaphorical interactions to engage with negative thoughts in virtual reality},
  author={Rasch, Julian and Zender, Michelle Johanna and Sakel, Sophia and Wagener, Nadine},
  booktitle={Proceedings of the 2024 ACM Designing Interactive Systems Conference},
  pages={2305--2318},
  year={2024}
}

@inproceedings{surani2026co,
  title={Co-designing MESA-Bot: Enhancing Accessibility, Privacy, Security, and Trust in a Mental Health Chatbot for Older Adults},
  author={Surani, Aishwarya Umeshkumar and Das, Sanchari},
  booktitle={Proceedings of the 2026 CHI Conference on Human Factors in Computing Systems},
  pages={1--21},
  year={2026}
}

@inproceedings{vossen2024effect,
  title={The effect of personalizing a psychotherapy conversational agent on therapeutic bond and usage intentions},
  author={Vossen, Wout and Szymanski, Maxwell and Verbert, Katrien},
  booktitle={Proceedings of the 29th International Conference on Intelligent User Interfaces},
  pages={761--771},
  year={2024}
}

@misc{who2025world,
  title={World mental health today: latest data},
  author={WHO},
  year={2025},
  publisher={World Health Organization Geneva, Switzerland}
}

@article{potts2025digital,
  title={Digital mental health interventions for young people aged 16-25 years: scoping review},
  author={Potts, Courtney and Kealy, Carmen and McNulty, Jamie M and Madrid-Cagigal, Alba and Wilson, Thomas and Mulvenna, Maurice D and O'Neill, Siobhan and Donohoe, Gary and Barry, Margaret M},
  journal={Journal of Medical Internet Research},
  volume={27},
  pages={e72892},
  year={2025},
  publisher={JMIR Publications Toronto, Canada}
}

@inproceedings{thieme2016challenges,
  title={Challenges for designing new technology for health and wellbeing in a complex mental healthcare context},
  author={Thieme, Anja and McCarthy, John and Johnson, Paula and Phillips, Stephanie and Wallace, Jayne and Lindley, Si{\^a}n and Ladha, Karim and Jackson, Daniel and Nowacka, Diana and Rafiev, Ashur and others},
  booktitle={Proceedings of the 2016 CHI Conference on Human Factors in Computing Systems},
  pages={2136--2149},
  year={2016}
}

@article{thieme2020machine,
  title={Machine learning in mental health: A systematic review of the HCI literature to support the development of effective and implementable ML systems},
  author={Thieme, Anja and Belgrave, Danielle and Doherty, Gavin},
  journal={ACM Transactions on Computer-Human Interaction (TOCHI)},
  volume={27},
  number={5},
  pages={1--53},
  year={2020},
  publisher={ACM New York, NY, USA}
}

@inproceedings{balcombe2022human,
  title={Human-computer interaction in digital mental health.},
  author={Balcombe, Luke and De Leo, Diego},
  booktitle={Informatics},
  volume={9},
  number={1},
  year={2022}
}

@article{knowles2014qualitative,
  title={Qualitative meta-synthesis of user experience of computerised therapy for depression and anxiety},
  author={Knowles, Sarah E and Toms, Gill and Sanders, Caroline and Bee, Penny and Lovell, Karina and Rennick-Egglestone, Stefan and Coyle, David and Kennedy, Catriona M and Littlewood, Elizabeth and Kessler, David and others},
  journal={PLoS one},
  volume={9},
  number={1},
  pages={e84323},
  year={2014},
  publisher={Public Library of Science San Francisco, USA}
}

@inproceedings{sanches2019hci,
  title={HCI and Affective Health: Taking stock of a decade of studies and charting future research directions},
  author={Sanches, Pedro and Janson, Axel and Karpashevich, Pavel and Nadal, Camille and Qu, Chengcheng and Daud{\'e}n Roquet, Claudia and Umair, Muhammad and Windlin, Charles and Doherty, Gavin and H{\"o}{\"o}k, Kristina and others},
  booktitle={Proceedings of the 2019 CHI Conference on Human Factors in Computing Systems},
  pages={1--17},
  year={2019}
}

@inproceedings{10.1145/3715336.3735795,
author = {Li, Yi and Ding, Xuanxuan and Chen, Yifan and Li, Yeye and Ma, Nan},
title = {Customizable AI for Depression Care: Improving the User Experience of Large Language Model-Driven Chatbots},
year = {2025},
isbn = {9798400714856},
publisher = {Association for Computing Machinery},
address = {New York, NY, USA},
url = {https://doi.org/10.1145/3715336.3735795},
doi = {10.1145/3715336.3735795},
booktitle = {Proceedings of the 2025 ACM Designing Interactive Systems Conference},
pages = {1844–1866},
numpages = {23},
location = {
},
series = {DIS '25}
}

@inproceedings{10.1145/3772318.3791817,
author = {Wu, Mengyuan and Jiang, Zhihan and Fan, Yuang and Feng, Richard and Dharmavaram, Sahiti and Polowitz, Mathew and Fallon, Shawn and Islam, Bashima and Benson, Lizbeth and Tung, Irene and Creswell, David and Xu, Xuhai},
title = {MindfulAgents: Personalizing Mindfulness Meditation via an Expert-Aligned Multi-Agent System},
year = {2026},
isbn = {9798400722783},
publisher = {Association for Computing Machinery},
address = {New York, NY, USA},
url = {https://doi.org/10.1145/3772318.3791817},
doi = {10.1145/3772318.3791817},
booktitle = {Proceedings of the 2026 CHI Conference on Human Factors in Computing Systems},
articleno = {1025},
numpages = {25},
location = {
},
series = {CHI '26}
}

@article{santomauro2026updated,
  title={Updated trends in the global prevalence and burden of mental disorders, 1990--2023: a systematic analysis for the Global Burden of Disease Study 2023},
  author={Santomauro, Damian F and Miller, Paul Anthony and Shadid, Jamileh and Hanson, Sarah Wulf and Vo, Anh and Roy, Darius Jake and Hagins, Hailey and Herrera, Ana M Mantilla and Scott, James G and Erskine, Holly E and others},
  journal={The Lancet},
  volume={407},
  number={10543},
  pages={2040--2064},
  year={2026},
  publisher={Elsevier}
}

@inproceedings{10.1145/3544548.3581188,
author = {Farrall, Alexz and Taylor, Jordan and Ainsworth, Ben and Alexander, Jason},
title = {Manifesting Breath: Empirical Evidence for the Integration of Shape-changing Biofeedback-based Artefacts within Digital Mental Health Interventions},
year = {2023},
isbn = {9781450394215},
publisher = {Association for Computing Machinery},
address = {New York, NY, USA},
url = {https://doi.org/10.1145/3544548.3581188},
doi = {10.1145/3544548.3581188},
booktitle = {Proceedings of the 2023 CHI Conference on Human Factors in Computing Systems},
articleno = {497},
numpages = {14},
location = {Hamburg, Germany},
series = {CHI '23}
}

@article{banos2022current,
  title={What is the current and future status of digital mental health interventions?},
  author={Ba{\~n}os, Rosa M{\textordfeminine} and Herrero, Roc{\'\i}o and Vara, M{\textordfeminine} Dolores},
  journal={The Spanish Journal of Psychology},
  volume={25},
  pages={e5},
  year={2022},
  publisher={Cambridge University Press}
}

@inproceedings{10.1145/3613904.3642937,
author = {Kim, Taewan and Bae, Seolyeong and Kim, Hyun Ah and Lee, Su-Woo and Hong, Hwajung and Yang, Chanmo and Kim, Young-Ho},
title = {MindfulDiary: Harnessing Large Language Model to Support Psychiatric Patients' Journaling},
year = {2024},
isbn = {9798400703300},
publisher = {Association for Computing Machinery},
address = {New York, NY, USA},
url = {https://doi.org/10.1145/3613904.3642937},
doi = {10.1145/3613904.3642937},
booktitle = {Proceedings of the 2024 CHI Conference on Human Factors in Computing Systems},
articleno = {701},
numpages = {20},
location = {Honolulu, HI, USA},
series = {CHI '24}
}

@inproceedings{10.1145/3613904.3642369,
author = {Yoo, Dong Whi and Woo, Hayoung and Nguyen, Viet Cuong and Birnbaum, Michael L. and Kruzan, Kaylee Payne and Kim, Jennifer G and Abowd, Gregory D. and De Choudhury, Munmun},
title = {Patient Perspectives on AI-Driven Predictions of Schizophrenia Relapses: Understanding Concerns and Opportunities for Self-Care and Treatment},
year = {2024},
isbn = {9798400703300},
publisher = {Association for Computing Machinery},
address = {New York, NY, USA},
url = {https://doi.org/10.1145/3613904.3642369},
doi = {10.1145/3613904.3642369},
booktitle = {Proceedings of the 2024 CHI Conference on Human Factors in Computing Systems},
articleno = {702},
numpages = {20},
location = {Honolulu, HI, USA},
series = {CHI '24}
}

@inproceedings{10.1145/3706598.3713699,
author = {Jha, Smriti and Chan, Gerry and Jewer, Seana and Agyapong, Vincent I.O. and Orji, Rita},
title = {“Bring them back to life”: LifeLink Application for Caregivers Dealing with Suicidality},
year = {2025},
isbn = {9798400713941},
publisher = {Association for Computing Machinery},
address = {New York, NY, USA},
url = {https://doi.org/10.1145/3706598.3713699},
doi = {10.1145/3706598.3713699},
booktitle = {Proceedings of the 2025 CHI Conference on Human Factors in Computing Systems},
articleno = {512},
numpages = {25},
location = {
},
series = {CHI '25}
}

@inproceedings{10.1145/3772318.3791591,
author = {Liu, Tony and Patil, Bhargavi and Ngo, Thu and Karr, Chris and Nguyen, Theresa and Kornfield, Rachel and Meyerhoff, Jonah},
title = {Framing Helper Therapy to Support User Engagement: Causal Evidence from a Public Deployment of a Mental Health Support Text Messaging Program},
year = {2026},
isbn = {9798400722783},
publisher = {Association for Computing Machinery},
address = {New York, NY, USA},
url = {https://doi.org/10.1145/3772318.3791591},
doi = {10.1145/3772318.3791591},
booktitle = {Proceedings of the 2026 CHI Conference on Human Factors in Computing Systems},
articleno = {916},
numpages = {17},
location = {
},
series = {CHI '26}
}

@inproceedings{10.1145/3706598.3714277,
author = {Zhang, Xuechen and He, Changyang and Zhang, Peng and Gu, Hansu and Gu, Ning and Shen, Qi and Hu, Zhan and Lu, Tun},
title = {RemiHaven: Integrating "In-Town" and "Out-of-Town" Peers to Provide Personalized Reminiscence Support for Older Drifters},
year = {2025},
isbn = {9798400713941},
publisher = {Association for Computing Machinery},
address = {New York, NY, USA},
url = {https://doi.org/10.1145/3706598.3714277},
doi = {10.1145/3706598.3714277},
booktitle = {Proceedings of the 2025 CHI Conference on Human Factors in Computing Systems},
articleno = {1034},
numpages = {20},
location = {
},
series = {CHI '25}
}

@inproceedings{10.1145/3563657.3595998,
author = {Miller, Noah and Stepanova, Ekaterina R. and Desnoyers-Stewart, John and Adhikari, Ashu and Kitson, Alexandra and Pennefather, Patrick and Quesnel, Denise and Brauns, Katharina and Friedl-Werner, Anika and Stahn, Alexander and Riecke, Bernhard E.},
title = {Awedyssey: Design Tensions in Eliciting Self-transcendent Emotions in Virtual Reality to Support Mental Well-being and Connection},
year = {2023},
isbn = {9781450398930},
publisher = {Association for Computing Machinery},
address = {New York, NY, USA},
url = {https://doi.org/10.1145/3563657.3595998},
doi = {10.1145/3563657.3595998},
booktitle = {Proceedings of the 2023 ACM Designing Interactive Systems Conference},
pages = {189–211},
numpages = {23},
location = {Pittsburgh, PA, USA},
series = {DIS '23}
}

@inproceedings{li2010stage,
  title={A stage-based model of personal informatics systems},
  author={Li, Ian and Dey, Anind and Forlizzi, Jodi},
  booktitle={Proceedings of the SIGCHI conference on human factors in computing systems},
  pages={557--566},
  year={2010}
}

@inproceedings{10.1145/3780045.3780052,
author = {Hashmati, Negin and Skoog, Ther{\'e}se and Obaid, Mohammad and Eriksson, Thommy},
title = {Friend or Therapist? A Systematic Literature Review on Chatbots and Mental Health in Young People},
year = {2026},
isbn = {9798400719486},
publisher = {Association for Computing Machinery},
address = {New York, NY, USA},
url = {https://doi.org/10.1145/3780045.3780052},
doi = {10.1145/3780045.3780052},
booktitle = {Proceedings of the 1st International Conference on Human-Computer Interaction in the Alps},
pages = {56–64},
numpages = {9},
location = {
},
series = {AlpCHI '26}
}

@article{10.1145/3805040,
author = {Butorac, Isobel and Mcnaney, Roisin and Thomas, Jacob and Flores-Herrera, Alexa and Paulo Seguin, Joshua and Northam, Jaimie C. and Carter, Adrian},
title = {A Queer Future for Digital Mental Health: LGBTQ+ young adult perspectives on engaging with digital mental health tools},
year = {2026},
publisher = {Association for Computing Machinery},
address = {New York, NY, USA},
url = {https://doi.org/10.1145/3805040},
doi = {10.1145/3805040},
note = {Just Accepted},
journal = {ACM Trans. Comput. Healthcare},
month = apr
}

@inproceedings{rooksby2014personal,
  title={Personal tracking as lived informatics},
  author={Rooksby, John and Rost, Mattias and Morrison, Alistair and Chalmers, Matthew},
  booktitle={Proceedings of the SIGCHI conference on human factors in computing systems},
  pages={1163--1172},
  year={2014}
}

@article{loke2018somatic,
  title={The somatic turn in human-computer interaction},
  author={Loke, Lian and Schiphorst, Thecla},
  journal={interactions},
  volume={25},
  number={5},
  pages={54--5863},
  year={2018},
  publisher={ACM New York, NY, USA}
}

@inproceedings{bodker2006second,
  title={When second wave HCI meets third wave challenges},
  author={B{\o}dker, Susanne},
  booktitle={Proceedings of the 4th Nordic conference on Human-computer interaction: changing roles},
  pages={1--8},
  year={2006}
}

@inproceedings{10.1145/3613904.3642662,
author = {Neupane, Sameer and Saha, Mithun and Ali, Nasir and Hnat, Timothy and Samiei, Shahin Alan and Nandugudi, Anandatirtha and Almeida, David M. and Kumar, Santosh},
title = {Momentary Stressor Logging and Reflective Visualizations: Implications for Stress Management with Wearables},
year = {2024},
isbn = {9798400703300},
publisher = {Association for Computing Machinery},
address = {New York, NY, USA},
url = {https://doi.org/10.1145/3613904.3642662},
doi = {10.1145/3613904.3642662},
booktitle = {Proceedings of the 2024 CHI Conference on Human Factors in Computing Systems},
articleno = {809},
numpages = {19},
location = {Honolulu, HI, USA},
series = {CHI '24}
}

@article{torous2018clinical,
  title={Clinical review of user engagement with mental health smartphone apps: evidence, theory and improvements},
  author={Torous, John and Nicholas, Jennifer and Larsen, Mark E and Firth, Joseph and Christensen, Helen},
  journal={Evidence Based Mental Health},
  volume={21},
  number={3},
  pages={116--119},
  year={2018},
  publisher={BMJ Publishing Group Ltd, Royal College of Psychiatrists and British~…}
}

@inproceedings{10.1145/3772318.3791615,
author = {Michelle Kim, Minsol and Low, Daniel and Lafond, David and Shim, Eugene and Han, Michelle and Kandil, Mohanad and Zhang, Chenyu and Kitsberg, Theo and Boccagno, Chelsea and Liang, Paul Pu and Maes, Pattie},
title = {Breaking Negative Cycles: A Reflection-to-Action System for Adaptive Change},
year = {2026},
isbn = {9798400722783},
publisher = {Association for Computing Machinery},
address = {New York, NY, USA},
url = {https://doi.org/10.1145/3772318.3791615},
doi = {10.1145/3772318.3791615},
booktitle = {Proceedings of the 2026 CHI Conference on Human Factors in Computing Systems},
articleno = {1438},
numpages = {22},
location = {
},
series = {CHI '26}
}

@inproceedings{wang2026caring,
  title={Caring about care: a meta-narrative review of HCI research on care},
  author={Wang, Zixuan and Guo, Yuanrong and Bowman, Eilidh and Zhai, Yuxiang and Shu, Xinhuan and Zhang, Shengchen and Helms, Karey and Capel, Tara and Vines, John},
  booktitle={Proceedings of the 2026 CHI Conference on Human Factors in Computing Systems},
  pages={1--31},
  year={2026}
}

@inproceedings{10.1145/3715336.3735845,
author = {Park, Joonyoung and Cho, Hyewon and Chu, Hyehyun and Lee, YeEun and Lim, Hajin},
title = {NoRe: Augmenting Journaling Experience with Generative AI for Music Creation},
year = {2025},
isbn = {9798400714856},
publisher = {Association for Computing Machinery},
address = {New York, NY, USA},
url = {https://doi.org/10.1145/3715336.3735845},
doi = {10.1145/3715336.3735845},
booktitle = {Proceedings of the 2025 ACM Designing Interactive Systems Conference},
pages = {2718–2737},
numpages = {20},
location = {
},
series = {DIS '25}
}

@inproceedings{10.1145/3706598.3713485,
author = {Choi, Ryuhaerang and Kim, Taehan and Park, Subin and Kim, Jennifer G. and Lee, Sung-Ju},
title = {Private Yet Social: How LLM Chatbots Support and Challenge Eating Disorder Recovery},
year = {2025},
isbn = {9798400713941},
publisher = {Association for Computing Machinery},
address = {New York, NY, USA},
url = {https://doi.org/10.1145/3706598.3713485},
doi = {10.1145/3706598.3713485},
booktitle = {Proceedings of the 2025 CHI Conference on Human Factors in Computing Systems},
articleno = {642},
numpages = {19},
location = {
},
series = {CHI '25}
}

@inproceedings{konrad2015finding,
  title={Finding the adaptive sweet spot: Balancing compliance and achievement in automated stress reduction},
  author={Konrad, Artie and Bellotti, Victoria and Crenshaw, Nicole and Tucker, Simon and Nelson, Les and Du, Honglu and Pirolli, Peter and Whittaker, Steve},
  booktitle={Proceedings of the 33rd Annual ACM Conference on Human Factors in Computing Systems},
  pages={3829--3838},
  year={2015}
}

@inproceedings{huang2015emotion,
  title={Emotion map: A location-based mobile social system for improving emotion awareness and regulation},
  author={Huang, Yun and Tang, Ying and Wang, Yang},
  booktitle={Proceedings of the 18th ACM Conference on Computer Supported Cooperative Work \& Social Computing},
  pages={130--142},
  year={2015}
}

@inproceedings{lee2017designing,
  title={Designing for self-tracking of emotion and experience with tangible modality},
  author={Lee, Kwangyoung and Hong, Hwajung},
  booktitle={Proceedings of the 2017 Conference on Designing Interactive Systems},
  pages={465--475},
  year={2017}
}

@inproceedings{yu2017stresstree,
  title={StressTree: A metaphorical visualization for biofeedback-assisted stress management},
  author={Yu, Bin and Funk, Mathias and Hu, Jun and Feijs, Loe},
  booktitle={Proceedings of the 2017 conference on designing interactive systems},
  pages={333--337},
  year={2017}
}

@inproceedings{roo2017inner,
  title={Inner garden: Connecting inner states to a mixed reality sandbox for mindfulness},
  author={Roo, Joan Sol and Gervais, Renaud and Frey, Jeremy and Hachet, Martin},
  booktitle={Proceedings of the 2017 CHI conference on human factors in computing systems},
  pages={1459--1470},
  year={2017}
}

@inproceedings{simm2016anxiety,
  title={Anxiety and autism: towards personalized digital health},
  author={Simm, Will and Ferrario, Maria Angela and Gradinar, Adrian and Tavares Smith, Marcia and Forshaw, Stephen and Smith, Ian and Whittle, Jon},
  booktitle={Proceedings of the 2016 CHI conference on human factors in computing systems},
  pages={1270--1281},
  year={2016}
}

@inproceedings{rubin2015towards,
  title={Towards a mobile and wearable system for predicting panic attacks},
  author={Rubin, Jonathan and Eldardiry, Hoda and Abreu, Rui and Ahern, Shane and Du, Honglu and Pattekar, Ashish and Bobrow, Daniel G},
  booktitle={Proceedings of the 2015 ACM International Joint Conference on Pervasive and Ubiquitous Computing},
  pages={529--533},
  year={2015}
}

@inproceedings{prpa2018attending,
  title={Attending to breath: exploring how the cues in a virtual environment guide the attention to breath and shape the quality of experience to support mindfulness},
  author={Prpa, Mirjana and Tatar, K{\i}van{\c{c}} and Fran{\c{c}}oise, Jules and Riecke, Bernhard and Schiphorst, Thecla and Pasquier, Philippe},
  booktitle={Proceedings of the 2018 designing interactive systems conference},
  pages={71--84},
  year={2018}
}

@inproceedings{schroeder2018pocket,
  title={Pocket skills: A conversational mobile web app to support dialectical behavioral therapy},
  author={Schroeder, Jessica and Wilkes, Chelsey and Rowan, Kael and Toledo, Arturo and Paradiso, Ann and Czerwinski, Mary and Mark, Gloria and Linehan, Marsha M},
  booktitle={Proceedings of the 2018 CHI Conference on Human Factors in Computing Systems},
  pages={1--15},
  year={2018}
}

@inproceedings{antle2019evaluating,
  title={Evaluating the impact of a mobile neurofeedback app for young children at school and home},
  author={Antle, Alissa N and McLaren, Elgin-Skye and Fiedler, Holly and Johnson, Naomi},
  booktitle={Proceedings of the 2019 CHI Conference on human factors in computing systems},
  pages={1--13},
  year={2019}
}

@inproceedings{antle2019design,
  title={Design for mental health: How socio-technological processes mediate outcome measures in a field study of a wearable anxiety app},
  author={Antle, Alissa N and McLaren, Elgin Skye and Fiedler, Holly and Johnson, Naomi},
  booktitle={Proceedings of the thirteenth international conference on tangible, embedded, and embodied interaction},
  pages={87--96},
  year={2019}
}

@inproceedings{doherty2019engagement,
  title={Engagement with mental health screening on mobile devices: Results from an antenatal feasibility study},
  author={Doherty, Kevin and Marcano-Belisario, Jos{\'e} and Cohn, Martin and Mastellos, Nikolaos and Morrison, Cecily and Car, Josip and Doherty, Gavin},
  booktitle={Proceedings of the 2019 CHI Conference on Human Factors in Computing Systems},
  pages={1--15},
  year={2019}
}

@inproceedings{lee2019caring,
  title={Caring for Vincent: a chatbot for self-compassion},
  author={Lee, Minha and Ackermans, Sander and Van As, Nena and Chang, Hanwen and Lucas, Enzo and IJsselsteijn, Wijnand},
  booktitle={Proceedings of the 2019 CHI conference on human factors in computing systems},
  pages={1--13},
  year={2019}
}

@inproceedings{rajcic2020mirror,
  title={Mirror ritual: An affective interface for emotional self-reflection},
  author={Rajcic, Nina and McCormack, Jon},
  booktitle={Proceedings of the 2020 CHI conference on human factors in computing systems},
  pages={1--13},
  year={2020}
}

@inproceedings{rohani2020mubs,
  title={MUBS: A personalized recommender system for behavioral activation in mental health},
  author={Rohani, Darius A and Quemada Lopategui, Andrea and Tuxen, Nanna and Faurholt-Jepsen, Maria and Kessing, Lars V and Bardram, Jakob E},
  booktitle={Proceedings of the 2020 CHI conference on human factors in computing systems},
  pages={1--13},
  year={2020}
}

@article{ryu2020simple,
  title={Simple and steady interactions win the healthy mentality: designing a chatbot service for the elderly},
  author={Ryu, Hyeyoung and Kim, Soyeon and Kim, Dain and Han, Sooan and Lee, Keeheon and Kang, Younah},
  journal={Proceedings of the ACM on human-computer interaction},
  volume={4},
  number={CSCW2},
  pages={1--25},
  year={2020},
  publisher={ACM New York, NY, USA}
}

@inproceedings{potapov2020lifemosaic,
  title={LifeMosaic: co-design of a personal informatics tool for youth},
  author={Potapov, Kyrill and Marshall, Paul},
  booktitle={Proceedings of the interaction design and children conference},
  pages={519--531},
  year={2020}
}

@inproceedings{schneeberger2021stress,
  title={Stress management training using biofeedback guided by social agents},
  author={Schneeberger, Tanja and Sauerwein, Naomi and Anglet, Manuel S and Gebhard, Patrick},
  booktitle={Proceedings of the 26th International Conference on Intelligent User Interfaces},
  pages={564--574},
  year={2021}
}

@inproceedings{park2021wrote,
  title={“I wrote as if I were telling a story to someone I knew.”: Designing Chatbot Interactions for Expressive Writing in Mental Health},
  author={Park, SoHyun and Thieme, Anja and Han, Jeongyun and Lee, Sungwoo and Rhee, Wonjong and Suh, Bongwon},
  booktitle={Proceedings of the 2021 ACM Designing Interactive Systems Conference},
  pages={926--941},
  year={2021}
}

@inproceedings{yan2022emoglass,
  title={EmoGlass: An end-to-end AI-enabled wearable platform for enhancing self-awareness of emotional health},
  author={Yan, Zihan and Wu, Yufei and Zhang, Yang and Chen, Xiang'Anthony'},
  booktitle={Proceedings of the 2022 CHI Conference on Human Factors in Computing Systems},
  pages={1--19},
  year={2022}
}

@inproceedings{kim2022prediction,
  title={Prediction for retrospection: Integrating algorithmic stress prediction into personal informatics systems for college students’ mental health},
  author={Kim, Taewan and Kim, Haesoo and Lee, Ha Yeon and Goh, Hwarang and Abdigapporov, Shakhboz and Jeong, Mingon and Cho, Hyunsung and Han, Kyungsik and Noh, Youngtae and Lee, Sung-Ju and others},
  booktitle={Proceedings of the 2022 CHI conference on human factors in computing systems},
  pages={1--20},
  year={2022}
}

@inproceedings{hamid2022you,
  title={What are you thinking?: Using CBT and storytelling to improve mental health among college students},
  author={Hamid, Aleesha and Arshad, Rabiah and Shahid, Suleman},
  booktitle={Proceedings of the 2022 CHI Conference on Human Factors in Computing Systems},
  pages={1--16},
  year={2022}
}

@inproceedings{wagener2022mood,
  title={Mood worlds: A virtual environment for autonomous emotional expression},
  author={Wagener, Nadine and Niess, Jasmin and Rogers, Yvonne and Sch{\"o}ning, Johannes},
  booktitle={Proceedings of the 2022 CHI Conference on Human Factors in Computing Systems},
  pages={1--16},
  year={2022}
}

@inproceedings{collins2022covid,
  title={Covid connect: Chat-driven anonymous story-sharing for peer support},
  author={Collins, Christopher and Arbour, Simone and Beals, Nathan and Yama, Shawn and Laffier, Jennifer and Zhao, Zixin},
  booktitle={Proceedings of the 2022 ACM Designing Interactive Systems Conference},
  pages={301--318},
  year={2022}
}

@inproceedings{speer2021mindfulnest,
  title={MindfulNest: Strengthening emotion regulation with tangible user interfaces},
  author={Speer, Samantha and Hamner, Emily and Tasota, Michael and Zito, Lauren and Byrne-Houser, Sarah K},
  booktitle={Proceedings of the 2021 International Conference on Multimodal Interaction},
  pages={103--111},
  year={2021}
}

@inproceedings{choi2026daily,
  title={From Daily Song to Daily Self: Supporting Emotional Growth of Deaf and Hard-of-Hearing Individuals through Generative AI Songwriting},
  author={Choi, Youjin and Yoo, JinYoung and Moon, JaeYoung and Kim, Yoonjae and Lee, Eun Young and Kim, Jennifer G and Hong, Jin-Hyuk},
  booktitle={Proceedings of the 2026 CHI Conference on Human Factors in Computing Systems},
  pages={1--26},
  year={2026}
}

@inproceedings{kim2023routineaid,
  title={Routineaid: Externalizing key design elements to support daily routines of individuals with autism},
  author={Kim, Bogoan and Kim, Sung-In and Park, Sangwon and Yoo, Hee Jeong and Hong, Hwajung and Han, Kyungsik},
  booktitle={Proceedings of the 2023 CHI Conference on Human Factors in Computing Systems},
  pages={1--18},
  year={2023}
}

@inproceedings{park2024collaborative,
  title={Collaborative School Mental Health System: Leveraging a Conversational Agent for Enhancing Children's Executive Function},
  author={Park, Doeun and Choo, Myounglee and Cho, Minseo and Kim, Jinwoo and Shin, Yee-Jin},
  booktitle={Proceedings of the 2024 CHI Conference on Human Factors in Computing Systems},
  pages={1--17},
  year={2024}
}

@inproceedings{10.1145/3613904.3643054,
author = {Staab, Phoebe A and Williams, A. Jess and Robertson, Mackenzie D. A. and Slovak, Petr},
title = {“Can you be with that feeling?”: Extending Design Strategies for Interoceptive Awareness for the Context of Mental Health},
year = {2024},
isbn = {9798400703300},
publisher = {Association for Computing Machinery},
address = {New York, NY, USA},
url = {https://doi.org/10.1145/3613904.3643054},
doi = {10.1145/3613904.3643054},
booktitle = {Proceedings of the 2024 CHI Conference on Human Factors in Computing Systems},
articleno = {695},
numpages = {21},
location = {Honolulu, HI, USA},
series = {CHI '24}
}

@article{10.1145/3718084,
author = {Rasouli, Samira and Ghafurian, Moojan and Dautenhahn, Kerstin},
title = {Co-Design and User Evaluation of a Robotic Mental Well-Being Coach to Support University Students’ Public Speaking Anxiety},
year = {2025},
issue_date = {June 2025},
publisher = {Association for Computing Machinery},
address = {New York, NY, USA},
volume = {32},
number = {3},
issn = {1073-0516},
url = {https://doi.org/10.1145/3718084},
doi = {10.1145/3718084},
journal = {ACM Trans. Comput.-Hum. Interact.},
month = jun,
articleno = {24},
numpages = {70}
}

@article{10.1145/3699761,
author = {Nepal, Subigya and Pillai, Arvind and Campbell, William and Massachi, Talie and Heinz, Michael V. and Kunwar, Ashmita and Choi, Eunsol Soul and Xu, Xuhai and Kuc, Joanna and Huckins, Jeremy F. and Holden, Jason and Preum, Sarah M. and Depp, Colin and Jacobson, Nicholas and Czerwinski, Mary P. and Granholm, Eric and Campbell, Andrew T.},
title = {MindScape Study: Integrating LLM and Behavioral Sensing for Personalized AI-Driven Journaling Experiences},
year = {2024},
issue_date = {December 2024},
publisher = {Association for Computing Machinery},
address = {New York, NY, USA},
volume = {8},
number = {4},
url = {https://doi.org/10.1145/3699761},
doi = {10.1145/3699761},
journal = {Proc. ACM Interact. Mob. Wearable Ubiquitous Technol.},
month = nov,
articleno = {186},
numpages = {44}
}

@inproceedings{10.1145/3544548.3581319,
author = {Tong, Xin and Mauriello, Matthew Louis and Mora-Mendoza, Marco Antonio and Prabhu, Nina and Kim, Jane Paik and Paredes Castro, Pablo E},
title = {Just Do Something: Comparing Self-proposed and Machine-recommended Stress Interventions among Online Workers with Home Sweet Office},
year = {2023},
isbn = {9781450394215},
publisher = {Association for Computing Machinery},
address = {New York, NY, USA},
url = {https://doi.org/10.1145/3544548.3581319},
doi = {10.1145/3544548.3581319},
booktitle = {Proceedings of the 2023 CHI Conference on Human Factors in Computing Systems},
articleno = {495},
numpages = {20},
location = {Hamburg, Germany},
series = {CHI '23}
}

@inproceedings{10.1145/3613904.3642766,
author = {Jung, Gyuwon and Park, Sangjun and Lee, Uichin},
title = {DeepStress: Supporting Stressful Context Sensemaking in Personal Informatics Systems Using a Quasi-experimental Approach},
year = {2024},
isbn = {9798400703300},
publisher = {Association for Computing Machinery},
address = {New York, NY, USA},
url = {https://doi.org/10.1145/3613904.3642766},
doi = {10.1145/3613904.3642766},
booktitle = {Proceedings of the 2024 CHI Conference on Human Factors in Computing Systems},
articleno = {1000},
numpages = {18},
location = {Honolulu, HI, USA},
series = {CHI '24}
}

@inproceedings{10.1145/3643834.3660702,
author = {Passler Bates, Danika and Dudek, Skyla Y. and Berzuk, James M. and Gonz{\'a}lez, Adriana Lorena and Young, James E.},
title = {SnuggleBot the Companion: Exploring In-Home Robot Interaction Strategies to Support Coping With Loneliness},
year = {2024},
isbn = {9798400705830},
publisher = {Association for Computing Machinery},
address = {New York, NY, USA},
url = {https://doi.org/10.1145/3643834.3660702},
doi = {10.1145/3643834.3660702},
booktitle = {Proceedings of the 2024 ACM Designing Interactive Systems Conference},
pages = {2972–2986},
numpages = {15},
location = {Copenhagen, Denmark},
series = {DIS '24}
}

@inproceedings{10.1145/3772318.3790933,
author = {Farrall, Alexz and Sharma, Adwait and Ainsworth, Ben and Jacobsen, Pamela and Alexander, Jason},
title = {Encouraging Breath: Increasing Out‑of‑Session DMHI Engagement using a Shape‑Changing Biofeedback Physicalization within a Longitudinal RCT},
year = {2026},
isbn = {9798400722783},
publisher = {Association for Computing Machinery},
address = {New York, NY, USA},
url = {https://doi.org/10.1145/3772318.3790933},
doi = {10.1145/3772318.3790933},
booktitle = {Proceedings of the 2026 CHI Conference on Human Factors in Computing Systems},
articleno = {1024},
numpages = {17},
location = {
},
series = {CHI '26}
}

@inproceedings{10.1145/3772318.3790406,
author = {Li, Yongming and Ling, Ling and Song, Yunpeng and Huang, Yun and Cai, Zhongmin},
title = {AReframedChair: Reframing the Empty Chair through Dyadic and Triadic AR-Mediated Self-Embodiment},
year = {2026},
isbn = {9798400722783},
publisher = {Association for Computing Machinery},
address = {New York, NY, USA},
url = {https://doi.org/10.1145/3772318.3790406},
doi = {10.1145/3772318.3790406},
booktitle = {Proceedings of the 2026 CHI Conference on Human Factors in Computing Systems},
articleno = {1563},
numpages = {20},
location = {
},
series = {CHI '26}
}

@inproceedings{10.1145/3643834.3661570,
author = {Wagener, Nadine and Kiesewetter, Arne and Reicherts, Leon and Wo{\'z}niak, Pawe{\l} W. and Sch{\"o}ning, Johannes and Rogers, Yvonne and Niess, Jasmin},
title = {MoodShaper: A Virtual Reality Experience to Support Managing Negative Emotions},
year = {2024},
isbn = {9798400705830},
publisher = {Association for Computing Machinery},
address = {New York, NY, USA},
url = {https://doi.org/10.1145/3643834.3661570},
doi = {10.1145/3643834.3661570},
booktitle = {Proceedings of the 2024 ACM Designing Interactive Systems Conference},
pages = {2286–2304},
numpages = {19},
location = {Copenhagen, Denmark},
series = {DIS '24}
}

@inproceedings{10.1145/3706598.3713453,
author = {Zheng, Xi and Li, Zhuoyang and Gui, Xinning and Luo, Yuhan},
title = {Customizing Emotional Support: How Do Individuals Construct and Interact With LLM-Powered Chatbots},
year = {2025},
isbn = {9798400713941},
publisher = {Association for Computing Machinery},
address = {New York, NY, USA},
url = {https://doi.org/10.1145/3706598.3713453},
doi = {10.1145/3706598.3713453},
booktitle = {Proceedings of the 2025 CHI Conference on Human Factors in Computing Systems},
articleno = {376},
numpages = {20},
location = {
},
series = {CHI '25}
}

@inproceedings{10.1145/3745900.3746099,
author = {Zulfikar, Wazeer and Chiaravalloti, Treyden and Shen, Jocelyn and Picard, Rosalind and Maes, Pattie},
title = {Resonance: Drawing from Memories to Imagine Positive Futures through AI-Augmented Journaling},
year = {2025},
isbn = {9798400715662},
publisher = {Association for Computing Machinery},
address = {New York, NY, USA},
url = {https://doi.org/10.1145/3745900.3746099},
doi = {10.1145/3745900.3746099},
booktitle = {Proceedings of the Augmented Humans International Conference 2025},
pages = {199–215},
numpages = {17},
location = {
},
series = {AHs '25}
}

@inproceedings{10.1145/3706598.3713883,
author = {Song, Inhwa and Park, SoHyun and Pendse, Sachin R and Schleider, Jessica Lee and De Choudhury, Munmun and Kim, Young-Ho},
title = {ExploreSelf: Fostering User-driven Exploration and Reflection on Personal Challenges with Adaptive Guidance by Large Language Models},
year = {2025},
isbn = {9798400713941},
publisher = {Association for Computing Machinery},
address = {New York, NY, USA},
url = {https://doi.org/10.1145/3706598.3713883},
doi = {10.1145/3706598.3713883},
booktitle = {Proceedings of the 2025 CHI Conference on Human Factors in Computing Systems},
articleno = {306},
numpages = {22},
location = {
},
series = {CHI '25}
}

@inproceedings{10.1145/3800645.3813019,
author = {Jin, Yuting and Wang, Tingting and Qin, Dantong and Zhou, Zhibin and Bi, Mengkun and Hua, Min and Wang, Pan},
title = {MIRA: A Human-AI Co-Creation Agent for Self-reflection through Squiggle Game},
year = {2026},
isbn = {9798400725630},
publisher = {Association for Computing Machinery},
address = {New York, NY, USA},
url = {https://doi.org/10.1145/3800645.3813019},
doi = {10.1145/3800645.3813019},
booktitle = {Proceedings of the 2026 Designing Interactive Systems Conference},
pages = {2972–2987},
numpages = {16},
location = {
},
series = {DIS '26}
}

@inproceedings{10.1145/3613904.3642761,
author = {Sharma, Ashish and Rushton, Kevin and Lin, Inna Wanyin and Nguyen, Theresa and Althoff, Tim},
title = {Facilitating Self-Guided Mental Health Interventions Through Human-Language Model Interaction: A Case Study of Cognitive Restructuring},
year = {2024},
isbn = {9798400703300},
publisher = {Association for Computing Machinery},
address = {New York, NY, USA},
url = {https://doi.org/10.1145/3613904.3642761},
doi = {10.1145/3613904.3642761},
booktitle = {Proceedings of the 2024 CHI Conference on Human Factors in Computing Systems},
articleno = {700},
numpages = {29},
location = {Honolulu, HI, USA},
series = {CHI '24}
}

@inproceedings{10.1145/3706598.3713730,
author = {Jung, Gyuwon and Lee, Uichin},
title = {CounterStress: Enhancing Stress Coping Planning through Counterfactual Explanations in Personal Informatics},
year = {2025},
isbn = {9798400713941},
publisher = {Association for Computing Machinery},
address = {New York, NY, USA},
url = {https://doi.org/10.1145/3706598.3713730},
doi = {10.1145/3706598.3713730},
booktitle = {Proceedings of the 2025 CHI Conference on Human Factors in Computing Systems},
articleno = {714},
numpages = {20},
location = {
},
series = {CHI '25}
}

@inproceedings{10.1145/3706598.3713269,
author = {Chandrasekaran, Aishwarya and Bielicke, London and Shah, Diya and Janakiraman, Harisha and Mauriello, Matthew Louis},
title = {"I spent 14 hours debugging just one assignment": Toward Computer-Mediated Personal Informatics for Computer Science Student Mental Health},
year = {2025},
isbn = {9798400713941},
publisher = {Association for Computing Machinery},
address = {New York, NY, USA},
url = {https://doi.org/10.1145/3706598.3713269},
doi = {10.1145/3706598.3713269},
booktitle = {Proceedings of the 2025 CHI Conference on Human Factors in Computing Systems},
articleno = {711},
numpages = {19},
location = {
},
series = {CHI '25}
}

@inproceedings{wei2025,
author = {Wei, Qianjie and Wei, Xiaoying and Liang, Yiqi and Lin, Fan and Si, Nuonan and Fan, Mingming},
title = {RemoteChess: Enhancing Older Adults' Social Connectedness via Designing a Virtual Reality Chinese Chess (Xiangqi) Community},
year = {2025},
isbn = {9798400713941},
publisher = {Association for Computing Machinery},
address = {New York, NY, USA},
url = {https://doi.org/10.1145/3706598.3714236},
doi = {10.1145/3706598.3714236},
booktitle = {Proceedings of the 2025 CHI Conference on Human Factors in Computing Systems},
articleno = {1128},
numpages = {16},
location = {
},
series = {CHI '25}
}

@inproceedings{guo2025exploring,
  title={Exploring the design of LLM-based agent in enhancing self-disclosure among the older adults},
  author={Guo, Yijie and Wang, Ruhan and Huang, Zhenhan and Jin, Tongtong and Yao, Xiwen and Feng, Yuan-Ling and Zhang, Weiwei and Yao, Yuan and Mi, Haipeng},
  booktitle={Proceedings of the 2025 CHI Conference on Human Factors in Computing Systems},
  pages={1--17},
  year={2025}
}

@inproceedings{jin2024exploring,
  title={Exploring the design of generative AI in supporting music-based reminiscence for older adults},
  author={Jin, Yucheng and Cai, Wanling and Chen, Li and Zhang, Yizhe and Doherty, Gavin and Jiang, Tonglin},
  booktitle={Proceedings of the 2024 CHI Conference on Human Factors in Computing Systems},
  pages={1--17},
  year={2024}
}

@inproceedings{seo2024chacha,
  title={Chacha: leveraging large language models to prompt children to share their emotions about personal events},
  author={Seo, Woosuk and Yang, Chanmo and Kim, Young-Ho},
  booktitle={Proceedings of the 2024 CHI Conference on Human Factors in Computing Systems},
  pages={1--20},
  year={2024}
}

@inproceedings{sun2025conversations,
  title={Conversations with the stressed body: Facilitating stress self-disclosure among adolescent girls through an embodied approach},
  author={Sun, Xinglin and Claisse, Caroline and Zhang, Runhua and Wu, Xinyu and Yuan, Jialin and Wang, Qi},
  booktitle={Proceedings of the 2025 ACM Designing Interactive Systems Conference},
  pages={3421--3439},
  year={2025}
}

@inproceedings{angelini2025speculating,
  title={Speculating deaf tech: Reimagining technologies centering deaf people},
  author={Angelini, Robin and Spiel, Katta and De Meulder, Maartje},
  booktitle={Proceedings of the 2025 CHI Conference on Human Factors in Computing Systems},
  pages={1--18},
  year={2025}
}

@article{brinkman2023shifting,
  title={Shifting the discourse on disability: Moving to an inclusive, intersectional focus.},
  author={Brinkman, Aurora H and Rea-Sandin, Gianna and Lund, Emily M and Fitzpatrick, Olivia M and Gusman, Michaela S and Boness, Cassandra L},
  journal={American Journal of Orthopsychiatry},
  volume={93},
  number={1},
  pages={50},
  year={2023},
  publisher={Educational Publishing Foundation}
}

@inproceedings{miri2022far,
  title={FAR: end-to-end vibrotactile distributed system designed to facilitate affect regulation in children diagnosed with autism spectrum disorder through slow breathing},
  author={Miri, Pardis and Arora, Mehul and Malhotra, Aman and Flory, Robert and Hu, Stephanie and Lowber, Ashley and Goyal, Ishan and Nguyen, Jacqueline and Hegarty, John P and Kohn, Marlo D and others},
  booktitle={Proceedings of the 2022 CHI Conference on Human Factors in Computing Systems},
  pages={1--20},
  year={2022}
}

@inproceedings{koch2021taking,
  title={Taking mental health \& well-being to the streets: An exploratory evaluation of in-vehicle interventions in the wild},
  author={Koch, Kevin and Tiefenbeck, Verena and Liu, Shu and Berger, Thomas and Fleisch, Elgar and Wortmann, Felix},
  booktitle={Proceedings of the 2021 CHI conference on human factors in computing systems},
  pages={1--15},
  year={2021}
}

@inproceedings{kou2019turn,
  title={Turn to the self in human-computer interaction: care of the self in negotiating the human-technology relationship},
  author={Kou, Yubo and Gui, Xinning and Chen, Yunan and Nardi, Bonnie},
  booktitle={Proceedings of the 2019 CHI Conference on Human Factors in Computing Systems},
  pages={1--15},
  year={2019}
}

@article{slovak2023designing,
  title={Designing for emotion regulation interventions: an agenda for HCI theory and research},
  author={Slovak, Petr and Antle, Alissa and Theofanopoulou, Nikki and Daud{\'e}n Roquet, Claudia and Gross, James and Isbister, Katherine},
  journal={ACM Transactions on Computer-Human Interaction},
  volume={30},
  number={1},
  pages={1--51},
  year={2023},
  publisher={ACM New York, NY}
}

@inproceedings{agapie2024hcipublichealth,
author = {Agapie, Elena and Karkar, Ravi and Aung, Tricia and Burgess, Eleanor R. and Chinguwa, Munyaradzi Joel and Graham, Andrea K and Klasnja, Predrag and Lyon, Aaron and McCall, Terika and Munson, Sean A. and Nunes, Francisco and Osterhage, Katie},
title = {Conducting Research at the Intersection of HCI and Health: Building and Supporting Teams with Diverse Expertise to Increase Public Health Impact},
year = {2024},
isbn = {9798400703317},
publisher = {Association for Computing Machinery},
address = {New York, NY, USA},
url = {https://doi.org/10.1145/3613905.3636298},
doi = {10.1145/3613905.3636298},
booktitle = {Extended Abstracts of the CHI Conference on Human Factors in Computing Systems},
articleno = {463},
numpages = {6},
location = {Honolulu, HI, USA},
series = {CHI EA '24}
}

@inproceedings{avellino2025envisioning,
  title={Envisioning the Future of Interactive Health},
  author={Avellino, Ignacio and Kuo, Pei-Yi and Foong, Pin Sym and Wiese, Jason and Mentis, Helena M and Munson, Sean A and Wallace, James R and Singh, Aneesha and Miller, Andrew D and Epstein, Daniel A and others},
  booktitle={Proceedings of the Extended Abstracts of the CHI Conference on Human Factors in Computing Systems},
  pages={1--5},
  year={2025}
}

@inproceedings{epstein2023symposium,
  title={Symposium: Workgroup on interactive systems in healthcare (WISH)},
  author={Epstein, Daniel A and O'Kane, Aisling Ann and Miller, Andrew D},
  booktitle={Extended Abstracts of the 2023 CHI Conference on Human Factors in Computing Systems},
  pages={1--4},
  year={2023}
}

@inproceedings{sengers2005reflective,
  title={Reflective design},
  author={Sengers, Phoebe and Boehner, Kirsten and David, Shay and Kaye, Joseph'Jofish'},
  booktitle={Proceedings of the 4th decennial conference on Critical computing: between sense and sensibility},
  pages={49--58},
  year={2005}
}

@inproceedings{bhattacharjee2023investigating,
  title={Investigating the Role of Context in the Delivery of Text Messages for Supporting Psychological Wellbeing},
  author={Bhattacharjee, Ananya and Williams, Joseph Jay and Meyerhoff, Jonah and Kumar, Harsh and Mariakakis, Alex and Kornfield, Rachel},
  booktitle={Proceedings of the 2023 CHI Conference on Human Factors in Computing Systems},
  year={2023},
  doi={10.1145/3544548.3580774}
}

@article{chow2023feeling,
  title={Feeling Stressed and Unproductive? A Field Evaluation of a Therapy-Inspired Digital Intervention for Knowledge Workers},
  author={Chow, Kevin and Fritz, Thomas and Holsti, Liisa and Barbic, Skye and McGrenere, Joanna},
  journal={ACM Transactions on Computer-Human Interaction},
  volume={31},
  number={1},
  year={2023},
  doi={10.1145/3609330}
}

@inproceedings{cai2023listen,
  title={{``Listen to Music, Listen to Yourself''}: Design of a Conversational Agent to Support Self-Awareness While Listening to Music},
  author={Cai, Wanling and Jin, Yucheng and Zhao, Xianglin and Chen, Li},
  booktitle={Proceedings of the 2023 CHI Conference on Human Factors in Computing Systems},
  year={2023},
  doi={10.1145/3544548.3581427}
}

@inproceedings{10.1145/3544548.3581332,
author = {Stefanidi, Evropi and Bentvelzen, Marit and Wo{\'z}niak, Pawe{\l} W. and Kosch, Thomas and Wo{\'z}niak, Miko{\l}aj P. and Mildner, Thomas and Schneegass, Stefan and M{\"u}ller, Heiko and Niess, Jasmin},
title = {Literature Reviews in HCI: A Review of Reviews},
year = {2023},
isbn = {9781450394215},
publisher = {Association for Computing Machinery},
address = {New York, NY, USA},
url = {https://doi.org/10.1145/3544548.3581332},
doi = {10.1145/3544548.3581332},
booktitle = {Proceedings of the 2023 CHI Conference on Human Factors in Computing Systems},
articleno = {509},
numpages = {24},
location = {Hamburg, Germany},
series = {CHI '23}
}

@inproceedings{woo2024notalone,
  title={{``I'm not alone in that battle''}: Designing Mobile AR for Mental Health Communication and Community Connectedness},
  author={Woo, Rachel and Harley, Daniel and Wallace, James R.},
  booktitle={Proceedings of the 2024 ACM Designing Interactive Systems Conference},
  year={2024},
  doi={10.1145/3643834.3661626}
}

@inproceedings{sakel2024social,
  title={The Social Journal: Investigating Technology to Support and Reflect on Meaningful Social Interactions},
  author={Sakel, Sophia and Blenk, Tabea and Schmidt, Albrecht and Haliburton, Luke},
  booktitle={Proceedings of the 2024 CHI Conference on Human Factors in Computing Systems},
  year={2024},
  doi={10.1145/3613904.3642411}
}

@article{barkercanler2024flexible,
  title={Flexible Minimalist Self-Tracking to Support Individual Reflection},
  author={Barker-Canler, Matthew and Gooch, Daniel and van der Linden, Janet and Petre, Marian},
  journal={ACM Transactions on Computer-Human Interaction},
  volume={31},
  number={3},
  pages={1--35},
  year={2024},
  doi={10.1145/3660339}
}

@inproceedings{kim2025lino,
  title={Lino: An Interactive System for Daily Mood Recordings Supporting Meaning-Making through Single Stroke Drawing Approach},
  author={Kim, Nanum and Jang, Sangsu and Kim, Hansol and Shin, Dayoung and Park, Young-Woo},
  booktitle={Proceedings of the 2025 ACM Designing Interactive Systems Conference},
  pages={2255--2269},
  year={2025},
  doi={10.1145/3715336.3735675}
}

@inproceedings{delatorre2025sonora,
  title={Sonora: Human-AI Co-Creation of 3D Audio Worlds and its Impact on Anxiety and Cognitive Load},
  author={De La Torre, Fernanda M. and Hernandez, Javier and Wilson, Andrew D. and Amores, Judith},
  booktitle={Proceedings of the 2025 CHI Conference on Human Factors in Computing Systems},
  year={2025},
  doi={10.1145/3706598.3713316}
}

@article{matheus2024ommie,
  title={Ommie: The Design and Development of a Social Robot for Anxiety Reduction},
  author={Matheus, Kayla and V{\'a}zquez, Marynel and Scassellati, Brian},
  journal={ACM Transactions on Human-Robot Interaction},
  volume={14},
  number={2},
  year={2024},
  doi={10.1145/3706122}
}

@article{spitale2025vita,
  title={VITA: A Multi-Modal LLM-Based System for Longitudinal, Autonomous, and Adaptive Robotic Mental Well-Being Coaching},
  author={Spitale, Micol and Axelsson, Minja and Gunes, Hatice},
  journal={ACM Transactions on Human-Robot Interaction},
  volume={14},
  number={2},
  pages={1--28},
  year={2025},
  doi={10.1145/3712265}
}

@inproceedings{yu2025creatively,
  title={Creatively Supporting Mental Wellbeing: A Tangible Toolkit to Scaffold Self-Tracking through Mindful Colouring},
  author={Yu, Jingxin and Ayobi, Amid and Marshall, Paul and O'Kane, Aisling Ann},
  booktitle={Proceedings of the Nineteenth International Conference on Tangible, Embedded, and Embodied Interaction},
  year={2025},
  doi={10.1145/3689050.3704944}
}

@inproceedings{chandra2026tech,
  title={{``Tech''} a Deep Breath: Technology-guided Breathing Practise With or Without a Social Robot for Psychological and Emotional Well-Being},
  author={Chandra, Shruti and Pasupuleti, Devasena and Chavez Castaneda, Gerardo and Zheng, Charlie and Avijeet, Priyank and Dixon, Mike J. and Dautenhahn, Kerstin},
  booktitle={Proceedings of the 2026 CHI Conference on Human Factors in Computing Systems},
  pages={1--27},
  year={2026},
  doi={10.1145/3772318.3790706}
}

@inproceedings{mccarren2026exploring,
  title={Exploring the Design of a LLM-Based AI Assistant for Mindfulness Practice With Older Adults},
  author={McCarren, Lucy and Eriksson, Ulrika and Ortiz Mengual, Laura and Kuoppam{\"a}ki, Sanna},
  booktitle={Proceedings of the 2026 CHI Conference on Human Factors in Computing Systems},
  pages={1--16},
  year={2026},
  doi={10.1145/3772318.3790734}
}

@inproceedings{zhang2026vrcalmplus,
  title={VR Calm Plus: Coupling a Squeezable Tangible Interaction with Immersive VR for Stress Regulation},
  author={Zhang, He and Li, Xinyang and Zhou, Xingyu and Fu, Xinyi},
  booktitle={Proceedings of the 2026 CHI Conference on Human Factors in Computing Systems},
  pages={1--22},
  year={2026},
  doi={10.1145/3772318.3790548}
}

@inproceedings{lucero2015using,
  title={Using affinity diagrams to evaluate interactive prototypes},
  author={Lucero, Andr{\'e}s},
  booktitle={IFIP conference on human-computer interaction},
  pages={231--248},
  year={2015},
  organization={Springer}
}

@article{bandura1982self,
  title={Self-efficacy mechanism in human agency.},
  author={Bandura, Albert},
  journal={American psychologist},
  volume={37},
  number={2},
  pages={122},
  year={1982},
  publisher={American Psychological Association}
}

@article{wang2026care,
  title={Care and Capability: How to Temper Excessive Social-Emotional Chatbot Use According to Analogous Professional Practices},
  author={Wang, Tony and Yang, Qian},
  year={2026}
}

@inproceedings{fang2025social,
  title={Social simulation for everyday self-care: Design insights from leveraging VR, AR, and LLMs for practicing stress relief},
  author={Fang, Anna and Chhabria, Hriday and Maram, Alekhya and Zhu, Haiyi},
  booktitle={Proceedings of the 2025 CHI Conference on Human Factors in Computing Systems},
  pages={1--23},
  year={2025}
}

@article{lawrence2024opportunities,
  title={The opportunities and risks of large language models in mental health},
  author={Lawrence, Hannah R and Schneider, Renee A and Rubin, Susan B and Matari{\'c}, Maja J and McDuff, Daniel J and Bell, Megan Jones},
  journal={JMIR Mental Health},
  volume={11},
  number={1},
  pages={e59479},
  year={2024},
  publisher={JMIR Publications Inc., Toronto, Canada}
}

@article{zimmerman2002becoming,
  title={Becoming a self-regulated learner: An overview},
  author={Zimmerman, Barry J},
  journal={Theory into practice},
  volume={41},
  number={2},
  pages={64--70},
  year={2002},
  publisher={Taylor \& Francis}
}

@book{mccarthy2007technology,
  title={Technology as experience},
  author={McCarthy, John and Wright, Peter},
  year={2007},
  publisher={MIT press}
  
}
